\documentclass[12pt,a4paper]{article}
 
\usepackage{jheppub}
\usepackage[usenames,dvipsnames]{xcolor}

\usepackage{array} 
\usepackage{amssymb} 
\usepackage{amsmath}
\usepackage{mathtools}
\usepackage{amsfonts}    
\usepackage{dsfont}
\usepackage{pdfpages}
\usepackage{verbatim}
\usepackage{stmaryrd}
\usepackage{graphicx}
\usepackage{bm}

\usepackage{tensor}
\usepackage{mathrsfs}
\usepackage{textgreek} 
\usepackage[mathscr]{euscript}
\usepackage[smalltableaux]{ytableau}
\ytableausetup{centertableaux}
\usepackage{tikz}
\usetikzlibrary{decorations.pathreplacing, decorations.markings,calc,shapes.misc,decorations.pathmorphing,patterns.meta, math}
\usetikzlibrary{decorations.pathmorphing}
\usetikzlibrary{decorations.markings}
\usepackage{slashed}
\usepackage{datetime}
\usepackage{hyperref}
\hypersetup{
    pdfencoding=unicode,
	colorlinks=true,
	urlcolor=Maroon,
	linkcolor=RoyalBlue,
	citecolor=Maroon,
	pdftitle={The complex Liouville string: the matrix integral},
	pdfauthor={Scott Collier, Lorenz Eberhardt, Beatrix M\"uhlmann, Victor Rodriguez},
	pdfdisplaydoctitle=true,
	pdfstartview=FitH,
	linktocpage=true
}

\usetikzlibrary{shapes}
\usetikzlibrary{arrows}
\usetikzlibrary{decorations.pathmorphing}
\usetikzlibrary{ decorations.markings}
\usetikzlibrary{shapes.misc}
\tikzset{cross/.style={cross out, draw=black, minimum size=2*(#1-\pgflinewidth), inner sep=0pt, outer sep=0pt},
cross/.default={1pt}}
\usetikzlibrary{shapes.geometric}
\tikzset{
    partial ellipse/.style args={#1:#2:#3}{
        insert path={+ (#1:#3) arc (#1:#2:#3)}
    }
}

\usepackage{subcaption}

\newcommand{\ba}{\begin{align}}

\newcommand{\be}{\begin{equation}}
\newcommand{\ee}{\end{equation}}
\def\bd{\begin{tikzpicture}}
\def\ed{\end{tikzpicture}}

\DeclareMathOperator\tr{tr}

\renewcommand\Re{\mathop{\text{Re}}}
\newcommand\Res{\mathop{\text{Res}}}
\newcommand{\bM}{\overline{\mathcal{M}}}
\allowdisplaybreaks

\def\XXint#1#2#3{{\setbox0=\hbox{$#1{#2#3}{\int}$}
     \vcenter{\hbox{$#2#3$}}\kern-.5\wd0}}

\definecolor{light-gray}{gray}{0.75}

\newcommand\SL{\text{SL}}
\newcommand\SU{\text{SU}}

\newcommand\Li{\text{Li}}

\newcommand\CC{\mathbb{C}}

\newcommand\ZZ{\mathbb{Z}}
\newcommand\RR{\mathbb{R}}

\newcommand\CP{\mathbb{CP}}

\newcommand\QQ{\mathbb{Q}}

\renewcommand\d{\text{d}}

\newcommand{\e}{\mathrm{e}}

\renewcommand{\le}{\leqslant}
\renewcommand{\ge}{\geqslant}
\renewcommand{\leq}{\leqslant}
\renewcommand{\geq}{\geqslant}

\newcommand{\xx}{\mathsf{x}}
\newcommand{\yy}{\mathsf{y}}

\DeclareMathOperator*\Disc{Disc}

\DeclareMathOperator{\im}{Im}

\definecolor{vert}{rgb}{0.1367 0.543 0.1367}

\title{The complex Liouville string:\\
the matrix integral}

\author[1]{Scott Collier}\emailAdd{sac@mit.edu}
\author[2]{\!\!, Lorenz Eberhardt}\emailAdd{l.eberhardt@uva.nl}
\author[3,4]{\!\!, Beatrix M\"uhlmann}\emailAdd{beatrix@ias.edu}
\author[5,6]{\!\!, Victor A. Rodriguez}\emailAdd{varodriguez@ucsb.edu}
\affiliation[1]{Center for Theoretical Physics, Massachusetts Institute of Technology, Cambridge, MA 02139, USA}
\affiliation[2]{Institute for Theoretical Physics, University of Amsterdam, PO Box 94485, 1090 GL Amsterdam, The Netherlands}
\affiliation[3]{Department of Physics, McGill University Montr\'eal, H3A 2T8, QC Canada 
}\affiliation[4]{School of Natural Sciences, Institute for Advanced Study, Princeton, NJ 08540, USA}
\affiliation[5]{Joseph Henry Laboratories, Princeton University, Princeton, NJ 08544, USA}\affiliation[6]{Department of Physics, University of California, Santa Barbara, CA 93106, USA}

\abstract{We propose a duality between the complex Liouville string and a two-matrix integral. The complex Liouville string is defined by coupling two Liouville theories with complex central charges $c = 13 \pm i \lambda$ on the worldsheet. The matrix integral is characterized by its spectral curve which allows us to compute the perturbative string amplitudes recursively via topological recursion. 
This duality constitutes a controllable instance of holographic duality. The leverage on the theory is provided by the rich analytic structure of the string amplitudes that we discussed in \cite{paper1} and allows us to perform numerous tests on the duality.
}

\begin{document}

\begin{flushright}
    \hfill{\tt MIT-CTP/5782}
\end{flushright}

\maketitle

\makeatletter
\g@addto@macro\bfseries{\boldmath}
\makeatother

\section{Introduction}

Low-dimensional string theories have proven to be invaluable theoretical laboratories for investigating fundamental aspects of string theory and of quantum gravity. They provide examples where holographic dualities may be derived and understood in complete detail from the string worldsheet. The last couple of years have experienced rapid progress in the derivation and exploration of such holographic dualities, such as in string theory on $\mathrm{AdS}_3$ with pure NS-NS flux \cite{Eberhardt:2018ouy, Eberhardt:2019ywk, Eberhardt:2021vsx}, the $c=1$ string \cite{Balthazar:2017mxh, Balthazar:2019rnh}, topological strings \cite{Gopakumar:2022djw}, and the Virasoro minimal string \cite{Collier:2023cyw}. Each of these instances teaches us new lessons about holography and sharpens our tools to understand richer instances of holography. In the present paper, we derive a new string theory/matrix integral duality. It is much richer than previous string theory/matrix integral dualities yet at the same time under good technical control. It thus represents a significant step up in complexity towards our quest to understand more realistic versions of holography.

\paragraph{The complex Liouville string.} 
In our previous paper \cite{paper1}, we introduced the complex Liouville string. This is a non-critical string theory defined by coupling two complex-conjugate copies of Liouville CFT together with the $\mathfrak{bc}$-ghosts on the worldsheet:
\begin{equation}
\begin{array}{c}
\text{Liouville CFT} \\ \text{$c^+= 13+i\lambda$}
\end{array}
\ \oplus\  
\begin{array}{c} (\text{Liouville CFT})^* \\ \text{$c^- = 13 - i\lambda$} \end{array} \ \oplus\  
\begin{array}{c} \text{$\mathfrak{b}\mathfrak{c}$-ghosts} \\ \text{$c= -26$} \end{array}\, ,
\label{eq:Liouville squared worldsheet definition}
\end{equation}
where $\lambda \in \mathbb{R}_+$.
This defines a fully consistent model of two-dimensional quantum gravity (both on the worldsheet and in target space). Moreover the integrals of worldsheet correlation functions over the moduli space of Riemann surfaces that define the perturbative string amplitudes $\mathsf{A}_{g,n}^{(b)}(p_1,\ldots,p_n)$ converge absolutely. Here $p_j$ labels the Liouville momentum of the external vertex operators. We may think of the string amplitudes as analytic functions of the momenta.

In our previous paper \cite{paper1} we focused on the worldsheet description of the theory. The exact solution of the worldsheet CFT (\ref{eq:Liouville squared worldsheet definition}) \cite{Zamolodchikov:1995aa, Teschner:1995yf} allowed us to deduce the rich analytic structure of the string amplitudes $\mathsf{A}_{g,n}^{(b)}$ viewed as analytic functions of the external momenta $p_j$. This inspired the initiation of a bootstrap program, which harnesses the analytic structure together with other constraints from the worldsheet description to pin down the string amplitudes without explicitly computing the moduli space integrals. We presented the explicit solution of this bootstrap program for low values of $(g,n)$, focusing in particular on the sphere four-point amplitude $\mathsf{A}_{0,4}^{(b)}$ and the torus one-point amplitude $\mathsf{A}_{1,1}^{(b)}$ as worked examples. 

In this paper we will demonstrate that this model admits an equivalent description in terms of a double-scaled \emph{two-matrix integral}. This will allow us to compute the perturbative string amplitudes algorithmically via the topological recursion of the matrix integral, and hence solve the model at the level of string perturbation theory. We will explore the duality at a non-perturbative level in a third paper in this series \cite{paper3}. This paper is an expanded version of the corresponding section of \cite{Collier:2024kmo}.

\paragraph{The $(p,q)$ minimal string and two-matrix integrals.}
An important benchmark and point of comparison that has been explored extensively in the literature is the $(p,q)$ minimal string. This model is defined by coupling the $(p,q)$ Virasoro minimal model to Liouville CFT and the $\mathfrak{bc}$-ghosts on the worldsheet, and we will see that it bears a number of similarities to the complex Liouville string, although there are some essential technical and conceptual differences between the two classes of models. 

For $p,q> 2$, the $(p,q)$ minimal string is conjecturally dual to a double-scaled two-matrix integral \cite{Kazakov:1986hu, Brezin:1989db, Douglas1991, Crnkovic:1989tn} (for reviews, see \cite{Ginsparg:1993is,DiFrancesco:1993cyw, Anninos:2020ccj}). Observables in the relevant class of two-matrix integrals are computed by integrating a pair of $N\times N$ Hermitian matrices weighted by potentials for the two matrices together with a minimal coupling
\begin{equation}
	\langle \cdot \rangle = \int_{\mathbb{R}^{2N^2}}[\d M_1][\d M_2]\, (\cdot)\, \mathrm{e}^{-N\tr\left(V_1(M_1)+V_2(M_2)-M_1M_2\right)}\, .
\end{equation}
In the double-scaling limit, we take $N\to\infty$ and zoom in on a particular region of the spectral curve that characterizes the eigenvalue distribution. In this limit observables in the matrix integral admit a topological genus expansion, and this perturbative expansion is completely fixed by the geometry of the spectral curve. This is facilitated by a recursion relation for the perturbative expansion of the matrix integral resolvents known as topological recursion \cite{Eynard:2007kz}, which is entirely determined by the spectral curve. This is analogous to how the leading density of eigenvalues determines the perturbative expansion of double-scaled single matrix integrals via topological recursion. Since the literature on the topological expansion of two-matrix integrals \cite{Eynard:2002kg, Chekhov:2006vd} is somewhat scattered and the derivation of topological recursion substantially more complicated compared to that of single-matrix integrals, we take some time to carefully review it in this paper.

The dual descriptions of the $(p,q)$ minimal string and its deformations correspond to a particular universality class of matrix integrals involving matrix potentials that are finite-degree polynomials subject to rational double-scalings. In particular, the spectral curve is algebraic and defines a Riemann surface of genus 0 and $\frac{(p-1)(q-1)}{2}$ nodal singularities.\footnote{This can also be seen as a surface of genus $\frac{(p-1)(q-1)}{2}$ in a degeneration limit where all cycles are collapsed to nodal singularities.} In this paper we will argue that the complex Liouville string is dual to a matrix integral characterized by a spectral curve with infinitely many nodal singularities and branch points. We will see that, in contrast to the $(p,q)$ minimal string, this can be engineered via an \emph{irrational} double-scaling of a two-matrix integral involving matrix potentials of infinite degree.

\paragraph{A two-matrix integral for the complex Liouville string.}
The central claim of this paper is that the complex Liouville string is dual to a double-scaled two-matrix integral characterized by the following spectral curve
\begin{equation}\label{eq:spectral curve intro}
	\xx(z) = -2\cos(\pi b^{-1}\sqrt{z}), \quad \yy(z) = 2\cos(\pi b \sqrt{z})\, .
\end{equation}
Here $b$ labels the central charge of one of the worldsheet Liouville CFTs via the usual parameterization $c = 1 + 6(b+b^{-1})^2$. The range of central charges of interest in (\ref{eq:Liouville squared worldsheet definition}) corresponds to $b\in \mathrm{e}^{\frac{\pi i}{4}}\mathbb{R}$.
In contrast to the spectral curve of the $(p,q)$ minimal string this is not algebraic and exhibits infinitely many nodal singularities (points $z^{\pm}$ that map to the same point on the spectral curve) and infinitely many branch points $z_m^*$ where $\d\xx(z_m^*) = 0$. The infinitely many branch points lead to an additional infinite sum in the topological recursion, which renders the resolvents significantly more complicated than those of the matrix integral duals of for example JT gravity \cite{Saad:2019lba} or the Virasoro minimal string \cite{Collier:2023cyw}. 

\paragraph{Feynman diagrams for string amplitudes.}
We claim that moreover there is a simple dictionary between the resolvents $\omega_{g,n}^{(b)}(z_1,\ldots, z_n)$ which are the natural observables of the matrix integral and the string amplitudes $\mathsf{A}_{g,n}^{(b)}(p_1,\ldots,p_n)$ of the complex Liouville string. The relation involves sums over the branch points of the spectral curve and is given in equation \eqref{eq:Agn omegagn relation}.

Theorems of \cite{Eynard:2011ga, Dunin-Barkowski:2012kbi} regarding topological recursion for spectral curves with multiple branch points allow us to express the resolvents in terms of intersection numbers on the moduli space of Riemann surfaces, which we may then translate to intersection theory expressions for the string amplitudes. The result takes the form of a sum over degenerations of the worldsheet Riemann surface (``stable graphs'') given in equation \eqref{eq:Agn through quantum volumes}. Remarkably, for each term in the sum the intersection theory data reassembles into a product of the corresponding ``quantum volumes'' $\mathsf{V}_{g,n}^{(b)}$ of the Virasoro minimal string, which were themselves shown to admit an intersection number representation in \cite{Collier:2023cyw}. We interpret this representation of the string amplitudes as a sum over Feynman diagrams of the closed string field theory of the complex Liouville string, with the VMS quantum volumes playing the role of the on-shell string vertices.

\paragraph{CohFT and TQFT.} We also explain that this structure is the one known as a cohomological field theory (CohFT) in the mathematical literature \cite{Kontsevich:1994qz}. The complex Liouville string thus provides an interesting CohFT of infinite rank. One can associate a 2d TQFT to any CohFT by restricting to the degree 0 piece in cohomology. This TQFT turns out to be $\SU(2)_q$ Yang-Mills theory, which in turn relates the theory to the Schur index of 4d class $\mathcal{S}$ theories.

\paragraph{Topological recursion for string amplitudes.}
Given the simple relation between the two observables, we then translate the topological recursion for the matrix integral resolvents into a recursion relation for the perturbative string amplitudes themselves. The recursion relation, given in equation \eqref{eq:Agn recursion relation}, expresses the string amplitude in terms of a sum of integrals of string amplitudes of lower complexity, corresponding to the different ways of excising a pair of pants with a particular external leg from the worldsheet surface. This may be viewed as a generalization of Mirzakhani's recursion relation for the Weil-Petersson volumes of the moduli space of Riemann surfaces \cite{Mirzakhani:2006fta}. Indeed the recursion relation is remarkably \emph{identical} to the recursion relation for the quantum volumes of the Virasoro minimal string presented in \cite{Collier:2023cyw} --- even the recursion kernel that appears in the integrals is the same. The \emph{only} difference is that the three-point function of the excised pair of pants also appears --- contrary to the case of the corresponding quantum volume $\mathsf{V}_{0,3}^{(b)}=1$, the sphere three-point amplitude $\mathsf{A}_{0,3}^{(b)}$ is a non-trivial function of the momenta that was studied in our previous paper \cite{paper1}.

\paragraph{Tests of the duality.}
Both sides of the proposed duality between the string theory \eqref{eq:Liouville squared worldsheet definition} and the two-matrix integral characterized by the spectral curve \eqref{eq:spectral curve intro} are sufficiently explicit that it is possible to perform many tests directly which collectively are close to constituting a proof of the duality. We list some of them here:
\begin{enumerate}
    \item The string amplitude Feynman rules directly reproduce the low-lying string amplitudes $\mathsf{A}_{0,3}^{(b)}$, $\mathsf{A}_{0,4}^{(b)}$, and $\mathsf{A}_{1,1}^{(b)}$ that were bootstrapped from the worldsheet definition in our first paper \cite{paper1}.
    \item Both the Feynman rules and the topological recursion facilitate the analytic continuation of the string amplitudes to general complex momenta. These representations of the general string amplitudes $\mathsf{A}_{g,n}^{(b)}$ viewed as analytic functions of the momenta manifest the analytic structure --- including an infinite set of poles and an infinite series of discontinuities  --- exactly as predicted from the worldsheet description \cite{paper1}. 
    \item Beyond reproducing the correct analytic structure, the matrix integral representations of the string amplitudes also satisfy the dilaton equation and exhibit the symmetry properties predicted from the worldsheet description. Intriguingly, the $b\to b^{-1}$ duality symmetry, which is a tautological symmetry of Liouville CFT in the worldsheet description, is non-trivial in the matrix integral representation --- it roughly amounts to a symmetry that exchanges $\xx(z)$ and $\yy(z)$ in the spectral curve, which is known as the $x$-$y$ symmetry in topological recursion \cite{Eynard:2007kz}. We will see that it nevertheless follows straightforwardly from the topological recursion for the string amplitudes.
\end{enumerate}
Collectively, we view these tests as even stronger evidence than has been amassed for the conventional $(p,q)$ minimal string/matrix integral dualities. 

\paragraph{Non-perturbative effects.} This paper treats the string theory/matrix integral duality perturbatively. The non-perturbative completion and instanton effects are interesting extensions that will be treated in the third installment of this series of papers \cite{paper3}.

\paragraph{Sine dilaton gravity.} The worldsheet theory can be viewed as a 2d theory of gravity. This theory of gravity is dilaton gravity with a periodic sine potential for the dilaton. As such perturbative string amplitudes can be seen as computing the gravitational path integral of this theory. One particularly interesting aspect of this 2d theory of gravity is that it hosts both AdS and dS vacua and thus the worldsheet theory can be viewed as a rigorous theory of 2d quantum gravity involving de Sitter vacua. We develop this intuition further in \cite{paper4} and show how the structure of the perturbative string amplitudes discussed in this paper can be reproduced from the gravitational path integral.

\paragraph{Integrated cosmological correlators and dS$_3$ holography.} There is yet another connection between the complex Liouville string and de Sitter quantum gravity. The worldsheet Liouville CFT partition functions may be interpreted as defining the wavefunctions of special states in the canonical quantization of pure three-dimensional Einstein gravity with positive cosmological constant. In \cite{paper5}  we will argue that the string amplitudes $\mathsf{A}_{g,n}^{(b)}$ may moreover be interpreted as cosmological correlators of massive particles in dS$_3$ integrated over the metric at future infinity, where the topology of future infinity is that of the worldsheet Riemann surface $\Sigma_{g,n}$. This establishes a precise holographic correspondence in the spirit of dS/CFT \cite{Strominger:2001pn, Maldacena:2002vr,  Anninos:2017eib, Witten:2001kn, Anninos:2011ui} between late-time integrated cosmological correlators in dS$_3$ and the double-scaled two-matrix integral that is the subject of the present paper.

\paragraph{Outline of the paper.}
The rest of this paper is organized as follows. We begin by a somewhat extensive review on two-matrix integrals in section~\ref{sec:two-matrix models}. Two-matrix integrals were completely solved in the mathematical literature in \cite{Eynard:2002kg, Chekhov:2006vd}. This happened after the surge of interest in string theory/matrix integral dualities in the 90's 
\cite{Kazakov:1986hu, Brezin:1989db, Douglas1991, Crnkovic:1989tn}
and in our view the physics literature has not fully caught up with these developments. We then discuss the specific double-scaled two-matrix integral of interest in section~\ref{sec:duality with the worldsheet theory} and explain the precise duality with the worldsheet observables. We discuss the above mentioned checks of the duality in section~\ref{sec:checks} and end with a number of open questions and future directions in \ref{sec:conclusion}. Some background and computations are relegated to the appendices \ref{app:two-matrix models}, \ref{app:intersection theory} and \ref{app:recursion derivation}.

\section{The two-matrix integral} \label{sec:two-matrix models}
The following section provides a significant amount of background on two-matrix integrals. It is not strictly necessary to understand the rest of the paper and readers just interested in the duality of the complex Liouville string to a matrix integral may safely skip to section~\ref{sec:duality with the worldsheet theory}.

\subsection{Why two-matrix integrals?}
Our main conjecture is that the complex Liouville string is dual to a \emph{two-matrix} integral of the form
\begin{equation}\label{eq:2MM}
    \int_{\RR^{2N^2}} [\d M_1][\d M_2] \, \mathrm{e}^{-N\, \tr(V_1(M_1) + V_2(M_2) - M_1 M_2)}~,
\end{equation}
where the integral is over Hermitian matrices $M_1$ and $M_2$ of size $N$. Here $V_1(M_1)$ and $V_2(M_2)$ are entire functions of $M_1$ and $M_2$. We also need to perform a double scaling limit on such a two-matrix integral.
Two-matrix integrals have appeared before as the dual description of the $(p,q)$-minimal string \cite{Kazakov:1986hu,Brezin:1989db, Douglas1991, Crnkovic:1989tn, Eynard:2002kg, Seiberg:2004at, Chekhov:2006vd}. While the specific two-matrix integral appearing here will share many similarities with the minimal string two-matrix integral, it will differ in some crucial ways.
In this paper, we will treat the two-matrix integral \eqref{eq:2MM} in an asymptotic genus expansion, while non-perturbative effects will be discussed in \cite{paper3}.

Let us first give some intuition why a two-matrix integral appears as the dual description of the bulk theory. One can loosely think of the two matrices as being associated to $b$ and $\frac{1}{b}$, and we will see in particular that the $b \to \frac{1}{b}$ duality symmetry is associated to the exchange of the two matrices. More technically, the two-matrix integral will live on the asymptotic boundaries of 2d spacetime. To define these boundaries, we have to specify FZZT boundary conditions in the worldsheet theory, which break the $b \to \frac{1}{b}$ symmetry. Observables will then be associated with single-trace operators in one or the other matrix.
A similar mechanism was described in \cite{Seiberg:2003nm} for the $(p,q)$ minimal string. Asymptotic boundaries will be discussed in more detail both from the 2d spacetime and the worldsheet BCFT points of view in \cite{paper3}.

While this is a nice motivation that one should look at two-matrix models, we could in principle also take suitable double scaling limits on say a three-matrix model or more generally a chain of matrices \cite{Eynard:2003kf}. However, one can already engineer the most general universality classes of random matrices for two-matrix models and thus it suffices to look at that case. 

Let us also notice that for the quadratic potential $V_2(M_2)=\frac{1}{2} M_2^2$, we can integrate out the matrix $M_2$ to reduce the integral to a single matrix integral. This happens in the $(2,p)$ minimal string which indeed can be described by a single matrix integral \cite{Gross:1989vs,Douglas:1989ve,Brezin:1990rb}. In the present case the integral \eqref{eq:2MM} cannot easily be reduced to a single matrix integral.

\subsection{Resolvents and all that}
Let us recall some basic notions of two-matrix integrals. Most of them are straightforward generalizations of the single matrix integral case.
\paragraph{Correlators.} We will define correlation functions of operators
\be 
\langle \mathcal{O}_1 \cdots \mathcal{O}_n \rangle=\int_{\RR^{2N^2}} [\d M_1][\d M_2] \, \prod_{i=1}^n \mathcal{O}_i(M_1, M_2) \, \mathrm{e}^{-N\, \tr(V_1(M_1) + V_2(M_2) - M_1 M_2)}~.
\ee
Assuming that $\mathcal{O}_i$ are single-trace operators, we can decompose such correlators into their connected part by summing over all partitions of the set $\{1,2,\dots,n\}$, e.g.
\begin{subequations}
\begin{align}
    \langle \mathcal{O}_1 \rangle&=\langle \mathcal{O}_1 \rangle_\text{c}~, \\
    \langle \mathcal{O}_1 \mathcal{O}_2 \rangle&=\langle \mathcal{O}_1\mathcal{O}_2 \rangle_\text{c}+\langle \mathcal{O}_1 \rangle_\text{c} \langle \mathcal{O}_2 \rangle_\text{c}~, \\
    \langle \mathcal{O}_1 \mathcal{O}_2 \mathcal{O}_3\rangle&=\langle \mathcal{O}_1\mathcal{O}_2 \mathcal{O}_3\rangle_\text{c}+\langle \mathcal{O}_1 \rangle_\text{c} \langle \mathcal{O}_2 \mathcal{O}_3 \rangle_\text{c}+\langle \mathcal{O}_2 \rangle_\text{c} \langle \mathcal{O}_1 \mathcal{O}_3 \rangle_\text{c}+\langle \mathcal{O}_3 \rangle_\text{c} \langle \mathcal{O}_1 \mathcal{O}_2 \rangle_\text{c}\nonumber\\
    &\quad+\langle \mathcal{O}_1 \rangle_\text{c}\langle \mathcal{O}_2 \rangle_\text{c}\langle \mathcal{O}_3 \rangle_\text{c}~,
\end{align} \label{eq:disconnected to connected correlators}%
\end{subequations}
etc. A connected correlator then has a $\frac{1}{N}$-expansion\footnote{This requires one to normalize the two-matrix integral by the Gaussian model.}
\be 
\langle \mathcal{O}_1 \cdots \mathcal{O}_n\rangle_\text{c}=\sum_{g=0}^\infty \langle \mathcal{O}_1 \cdots \mathcal{O}_n\rangle_g \, N^{2-2g-n}~. \label{eq:genus expansion}
\ee
This can be confirmed by the usual large-$N$ 't Hooft counting. 

\paragraph{Reducing to eigenvalues.} One can reduce the integral \eqref{eq:2MM} to an integral over eigenvalues by diagonalizing the two matrices as $M_j=U_j D_j U_j^{-1}$ with $D_j$ diagonal and $U_j$ unitaries. The integral over the relative unitary $U_1 U_2^{-1}$ is non-trivial. If there are no operator insertions as in \eqref{eq:2MM}, this is the Harish-Chandra-Itzykson-Zuber integral \cite{Itzykson:1979fi}, which can be performed explicitly with the result
\begin{multline} 
\int_{\RR^{2N^2}} [\d M_1][\d M_2] \, \mathrm{e}^{-N\, \tr(V_1(M_1) + V_2(M_2) - M_1 M_2)} \\
\sim \int_{\mathbb{R}^{2N}} \prod_{i=1}^N (\d \lambda_i\, \d \mu_i)\, \Delta_N(\lambda) \Delta_N(\mu) \mathrm{e}^{-N\sum_i (V_1(\lambda_i)+V_2(\mu_i)-\lambda_i \mu_i)}~, \label{eq:HCIZ}
\end{multline}
with $\lambda_i$ and $\mu_i$ the eigenvalues of the two matrices. There is an overall $N$-dependent normalization factor that we suppressed. Finally $\Delta_N(\lambda)$ is the Vandermonde determinant
\be 
\Delta(\lambda)=\prod_{1 \le i<j\le N} (\lambda_i-\lambda_j)~.
\ee
Notice that contrary to the single matrix integral there is only a single power of the Vandermonde determinant for each matrix. \eqref{eq:HCIZ} also holds in the presence of operators which are invariant under separate conjugation of the two matrices,
\be 
\mathcal{O}(U_1 M_1 U_1^{-1},U_2 M_2 U_2^{-1})=\mathcal{O}(M_1,M_2)~,
\ee
but becomes much more complicated for more general observables.

\paragraph{Resolvents.} The main observables we will be interested in are resolvents in one matrix, which we take to be the first,
\be 
R(x)=\tr\frac{1}{x-M_1}~.
\ee
When we have to distinguish quantities in the first matrix, we write $R^{(1)}$.
We will also consider products
\be 
R(x_1,\dots,x_n) \equiv \prod_{i=1}^n R(x_i)~,
\ee
for which following \cite{Stanford:2019vob} we also employ the short-hand notation $R(I)$ with $I=\{x_1,\dots,x_n\}$. We then denote the terms in the genus expansion as
\be 
\langle R(x_1,\dots,x_n) \rangle_\text{c}=\sum_{g=0}^\infty R_{g,n}(x_1,\dots,x_n) N^{2-2g-n}~. \label{eq:resolvents genus expansion}
\ee
\paragraph{Cuts.} For finite values of $N$, $R(x)$ has poles whenever $x$ coincides with one of the eigenvalues of $M_1$. Integrating over $M_1$ will smear out these poles into branch cuts located at the support of the eigenvalues of $M_1$. Thus $R_{g,n}(x_1,\dots,x_n)$ is a multi-valued function in all its arguments.
In particular, the discontinuity of $\langle R(x) \rangle$ gives the density of eigenvalues of the first matrix. To leading order in $\frac{1}{N}$,
\be 
\rho_0(x)=-\frac{1}{2\pi i} (R_{0,1}(x+i \varepsilon)-R_{0,1}(x-i \varepsilon))~, \quad \rho_0(x)=\sum_i \delta(x-\lambda_i)~. \label{eq:density of states discontinuity R01}
\ee
Since we are discussing integrals over Hermitian matrices, the cuts must be located on the real axis.
Thus, $x$ will naturally live on a multi-sheeted cover of the complex plane with potentially several cuts on the real axis. This defines a Riemann surface $\Sigma$, called the spectral curve. We will discuss it further below. We get one distinguished sheet where $x$ initially took values, which is the physical sheet. 

In principle, we can have several cuts and assign some proportion of the eigenvalues to the first cut, some proportion to the second cut, etc. These proportions are the filling fractions. To get a well-defined $\frac{1}{N}$ expansion, one needs to prescribe the values of the filling fractions. Given that the discontinuity of $R_{0,1}$ gives the density of states \eqref{eq:density of states discontinuity R01}, we can measure the filling fraction by integrating $R_{0,1}$ counterclockwise around the cut,
\be 
\text{filling fraction}=\frac{1}{2\pi i} \int_\text{cut} \d x \, R_{0,1}(x)~. \label{eq:filling fractions}
\ee
\paragraph{Large $N$ saddle-point equations.} Let us further discuss the distribution of eigenvalues. At large $N$, we have an effective potential for a pair of eigenvalues $(\lambda_i,\mu_i)$:
\be 
V_\text{eff}(\lambda_i,\mu_i)=V_1(\lambda_i)+V_2(\mu_i)-\lambda_i\mu_i-\frac{1}{N} \sum_{j,j \ne i} \log(\lambda_i-\lambda_j)-\frac{1}{N} \sum_{j,j \ne i} \log(\mu_i-\mu_j)~.
\ee
The saddle-point equations are hence obtained by putting the derivative to zero,
\begin{subequations}
\begin{align} 
\mu_i&=V_1'(\lambda_i)-\frac{1}{N}\sum_{j,j \ne i} \frac{1}{\lambda_i-\lambda_j}~, \\
\lambda_i&=V_2'(\mu_i)-\frac{1}{N}\sum_{j,j \ne i} \frac{1}{\mu_i-\mu_j}~.
\end{align}
\end{subequations}
We recognize the definition of the resolvent and obtain in the continuum limit
\begin{subequations}
\begin{align}
    y&=V_1'(x)-\text{P} \int \frac{\rho_0^{(1)}(x')\, \d x'}{x-x'}~, \\
    x&=V_2'(y)-\text{P} \int \frac{\rho_0^{(2)}(y')\, \d y'}{y-y'}~,
\end{align} \label{eq:saddle point equations}%
\end{subequations}
where P denotes the principal value of the integral. 
Here $x$ and $y$ lie on the eigenvalue support of the two matrices, with $\rho_0^{(1)}(x)$ and $\rho_0^{(2)}(y)$ the corresponding leading densities of eigenvalues. 
We can also rewrite this as
\begin{align}
    2y&=Y(x+i \varepsilon)+Y(x-i \varepsilon)~, \\
    2x&=X(y+i \varepsilon)+X(y-i \varepsilon)~,
\end{align}
with
\be \label{eq: Yz and Xy}
Y(x)=V_1'(x)-R_{0,1}^{(1)}(x)~, \quad X(y)=V_2'(y)-R_{0,1}^{(2)}(y)~.
\ee
These equations are solved with the help of the loop equations.

\subsection{Loop equations} \label{subsec:loop equations}
It is possible to solve the matrix integral perturbatively in $\frac{1}{N}$ thanks to the loop equations. The loop equations can be derived by using that total derivatives integrate to zero; or, alternatively that the matix integral is invariant under change of variables of the matrices $M_1$ and $M_2$. The loop equations take the form \cite{Eynard:2002kg}
\begin{multline}
    \big\langle P(x,y) R(I) \big\rangle=\left\langle \left(y-V_1'(x)+\frac{1}{N} R(x)\right) U(x,y) R(I) \right\rangle\\
    +\frac{1}{N} \sum_{k=1}^n \partial_{x_k} \left \langle \frac{U(x,y)-U(x_k,y)}{x-x_k} R(I \setminus x_k) \right \rangle ~. \label{eq:master loop equation}
\end{multline}
Here,
\begin{subequations}
\begin{align}
    U(x,y)&=\tr \left(\frac{1}{x-M_1} \frac{V_2'(y)-V_2'(M_2)}{y-M_2}\right)+N(x-V_2'(y))~, \label{eq:def U} \\
    P(x,y)&=N(V_2'(y)-x)(V_1'(x)-y)-\tr \left(\frac{V_1'(x)-V_1'(M_1)}{x-M_1} \frac{V_2'(y)-V_2'(M_2)}{y-M_2}\right)+N~.\label{eq:def P}
\end{align}
\end{subequations}
\eqref{eq:master loop equation} is called the master loop equation. For completeness, we included a derivation of \eqref{eq:master loop equation} in appendix \ref{subapp:loop equations}. It can be derived by requiring invariance of the matrix integral \eqref{eq:2MM} under reparametrization of the matrices $M_1$ and $M_2$.
\paragraph{Spectral curve.} Let us consider the case with $I=\emptyset$ and take the large $N$ limit of \eqref{eq:master loop equation}. This gives with the help of the definition \eqref{eq: Yz and Xy}
\be 
    P_0(x,y)=(y-Y(x))U_0(x,y)~. \label{eq:spectral curve loop equations}
\ee
We denoted the genus 0 contribution to $U$ and $P$ by the subscript 0. Notice that the additional terms proportional to $N$ in \eqref{eq:def U} and \eqref{eq:def P} only contribute to the genus 0 piece. The crucial observation is now that $P_0(x,y)$ is an entire function and thus no branch cuts appear after integrating over the matrices. For polynomial potentials, $P_0(x,y)$ is in fact a polynomial. Indeed, $P(x,y)$ does not have any poles in its definition \eqref{eq:def P}. Since the right-hand side vanishes for $y=Y(x)$, we find in particular that
\be 
P_0(x,Y(x))=0~. \label{eq:PxY}
\ee
Recall that $Y(x)$ (\ref{eq: Yz and Xy}) defines a multi-sheeted cover of the $x$-plane. This equation precisely describes $Y(x)$. Notice in particular that for $V_2(M_2)=\frac{1}{2}M_2^2$, $P_0(x,y)$ is quadratic in $y$, which means that the spectral curve is a two-fold cover of the complex plane.

We could have reversed the roles of the two matrices in the derivation. Since the definition of $P_0(x,y)$ is symmetric in the two matrices, we also find that
\be 
P_0(X(y),y)=0~. \label{eq:PXy}
\ee
Thus both of the points $(x,Y(x))$ and $(X(y),y)$ lie on the spectral curve. However, this does not mean that $X$ and $Y$ are inverse functions of each other since they are multivalued. We will get back to this point below. 
\paragraph{Explicit parametrization.} \eqref{eq:PxY} and \eqref{eq:PXy} are implicit parametrizations of the spectral curve. We can choose some direct parametrization by writing $\xx(z)$ and $\yy(z)$ where $z \in \Sigma$. We then have by definition
\be 
P_0(\xx(z),\yy(z))=0~. \label{eq:Pxy}
\ee
$\xx(z)$ and $\yy(z)$ are maps $\xx:\Sigma \to \CP^1$, $\yy:\Sigma \to \CP^1$.
We use also the following notation below. For $z$ on the physical sheet, we write $z^0=z$ and $z^i$ with $i=1,\dots,\mathrm{deg}(V_2)-1$ for the other preimages of $\xx^{-1}(\xx(z))$, i.e. $\xx(z^i)=\xx(z)$.
$\xx$ has a number of branch points $ \d \xx(z_m^*)=0$ labelled by $m$. These will play an important role below. In particular, two branches meet at the branch points, which given \eqref{eq:density of states discontinuity R01} implies that the support of the eigenvalues starts and ends on the branch points.\footnote{We assume that there are only simple branch points.}

\paragraph{Genus and filling fractions.} Let us consider the case where $V_1'$ and $V_2'$ are polynomials of degree $d_1$ and $d_2$, respectively and write
\be 
V_1'(x)=\sum_{k=0}^{d_1} a_k x^k~, \quad V_2'(y)=\sum_{k=0}^{d_2} b_k y^k~. \label{eq:V1p V2p Taylor coefficients}
\ee
Then $P_0(x,y)$ is a polynomial of degree $d_1+1$ in $x$ and $d_2+1$ in $y$. 
Notice that in view of the definition \eqref{eq:def P}, knowledge of $P_0(x,y)$ completely determines the potentials from the coefficients of $x^{d_1}$ and $y^{d_2}$. 
Also, the coefficients of $x^{d_1+1}y^{j\neq 0}$ and of $x^{i\neq 0}y^{d_2+1}$ vanish by definition. 
The rest of $P_0$ is new data, except for the coefficient of $x^{d_1-1}y^{d_2-1}$, which follows from the definition \eqref{eq:def P}. Thus $P_0$ contains $d_1 d_2-1$ undetermined coefficients. As we shall now explain, they corresponds to the additional data of the filling fractions \eqref{eq:filling fractions}.

For generic choices of potentials and $P_0$, the resulting spectral curve has genus $d_1 d_2-1$. However, for special choices of the potentials and $P_0$, the curve can be singular and the topological genus can be lower. This was first observed in \cite{Kazakov:2002yh} and is an application of Baker's theorem \cite{Baker}. It states that the geometric genus of a projective plane curve generically is the number of integer lattice points in the interior of the Newton polygon of the irreducible polynomial defining it. In the present case, the Newton polytope is the convex polytope spanned by the vertices
\be 
\{(0,0),\, (d_1+1,0),\, (d_1,d_2),\, (0,d_2+1) \}~, \label{eq:Newton polytope}
\ee
which contains the lattice points $(m,n)$ with $1 \le m \le d_1$, $1 \le n \le d_2$, except for $(d_1,d_2)$. Thus there are $d_1 d_2-1$ interior lattice points and the result follows.

We can introduce a canonical homology basis of $A_I$ and $B_I$ cycles with $I=1,2,\dots,\textsl{g}=d_1 d_2-1$ satisfying the standard intersection relations
\be 
A_I \cap A_J=0~, \quad B_I \cap B_J=0~, \quad A_I \cap B_J=\delta_{IJ}~.
\ee
We use a different font for the genus of the spectral curve to avoid confusions with the genus appearing in the expansion \eqref{eq:genus expansion}.
As described above, the filling fractions are obtained by integrating $R_{0,1}(x)$ around a cut. Alternatively, we can integrate $Y(x)$ around a cut since $V_1'(x)$ does not have a discontinuity. We can choose a basis of $A_I$ cycles that encircle the cuts counterclockwise and compute the filling fractions via 
\be 
\varepsilon_I=-\frac{1}{2\pi i} \int_{A_I} \yy(z) \d \xx(z)~.
\ee
Thus, there are $\textsl{g}=d_1d_2-1$ many filling fractions and the data of specifying $P_0$ precisely corresponds to the data of the filling fractions.

Let us also note that the filling fractions are set at leading order in $N$ and are not corrected at subleading orders. This means that 
\be 
\int_{A_I} R_{g,n}(\xx(z_1),\dots,\xx(z_n)) \, \d \xx(z_1)=0 \label{eq:A cycle integral trivial}
\ee
except for $(g,n)=(0,1)$.

\paragraph{Rational parametrization and nodal singularities.} In the case of interest, the spectral curve will turn out to have genus 0 and all cycles are collapsed. This means that there are $d_1d_2-1$ nodal singularities, i.e.\ $d_1d_2-1$ solutions to the equation
\be
P_0(x,y)=\partial_x P_0(x,y)=\partial_y P_0(x,y)=0~. \label{eq:singular points}
\ee
These conditions determine $P_0$ already completely. This in particular implies that there exists a rational parametrization of the spectral curve, i.e. $z$ takes value in $\CP^1$. $\xx(z)$ and $\yy(z)$ are then degree $d_2+1$ and degree $d_1+1$ maps from $\CP^1$ to itself. Notice that $P_0(x,y)$ has the form
\be \label{eq:P0 poly}
P_0(x,y)=-a_{d_1} x^{d_1+1}-b_{d_2} y^{d_2+1}+a_{d_1}b_{d_2} x^{d_1} y^{d_2}+\sum_{\begin{subarray}{c} 0 \le m \le d_1 \\ 0 \le n \le d_2 \\ (m,n) \ne (d_1,d_2) \end{subarray}} a_{m,n} x^m y^n~,
\ee
where the appearing exponents all lie inside the Newton polygon \eqref{eq:Newton polytope}. Suppose $\xx(z)$ has a pole of order $p_1$ at $z_\text{pole}$ and $\yy(z)$ a pole of order $p_2$ at $z_\text{pole}$. Then the first three terms in $P_0(\xx(z),\yy(z))$ have poles of order $(d_1+1)p_1$, $(d_2+1)p_2$ and $d_1 p_1+d_2 p_2$, while all other terms have subleading poles. The leading pole order has to cancel, which implies that $(d_1+1)p_1=d_1 p_1+d_2 p_2$ or $(d_2+1)p_2=d_1 p_1+d_2 p_2$. Using that $p_1 \le d_2+1$ and $p_2 \le d_1+1$ we find in the former case $p_1=d_2$ and $p_2=1$, while in the latter case $p_1=1$ and $p_2=d_1$. Since the degree of the maps is $d_2+1$ and $d_1+1$ respectively, there are hence precisely two poles, one of each kind. We can choose the coordinate $z$ such that these two poles are at 0 and $\infty$. The most general such maps take the form
\be 
\xx(z)=\gamma z+\sum_{k=0}^{d_2} \alpha_k z^{-k}~, \quad \yy(z)=\frac{\gamma}{z}+\sum_{k=0}^{d_1} \beta_k z^k~. \label{eq:x and y expansion}
\ee
We used the remaining scaling symmetry in $z$ to put the two coefficients $\gamma$ equal. 

There is one more condition on the coefficients. Consider the holomorphic differential $\yy(z) \d \xx(z)$. We can use that $z \to \infty$ implies $\xx(z) \to \infty$. But the resolvent decays as $\frac{1}{x}$ for large $x$, leading to
\begin{align}
    \yy(z)\sim V_1'(\xx(z))-\frac{1}{\xx(z)}~, \quad \text{as}\ z \to \infty~.
\end{align}
This implies that
\be 
\Res_{z=\infty} \yy(z) \d \xx(z) = - \Res_{z=\infty} \frac{\d \xx(z)}{\xx(z)}=-\Res_{x=\infty} \frac{\d x}{x}=1~.
\ee
Writing out the left-hand-side explicitly leads to
\be 
\gamma^2-\sum_{k=1}^{\min(d_1,d_2)} k \alpha_k \beta_k=-1~.
\ee
At this point, the $\alpha_k$'s and $\beta_k$'s determine both $P_0(x,y)$ and the potentials completely by inserting in \eqref{eq:Pxy}.

\paragraph{Propagator.} Consider next the special case $I=\{x'\}$ in the loop equations. We put $x=\xx(z)$, $x'=\xx(z')$ and $y=\yy(z)$ and restrict to connected parts. Extracting the genus 0 part then leads to
\begin{multline} 
R_{0,2}(\xx(z),\xx(z')) U_0(\xx(z),\yy(z))+\left(\frac{\partial \xx(z')}{\partial z'} \right)^{-1}\frac{\partial}{\partial z'} \left(\frac{U_0(\xx(z),\yy(z))-U_0(\xx(z'),\yy(z))}{\xx(z)-\xx(z')}\right)\\
=\text{analytic in $\xx(z)$}~.\label{eq:propagator loop equation large N}
\end{multline}
We see from \eqref{eq:propagator loop equation large N} that $R_{0,2}(\xx(z),\xx(z'))$ has a double pole when $\xx(z)=\xx(z')$ but $z \ne z'$. Let us note that \eqref{eq:spectral curve loop equations} implies that $U_0(\xx(z'),\yy(z))=0$ when $\xx(z)=\xx(z')$ but $z \ne z'$, since putting $x=\xx(z')$ and $y=\yy(z)$ leads to
\be 
0=P_0(\xx(z),\yy(z))=P_0(\xx(z'),\yy(z))=(\yy(z)-\yy(z')) U_0(\xx(z'),\yy(z))~,
\ee
where the LHS vanishes by construction \eqref{eq:Pxy}.
Since $\yy(z) \ne \yy(z')$, it follows that $U_0(\xx(z'),\yy(z))=0$. Thus \eqref{eq:propagator loop equation large N} implies that
\be 
R_{0,2}(\xx(z),\xx(z')) \sim -\frac{1}{(\xx(z)-\xx(z'))^2} \label{eq:double pole R02}
\ee
when $\xx(z) \to \xx(z')$ but $z \ne z'$. Let us also discuss what happens when $\xx(z)$ approaches a branch point. Then $U_0$ can have square root singularities in $\xx(z)$, just like the resolvent. This means that we should look at the quantity
\be 
R_{0,2}(\xx(z),\xx(z')) \, \d \xx(z) \, \d \xx(z')
\ee
which is a well-defined meromorphic differential on the spectral curve in both arguments. This then also makes the singularity \eqref{eq:double pole R02} coordinate independent. The combination
\be 
\left(R_{0,2}(\xx(z),\xx(z'))+\frac{1}{(\xx(z)-\xx(z'))^2}\right) \d \xx(z) \, \d \xx(z')
\ee
then only has a singularity at $z=z'$, which behaves like $\frac{\d z \, \d z'}{(z-z')^2}$. Furthermore all its $A$-cycle integrals vanish according to \eqref{eq:A cycle integral trivial}. This object is known as the Bergman kernel $B(z,z')$ on the spectral curve and is uniquely determined by these conditions. 
In the case where the curve has genus 0, this kernel is simply
\be 
B(z,z')=\frac{\d z\, \d z'}{(z-z')^2}~.
\ee
Thus we conclude
\be 
R_{0,2}(\xx(z),\xx(z'))\d \xx(z) \, \d \xx(z')=\frac{\d z\, \d z'}{(z-z')^2}-\frac{\d \xx(z) \, \d \xx(z')}{(\xx(z)-\xx(z'))^2}~. \label{eq:propagator}
\ee

\paragraph{Uniqueness.} The loop equations can be solved in principle by induction over $2g+|I|$. To see this, rewrite \eqref{eq:master loop equation} first in terms of connected quantities, which takes the form
\begin{align}
    \left\langle P(x,y) R(I) \right\rangle_\text{c}&=\frac{1}{N}\left\langle  U(x,y) R(x,I) \right\rangle_\text{c}\nonumber\\
    &\quad+ \sum_{J \subseteq I} \left\langle \big(y-V_1'(x)+\tfrac{1}{N} R(x)\big) R(J) \right \rangle_\text{c} \left \langle U(x,y) R(J^c) \right \rangle_\text{c} \nonumber\\
    &\quad+\frac{1}{N} \sum_{k=1}^n \partial_{x_k} \left \langle \frac{U(x,y)-U(x_k,y)}{x-x_k} R(I \setminus x_k) \right \rangle_\text{c} ~.  \label{eq:master loop equation connected}
\end{align}
This equation can be proved from \eqref{eq:master loop equation} by induction over $|I|$. If we expand \eqref{eq:master loop equation} into connected components, many terms can be removed thanks to \eqref{eq:master loop equation connected} for sets $|J| <|I|$ which holds by induction. The remaining equation is \eqref{eq:master loop equation connected}. 

We can then further expand the quantities in $\frac{1}{N}$
\begin{subequations}
\begin{align} 
\left\langle P(x,y) R(I) \right\rangle_\text{c}&=\sum_{g=0}^\infty N^{1-2g-|I|} P_g(x,y,I)~, \\
\left\langle U(x,y) R(I) \right\rangle_\text{c}&=\sum_{g=0}^\infty N^{1-2g-|I|} U_g(x,y,I)~, \\
\left\langle R(I) \right\rangle_\text{c}&=\sum_{g=0}^\infty N^{2-2g-|I|} R_g(I)~.
\end{align}
\end{subequations}
Notice that by definition $P_g(x,y,I)$ is a polynomial of order $d_2-1$ in $y$ (except for $g=0$ and $I=\emptyset$, where it is of order $d_2+1$ as discussed above).
Inserting this into the connected loop equations gives
\begin{align}
    P_g(x,y,I)&=U_{g-1}(x,y,I \cup x)+\sum_{h=0}^g\sum_{J \subseteq I} \big(R_h(x,J)+\delta_{h,0}\delta_{J,\emptyset}(y-V_1'(x))\big) U_{g-h}(x,y,J^c)  \nonumber\\
    &\quad+\sum_{k=1}^n \partial_{x_k} \frac{U_g(x,y,I \setminus x_k)-U_g(x_k,y,I \setminus x_k)}{x-x_k} ~. \label{eq:master loop equation genus expansion}
\end{align}
Let us now show that this is a recursion relation for $U_g(x,y,I)$ and $R_g(x,I)$. Suppose that we know $U_h(x,y,J)$ and $R_h(x,J)$ for all $2h+|J| < 2g+|I|$. Let us write $x=\xx(z)$ and $y=\yy(z')$. \eqref{eq:master loop equation genus expansion} becomes then schematically
\begin{multline} 
P_g(\xx(z),\yy(z'),I)=\text{known}+(\yy(z')-\yy(z))U_g(\xx(z),\yy(z'),I)\\ +R_g(\xx(z),I) U_0(\xx(z),\yy(z'))~, \label{eq:Ug recursion schematic}
\end{multline}
where `known' stands for expressions that are known by recursion. 
 We can then compute $U_g(x,y,|I|)$ and $R_g(x,|I|)$ via the following steps:
\begin{enumerate}
    \item We first determine $R_g(\xx(z),I)$. For this purpose, put $z'=z$. Then
    \be 
        R_g(\xx(z),I)U_0(\xx(z),\yy(z))=P_g(\xx(z),\yy(z),I)+\text{known}~. \label{eq:resolvent recursion schematic}
    \ee
    We have an explicit formula for $U_0(\xx(z),\yy(z))$ in terms of the spectral curve, see eq.~\eqref{eq:spectral curve loop equations}. It in particular implies that $U_0(\xx(z),\yy(z)) \to 0$ for $z \to z_m^*$ a branch point. This means that $R_g(\xx(z),I)$ will only have singularities at branch points. It also means that $R_g(\xx(z),I) \d \xx(z)$ is a well-defined meromorphic differential on the spectral curve. One can compute all the singular pieces of this differential from \eqref{eq:resolvent recursion schematic}. The regular piece is fixed by requiring that the $A$-cycle integrals vanish, \eqref{eq:A cycle integral trivial}. This is explained more systematically in \cite{Eynard:2005kc}.
    \item Once $R_g(\xx(z),I)$ is known, we can solve \eqref{eq:resolvent recursion schematic} for $P_g(\xx(z),\yy(z),I)$. Since $P_g(\xx(z),\yy(z),I)$ is a polynomial of degree $d_2-1$ in $y$, this is actually more than enough to determine it completely. Indeed we tautologically also know $P_g(\xx(z),\yy(z^i),I)=P_g(\xx(z^i),\yy(z^i),I)$ since $P_g$ is single-valued and $\xx(z^i)=\xx(z)$. This gives $d_2+1$ values of $y$ for which we know $P_g(x,y,I)$, which is enough to reconstruct the $d_2$ coefficients of the polynomial in $y$.
    \item Finally, it is trivial to solve \eqref{eq:Ug recursion schematic} for general $x$ and $y$ for $U_g(x,y,I)$.
\end{enumerate}
Let us note that we got slightly more than what we needed. We did not need to assume the $P_g(x,y,I)$ is a polynomial of degree $d_2-1$, but only of degree $d_2$ in $y$. This is important in the derivation of the analyticity properties required for topological recursion, see appendix~\ref{subapp:analyticity resolvents}.

\subsection{Topological recursion} \label{subsec:top rec review}
A remarkable property of two-matrix integrals is that the resolvents \eqref{eq:resolvents genus expansion} can be recursively determined from the knowledge of the spectral curve and one can bypass actually solving the loop equations also for $U_g(x,y,I)$ and $P_g(x,y,I)$ in which we are ultimately not interested. The resulting recursion relation is topological recursion. We now explain this recursion and the detailed derivation can be found in appendix~\ref{subapp:analyticity resolvents} and \ref{subapp:top rec derivation}.
\paragraph{Definition of $\omega_{g,n}$.} We have already seen how $R_{0,2}$ is completely determined from the spectral curve, see \eqref{eq:propagator}. A crucial observation is that
\be 
\omega_{g,n}(z_1,\dots,z_n)\equiv R_{g,n}(\xx(z_1),\dots,\xx(z_n))\,  \d \xx(z_1) \cdots \d \xx(z_n) \label{eq:omegagn definition}
\ee
is a well-defined meromorphic multi-differential on the spectral curve $\Sigma$. We define $\omega_{0,1}(z)$ and $\omega_{0,2}(z_1,z_2)$ slightly differently as follows,
\begin{subequations}
\begin{align} 
\omega_{0,1}(z)&=\big(R_{0,1}(\xx(z))-V_1'(\xx(z))\big)\, \d \xx(z)=-\yy(z) \d \xx(z)~, \label{eq:omega01 definition}\\
\omega_{0,2}(z,z')&= \left(R_{0,2}(\xx(z),\xx(z'))+\frac{1}{(\xx(z)-\xx(z'))^2}\right)\, \d \xx(z)\, \d \xx(z')~. \label{eq:omega02 definition}
\end{align}  \label{eq:omega01 02 definition}%
\end{subequations}
The fact that $\omega_{g,n}$ is a differential on the spectral curve follows recursively through the master loop equation \eqref{eq:master loop equation}. We saw this explicitly for $\omega_{0,2}$. We explain this for completeness in appendix~\ref{subapp:analyticity resolvents} following \cite{Chekhov:2006vd}. 

\paragraph{The recursion kernel.} 
A crucial ingredient in the topological recursion formula is the recursion kernel. Let $z_m^*$ be an enumeration of the branch points of $\xx(z)$, i.e.
\be 
\d \xx(z_m^*)=0~.
\ee
We assume that all branch points are simple. By definition, two branches of $\xx(z)$ meet at $z=z_m^*$. This means that there is a second $z^i$ for some $i$ on a different sheet that also tends to $z_m^*$. Let us write $z^i=\sigma_m(z)$. $\sigma_m$ is called the local Galois involution at the branch point $z_m^*$. It is defined by the two properties
\be \label{eq: Galois inversion}
\xx(z)=\xx(\sigma_m(z))~, \quad \sigma_m(z_m^*)=z_m^*
\ee
for $z$ in a neighborhood of $z_m^*$.

We then define the recursion kernel as
\be 
K_m(z_1,z)\equiv\frac{\frac{1}{2} \int_{z'=\sigma_m(z)}^{z}\omega_{0,2}(z_1,z')}{\omega_{0,1}(z)-\omega_{0,1}(\sigma_m(z))}=\frac{\frac{1}{z-z_1}-\frac{1}{\sigma_m(z)-z_1}}{\yy(z)-\yy(\sigma_m(z))} \frac{\d z_1}{2 \d \xx(z)}~, \label{eq:definition recursion kernel}
\ee
 where we decoded the definition in the second expression explicitly for the genus 0 case. 

 \paragraph{Recursion relation.} The statement of topological recursion is that the differentials $\omega_{g,n}$ can be recursively determined from $\omega_{0,1}$ and $\omega_{0,2}$ via the topological recursion formula
\begin{multline}
    \omega_{g,n}(z_1,\dots,z_n)=\sum_{m\text{ branch points}}\Res_{z=z^*_m} K_m(z_1,z) \bigg(\omega_{g-1,n+1}(z,\sigma_m(z),z_2,\dots,z_n)\\
    +\sum_{h=0}^g \sum_{{\substack{\mathcal{I}\cup \mathcal{J}=\{z_2,\ldots z_n\}\\ \{h,\mathcal{I}\}\neq \{0,\emptyset\}\\ \{h,\mathcal{J}\}\neq \{g,\emptyset\}}}}\omega_{h,1+|\mathcal{I}|}(z,\mathcal{I})\omega_{g-h, 1+|\mathcal{J}|}(\sigma_m(z),\mathcal{J})\big]\bigg)~.\label{eq:topological recursion}
\end{multline}
Note that this is much simpler than the procedure outlined above for solving the loop equations recursively.
\paragraph{Dilaton and string equation.} The differentials $\omega_{g,n}$ satisfy two simple relations. They are consequences of the topological recursion \eqref{eq:topological recursion} and take the form 
\begin{subequations}
\begin{align}
    \sum_{m} \Res_{z_{n+1}=z_m^*} F_{0,1}(z_{n+1}) \omega_{g,n+1}(\boldsymbol{z},z_{n+1})&=(2-2g-n) \omega_{g,n}(\boldsymbol{z})~,  \label{eq:topological recursion dilaton equation}\\
    \sum_{m} \Res_{z_{n+1}=z_m^*} \xx(z_{n+1})^k \yy(z_{n+1}) \omega_{g,n+1}(\boldsymbol{z},z_{n+1})&=-\sum_{j=1}^n \d z_j \, \partial_{z_j} \left(\frac{\xx(z_j)^k \omega_{g,n}(\boldsymbol{z})}{\d \xx(z_j)}\right)~. \label{eq:topological recursion string equation}
\end{align} \label{eq:topological recursion dilaton and string equations}%
\end{subequations}
Here, $\d F_{0,1}=\omega_{0,1}$ and $k=0,1$. We also wrote $\boldsymbol{z}=\{z_1,\dots,z_n\}$. These equations are known as the dilaton and string equation, respectively. In particular, we can use \eqref{eq:topological recursion dilaton equation} to define $\omega_{g,0}$ for $g \ge 2$. The definition of $\omega_{1,0}$ and $\omega_{0,0}$ is more subtle \cite{Eynard:2007kz}. A proof of these two equations can be found in \cite[Corrolary 4.1, Theorem 4.7]{Eynard:2007kz}.\footnote{Notice that \cite[Corrolary 4.1]{Eynard:2007kz} is stated incorrectly in the main text, but the proof is correct.}

\paragraph{$x$-$y$ symmetry.} Consider $\omega_{g,0}$, which are the genus $g$ free energies of the two-matrix integral. The definition through the matrix integral treats $\xx(z)$ and $\yy(z)$ on completely equal footing, which means that $\omega_{g,0}$ could be computed from the topological recursion as described above, or alternatively through the topological recursion with the roles of $\xx(z)$ and $\yy(z)$ exchanged. This property is highly non-obvious from the topological recursion \eqref{eq:topological recursion}. It was formally proven in \cite{Eynard:2007nq}. We will see below that for the case of interest, this symmetry extends to a certain integral transform of $\omega_{g,n}$, which will be identified with the string amplitudes.

\subsection{Double scaling} \label{subsec:double scaling}
The two-matrix integral of interest is a double-scaled two-matrix integral. This means that we zoom in on a particular region of the spectral curve.

\paragraph{Rational double scalings.} Suppose that we tune the coefficients in the potential and the filling fractions such that there is a special point where the relation between $x$ and $y$ locally reads
\be 
(x-x_*)^p =\text{const.}\times  (y-y_*)^q+ \dots
\ee
for $p$ and $q$ coprime positive integers. This requires that $\frac{(p-1)(q-1)}{2}$ nodal points collide on the spectral curve.\footnote{One can check this by parametrizing locally $\xx(z)=x_*+\text{const.}\, z^q$, $\yy(z)=y_*+\text{const.}\, z^p$. One then perturbs this equation slightly so that $\xx(z)$ and $\yy(z)$ become generic polynomials of degree $q$ and $p$ respectively. Nodal points correspond to pair of points with $z_1 \ne z_2$ such that $\xx(z_1)=\xx(z_2)$ and $\yy(z_1)=\yy(z_2)$. Hence we are searching for simultaneous solutions to the system of equations
\be 
\frac{\xx(z_1)-\xx(z_2)}{z_1-z_2}=0~, \quad \frac{\yy(z_1)-\yy(z_2)}{z_1-z_2}=0~,
\ee 
which are polynomials of degree $q-1$ and $p-1$ respectively. By Bezout's theorem, there will be generically $(p-1)(q-1)$ solutions. Since $(z_1,z_2)$ and $(z_2,z_1)$ are two different solutions that describe the nodal singularities we find $\frac{1}{2}(p-1)(q-1)$ when we slightly perturb away from the singularity. \label{footnote:nodal singularities minimal string}
}

We want to zoom into such a singular region of the spectral curve. Mathematically, we are taking a blow up. Physically, we are taking a one-parameter family of potentials described by $t$ such that for $t \to t_\text{c}$, the potential exhibits such a singular behaviour. We then expand for $t \sim t_\text{c}$ and $z\sim z_*$ in a coordinated way. To get something reasonable, we put $z=z_*+(t-t_\text{c})^\nu \zeta$ for some critical exponent $\nu$ and a new coordinate $\zeta$.\footnote{$t$ is usually taken to be the coefficient of the mixed term $M_1 M_2$ in the exponent \eqref{eq:2MM}. In that case one can show that $\nu=\frac{1}{p+q-1}$.} This gives
\begin{subequations}
\begin{align}
    \xx(z)&=x_*+(t-t_\text{c})^{q \nu} Q(\zeta)+\dots ~, \\
    \yy(z)&=y_*+(t-t_\text{c})^{p \nu} P(\zeta)+\dots ~,    
\end{align} \label{eq:x y rational double scaling limit}%
\end{subequations}
for two polynomials $Q$ and $P$ or degree $q$ and $p$. One can easily verify the degree by noticing that this double scaled spectral curve has the right number of double points. These polynomials are in principle undetermined since they depend precisely on how we take the double scaling limit. This is not surprising since we get a whole family of possible spectral curves that are dual to the minimal string perturbed by the $\frac{(p-1)(q-1)}{2}$ operators of the theory. There will be a special choice known as the conformal background where $Q$ and $P$ are Chebychev polynomials of order $q$ and $p$, respectively.

Notice that $\omega_{0,1}(z)$ scales like $(t-t_\text{c})^{(p+q)\nu}$ and hence by topological recursion, $\omega_{g,n}$ scales like $(t-t_\text{c})^{-(p+q)\nu(2g-2+n)}$. The $\frac{1}{N}$ expansion of this theory takes the form
\be 
\sum_{g=0}^\infty \omega_{g,n}(z_1,\dots,z_n) \big(N (t-t_\text{c})^{(p+q)\nu}\big)^{2-2g-n}~.
\ee
In order to get a good limit, we also need to send $N \to \infty$ in a coordinated way such that
\be 
\mathrm{e}^{S_0} \equiv N (t-t_\text{c})^{(p+q)\nu}
\ee
remains finite. This explains the name double scaling.
\paragraph{Irrational double scalings.} The spectral curve that we will find is not of this type: it has \emph{infinitely} many nodal points. To engineer such a spectral curve via a double scaling limit, we have to start with a more drastic singularity which requires the collision of infinitely many nodal points in the unscaled spectral curve. This is of course only possible with potentials of infinite degree since the number of nodal points is bounded by $d_1d_2$ where $d_1$ and $d_2$ are the degrees of $V_1'(x)$ and $V_2'(y)$, respectively.

The discussion of topological recursion etc above however more or less straightforwardly goes through provided that there are no convergence problems since one can approximate the potential arbitrarily well by a polynomial of very high degree. In any case, we will be interested in a local singularity of the form
\be 
(x-x_*)^{b^2}=\text{const.} \times (y-y_*)+\dots~,
\ee
where $b^2$ is \emph{purely imaginary}. The reason for the notation $b^2$ is to connect to the bulk string theory. Clearly such a singularity requires infinitely many nodal singularities to collide and hence $\xx(z)$ and $\yy(z)$ will have an essential singularity at $z=z_*$. We can locally engineer such a singularity for example by setting
\be 
\xx(z)=x_*+\mathrm{e}^{z}~, \quad \yy(z)=y_*+\mathrm{e}^{z\, b^2}~.
\ee
We have $\xx(z)=x_*$ and $\yy(z)=y_*$ for $z \to \infty$, provided that we approach infinity from the correct direction.

Since we want to zoom into the region $z \to \infty$, the way to introduce a new coordinate is to set $z=\zeta+\nu \log(t-t_\text{c})$, so that for fixed $\zeta$, $z$ diverges as $t \to t_\text{c}$. Plugging this into $\xx(z)$ and $\yy(z)$ leads to a spectral curve of the form
\be 
\xx(z)=x_*+(t-t_\text{c})^\nu F (\zeta)~, \quad \yy(z)=y_*+(t-t_\text{c})^{\nu b^2}  G(\zeta)~,
\ee
where $F$ and $G$ are entire functions. $F$ and $G$ are not completely arbitrary: they are still both of exponential type, i.e.\ grow at most like an exponential function near infinity. Moreover, we know that
\be 
\lim_{\zeta \to \infty} \frac{\log G(\zeta)}{\log F(\zeta)}=b^2~, \label{eq:F and G growth condition}
\ee
at least in some directions in the complex plane. This is the analogue of the corresponding functions being polynomials of degree $q$ and $p$ in the rational case \eqref{eq:x y rational double scaling limit}. For practical purposes, we notice that essentially all the formulas from the rational case will carry over. We can first assume $b^2 \in \RR$ and approximate it arbitrarily well by rational numbers. We can then often simply analytically continue to $b^2 \in i \RR$. 
The rest of the double scaling limit is completely analogous to the rational case. 

\subsection{Relation to 2d gravity} \label{subsec:relation 2d gravity}

Two-matrix integrals compute 2d gravity amplitudes in the double scaling limit. The intuition for this is well-known: two-matrix integrals count certain triangulations of 2d surfaces. Upon taking the double scaling limit, the dominant contributions come from very fine triangulations which define the 2d gravity path integral.\footnote{ 
This construction has actually been made rigorous in the mathematical literature in recent years in the form of Brownian surfaces, see e.g.\ \cite{Miller:2015qaa}.}

Starting with Witten's conjecture \cite{Witten:1990hr, Kontsevich:1992ti}, this relation has been made very precise. Observables in 2d theories of gravity can be realized as intersection numbers on the moduli space of surfaces and hence the differentials $\omega_{g,n}$ can be expressed in terms of such intersection numbers. 

\paragraph{Relation to intersection numbers.} The general formula for a topological recursion with $N$ branch points labelled by $\{1,\dots,N\}$ is \cite{Eynard:2011ga, Dunin-Barkowski:2012kbi}
\begin{multline}
    \omega_{g,n}(z_1,\dots,z_n)=2^{3g-3+n} \sum_{\Gamma \in \mathcal{G}^N_{g,n}} \frac{1}{|\text{Aut}(\Gamma)|} \int_{\bM_\Gamma} \prod_{v \in \mathcal{V}_\Gamma} \mathrm{e}^{\sum_{k \ge 0} \hat{t}_{m_v,k} \kappa_k} \\ \times \prod_{(\bullet,\circ) \in \mathcal{E}_\Gamma} \sum_{r,s=0}^\infty B_{m_\bullet, 2r,m_\circ, 2s} \psi_\bullet^r \psi_\circ^s \prod_{i=1}^n \sum_{\ell \ge 0} \psi_i^\ell \d \eta_{m_i,\ell}(z_i)~. \label{eq:omegagn general intersection number formula}
\end{multline}
The equation is a sum over stable graphs of colored Riemann surfaces, whose set we denote by $\mathcal{G}_{g,n}^N$. The colors are indexed by natural numbers $m \in \{1,\dots,N\}$. A graph in $\mathcal{G}_{g,n}^N$ has vertices labelled by genera $g_v$ as well as a color $m_v \in \ZZ_{\ge 1}$. There are $n$ labelled external legs. Let also $n_v$ be the number of outgoing edges from every vertex $v$. Then stability of the graph means that every vertex satisfies $n_v \ge 3$ for $g_v=0$ and $n_v \ge 1$ for $g_v=1$. We denote the set of vertices by $\mathcal{V}_\Gamma$ and the set of edges by $\mathcal{E}_\Gamma$.
Such graphs describe degenerations of Riemann surfaces into $|\mathcal{V}_\Gamma|$ components connected at various nodal points that correspond to the edges of the graph. Every component can have a different color, and the sum in (\ref{eq:omegagn general intersection number formula}) runs over all possible combinations. 

Furthermore, every such stable graph has some number of automorphisms. These are not allowed to permute external lines (which are labelled by $\{1,\dots,n\})$, but can arbitrarily permute internal lines. Just like in Feynman diagram computations, we have to divide by the order of the automorphism group.
For example, let us list all the stable graphs $\Gamma \in \mathcal{G}_{1,2}^N$:
\begingroup
\setlength{\tabcolsep}{6pt}
\renewcommand{\arraystretch}{1.5}
\begin{center}
\begin{tabular}{c|ccccc}
$\Gamma$ & \begin{tikzpicture}[baseline={([yshift=-.5ex]current bounding box.center)},scale=.6]
    \node[shape=circle,draw=black, very thick] (A) at (0,0) {$1$};
    \draw[very thick] (A) to (1,1) node[right] {1};
    \draw[very thick] (A) to (1,-1) node[right] {2};
\end{tikzpicture} &\begin{tikzpicture}[baseline={([yshift=-.5ex]current bounding box.center)},scale=.6]
    \node[shape=circle,draw=black, very thick] (A) at (0,0) {$0$};
    \node[shape=circle,draw=black, very thick] (B) at (-1.5,0) {$1$};
    \draw[very thick] (A) to (1,1) node[right] {1};
    \draw[very thick] (A) to (1,-1) node[right] {2};
    \draw[very thick] (A) to (B);
\end{tikzpicture} &\begin{tikzpicture}[baseline={([yshift=-.5ex]current bounding box.center)},scale=.6]
    \node[shape=circle,draw=black, very thick] (A) at (0,0) {$0$};
    \draw[very thick] (A) to (1,0) node[above] {1};
    \draw[very thick] (A) to (-1,0) node[above] {2};
    \draw[very thick, out=50, in=130, looseness=3] (A) to (A);
\end{tikzpicture}&\begin{tikzpicture}[baseline={([yshift=-.5ex]current bounding box.center)},scale=.6]
    \node[shape=circle,draw=black, very thick] (A) at (0,0) {$0$};
    \node[shape=circle,draw=black, very thick] (B) at (-1.5,0) {$0$};
    \draw[very thick] (A) to (1,1) node[right] {1};
    \draw[very thick] (A) to (1,-1) node[right] {2};
    \draw[very thick] (A) to (B);
    \draw[very thick, out=145, in=-145, looseness=3] (B) to (B);
\end{tikzpicture} & \begin{tikzpicture}[baseline={([yshift=-.5ex]current bounding box.center)},scale=.6]
    \node[shape=circle,draw=black, very thick] (A) at (0,0) {$0$};
    \node[shape=circle,draw=black, very thick] (B) at (-1.5,0) {$0$};
    \draw[very thick] (A) to (1,0) node[above] {1};
    \draw[very thick] (B) to (-2.5,0) node[above] {2};
    \draw[very thick, bend left=30] (A) to (B);
    \draw[very thick, bend right=30] (A) to (B);
\end{tikzpicture} \\
\hline
$|\text{Aut}(\Gamma)|$ & 1 & 1 & 2 & 2 & 2 
\end{tabular}
\end{center}
\endgroup
The number inside each component of the graph indicates the genus $g_v$ of the vertex. We suppressed the color label $m_v$.

The integral appearing on the right hand side of \eqref{eq:omegagn general intersection number formula} is over $\bM_\Gamma=\prod_{v} \bM_{g_v,n_v}$ and involves the standard kappa- and psi-classes on moduli space. Every internal edge is associated to two punctures on the adjacent vertices. Thus every edge is associated to two psi-classes which we denote by $\psi_\bullet$ and $\psi_\circ$ and we may label the edge by the pair of psi-classes $(\bullet,\circ) \in \mathcal{E}_\Gamma$. Finally, we also have psi-classes $\psi_i$ of the external legs entering the formula. The quantities $\hat{t}_{m_v,k}$, $B_{m,2r,m',2s}$ and the differentials $\d \eta_{m,\ell}(z)$ are determined through the data of the spectral curve. We refer to appendix~\ref{app:intersection theory} for the precise formula. One can also refine the intersection number data and define a so-called cohomological field theory (CohFT), which keeps track of the full integrand in \eqref{eq:omegagn general intersection number formula} and not only its intersection number. We will discuss this for the case of interest briefly in section~\ref{subsec:CohFT}.

\paragraph{Continuum description.} The intuition above should also mean that there is a continuum description of such double scaling limits in terms of a string worldsheet theory. However, such a relation is much harder to make precise rigorously and is not known in great generality. The cases under control are
\begin{enumerate}
    \item Rational models coming from a rational double scaling limit as described in section~\ref{subsec:double scaling}. These are dual to the $(p,q)$-minimal string consisting of Liouville theory coupled to a $(p,q)$-Virasoro minimal model. To describe general spectral curves, it is necessary to deform this theory by the marginal operators of the theory. The case of $q=2$ can also be described in terms of a single matrix integral.
    \item Irrational single-matrix integrals. The Virasoro minimal string \cite{Collier:2023cyw} is dual to such a spectral curve with $\xx(z)=-z^2$, $\yy(z)=z^{-1}\sin(bz) \sin(b^{-1}z)$ and $b \in \RR$. Under some restrictions, one can presumably also deform by marginal operators to obtain different spectral curves as was done in the language of dilaton gravity in \cite{Witten:2020wvy, Maxfield:2020ale}.
\end{enumerate}
Removing the nodal singularities conjecturally requires putting the bulk theory in a background of a non-perturbatively large number of ZZ-instantons \cite{Seiberg:2003nm, Kutasov:2004fg}, but this is not under computational control from the bulk. We will further comment on this in the discussion \ref{sec:conclusion}.

\section{The duality with the worldsheet theory} \label{sec:duality with the worldsheet theory}
\subsection{The spectral curve}
As already mentioned in the introduction, our main claim is that the 2d gravity theory is dual to a double-scaled two-matrix integral with spectral curve
\be 
\xx(z)=-2\cos(\pi b^{-1} \sqrt{z})~, \quad \yy(z)=2 \cos(\pi b \sqrt{z})~. \label{eq:spectral curve}
\ee
This is a curve of genus 0 and $z$ provides the rational parametrization.\footnote{It is computationally often more useful to use a parametrization of the spectral curve in terms of the parameter $w=\sqrt{z}$, but conceptually the use of $z$ is much cleaner. For various computations below we will use $w$.}
$z$ plays the role of $\zeta$ in section~\ref{subsec:double scaling}, but we write $z$ for notational simplicity. It satisfies the condition \eqref{eq:F and G growth condition} and hence can be realized as a double scaling limit around an essential singularity in the spectral curve.

\paragraph{Sheets.} The spectral curve has \emph{infinitely} many sheets since $\xx(z)=\xx((\sqrt{z}+2b n)^2)$ for all $n \in \ZZ$. There are also infinitely many branch points
\be 
z_m^*=(m b)^2 \label{eq:branch points}
\ee
with $m \in \ZZ_{\ge 1}$. As we shall see, the sum appearing on the RHS of \eqref{eq:topological recursion} is always very rapidly converging and the infinite number of branch points does not create convergence problems.

The perturbative expansion is fully controlled by topological recursion, which we reviewed in section~\ref{subsec:top rec review}. The spectral curve leads us to the differential
\be \label{eq: omega01}
\omega^{(b)}_{0,1}(z)=-\yy(z) \, \d \xx(z)=-\frac{2\pi \sin(\pi b^{-1} \sqrt{z}) \cos(\pi b \sqrt{z})\, \d z}{b \sqrt{z}}  ~.
\ee
$\omega^{(b)}_{0,2}(z_1,z_2)$ takes the form 
\be 
\omega^{(b)}_{0,2}(z_1,z_2)=B(z_1,z_2)=\frac{\d z_1\, \d z_2}{(z_1-z_2)^2} ~. \label{eq:omega02}
\ee
As explained above, $B(z_1,z_2)$ is the Bergman kernel on the spectral curve and its form is dictated by the two-matrix integral, see section \ref{subsec:loop equations}.

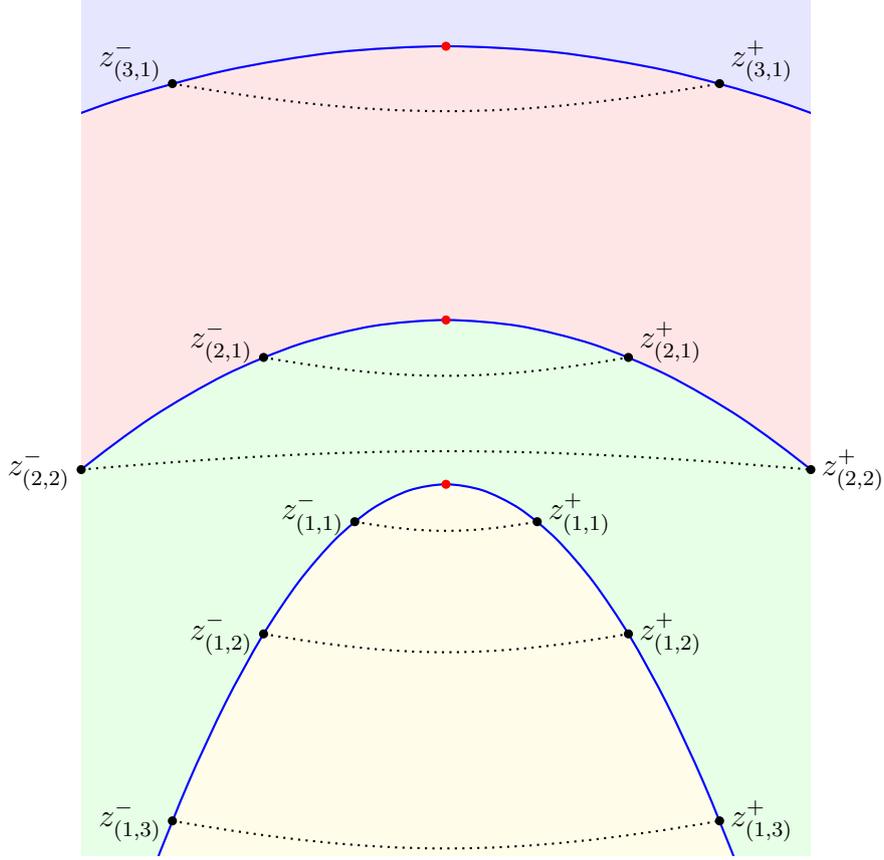
\begin{figure}
    \centering
    \begin{tikzpicture}[scale=.6]
        \begin{scope}
            \clip (-8,-7) rectangle (8,12);
            \foreach \m in {4}
            \fill[domain=-4:4, smooth, variable=\x, blue!10!white] plot ({2.42*\x *\m}, {1.21*\m*\m-1.21*\x*\x});
        \end{scope}
        \begin{scope}
            \clip (-8,-7) rectangle (8,12);
            \foreach \m in {3}
            \fill[domain=-4:4, smooth, variable=\x, red!10!white] plot ({2.42*\x *\m}, {1.21*\m*\m-1.21*\x*\x});
        \end{scope}
        \begin{scope}
            \clip (-8,-7) rectangle (8,12);
            \foreach \m in {2}
            \fill[domain=-4:4, smooth, variable=\x, green!10!white] plot ({2.42*\x *\m}, {1.21*\m*\m-1.21*\x*\x});
        \end{scope}
        \begin{scope}
            \clip (-8,-7) rectangle (8,12);
            \foreach \m in {1}
            \fill[domain=-4:4, smooth, variable=\x, yellow!10!white] plot ({2.42*\x *\m}, {1.21*\m*\m-1.21*\x*\x});
        \end{scope}
        \begin{scope}
            \clip (-8,-7) rectangle (8,12);
            \foreach \m in {1,2,3}
            \draw[domain=-4:4, smooth, variable=\x, blue, thick] plot ({2.42*\x *\m}, {1.21*\m*\m-1.21*\x*\x});
        \end{scope}
        \fill (2., 0.383554) circle (.1) node[right, shift={(0,.1)}] {$z_{(1,1)}^+$};
        \fill (-2., 0.383554) circle (.1) node[left, shift={(0,.1)}] {$z_{(1,1)}^-$};
        \draw[dotted, thick, bend right=10] (-2., 0.383554) to (2., 0.383554);
        \fill (4., -2.09579) circle (0.1) node[right] {$z_{(1,2)}^+$};
        \fill (-4., -2.09579) circle (0.1) node[left] {$z_{(1,2)}^-$};
        \draw[dotted, thick, bend right=10] (-4., -2.09579) to (4., -2.09579);
        \fill (6., -6.22802) circle (0.1) node[right] {$z_{(1,3)}^+$};
        \fill (-6., -6.22802) circle (0.1) node[left] {$z_{(1,3)}^-$};
        \draw[dotted, thick, bend right=10] (-6., -6.22802) to (6., -6.22802);
        \fill (4., 4.01355) circle (0.1) node[right, shift={(0,.2)}] {$z_{(2,1)}^+$};
        \fill (-4., 4.01355) circle (0.1) node[left, shift={(0,.2)}] {$z_{(2,1)}^-$};
        \draw[dotted, thick, bend right=10] (-4., 4.01355) to (4., 4.01355);
        \fill (8., 1.53421) circle (0.1) node[right] {$z_{(2,2)}^+$};
        \fill (-8., 1.53421) circle (0.1) node[left] {$z_{(2,2)}^-$};
        \draw[dotted, thick, bend left=5] (-8., 1.53421) to (8., 1.53421);
        \fill (6., 10.0636) circle (0.1) node[right, shift={(0,.3)}] {$z_{(3,1)}^+$};
        \fill (-6., 10.0636) circle (0.1) node[left, shift={(0,.3)}] {$z_{(3,1)}^-$};
        \draw[dotted, thick, bend right=10] (-6., 10.0636) to (6., 10.0636);
        \fill[red] (0,1.21) circle (0.1);
        \fill[red] (0,4.84) circle (0.1);
        \fill[red] (0,10.89) circle (0.1);
    \end{tikzpicture}
    \caption{The structure of the spectral curve plotted for $b=\frac{11}{10} \, \mathrm{e}^{\frac{\pi i}{4}}$. The points $z_{(m,n)}^\pm$ are nodal singularities and correspond to the same point on the spectral curve. We denoted this by a dotted line, which can be viewed as a pinched handle of the surface. The red dots correspond to the branch points $z_m^*$ of $\xx(z)$. The differently colored regions correspond to the different sheets of $\xx(z)$. The yellow and red regions map to $\CC \setminus [2,\infty)$ under $\xx(z)$, while the green and blue regions map to  $\CC \setminus (-\infty,-2]$. The yellow region corresponds to the physical sheet. The support of the eigenvalues is the lowest blue parabola which delineates the boundary of the physical sheet. It maps to $[2,\infty)$ under $\xx(z)$.}
    \label{fig:structure spectral curve}
\end{figure}

\paragraph{Singular points.} The spectral curve has a number of singular points of the form \eqref{eq:singular points}. They are located at
\be 
z_{(r,s)}^\pm=(rb \pm s b^{-1})^2 \label{eq:singular points spectral curve}
\ee
with $r,\, s \in \ZZ_{\ge 1}$. Both choices of sign map to the same point under $(\xx(z),\yy(z))$, i.e.
\be 
\xx(z_{(r,s)}^+)=\xx(z_{(r,s)}^-)~, \quad \yy(z_{(r,s)}^+)=\yy(z_{(r,s)}^-)~.
\ee
This means that the spectral curve self-intersects at these points which is the definition of a nodal singularity. One can hence picture the spectral curve as in figure~\ref{fig:structure spectral curve}. In particular, the yellow region is the physical sheet and the eigenvalues are supported on $\xx(z) \in [2,\infty)$. 
\paragraph{Density of states.} From the definition of $\omega_{0,1}^{(b)}$ (\ref{eq:omega01 definition}) we infer
\begin{equation}
  R_{0,1}(\xx(z))\d \xx(z)=\big(V_1'(\xx(z)) -\yy(z)\big)\, \d \xx(z)~.
\end{equation}
It is natural to interpret the first matrix as a Hamiltonian; in this sense, we define the energy $E\equiv \xx(z)$. The eigenvalue density of the Hamiltonian can be computed from \eqref{eq:density of states discontinuity R01}.
The region slightly above and below the branch cut is mapped to 
\be 
z_+=-\frac{b^2}{\pi^2}\left(\text{arccosh}\left(\frac{E}{2}\right)-\pi i \right)^2~, \quad z_-=-\frac{b^2}{\pi^2}\left(\text{arccosh}\left(\frac{E}{2}\right)+\pi i \right)^2~,
\ee
respectively. Hence the density of states becomes
\begin{align} \nonumber
\rho_0(E)&=\frac{1}{2\pi i}\big(\yy(z_+)-\yy(z_-)\big) \\
&=\frac{2}{\pi} \sinh\left(-\pi i b^2\right)\sin \left(-i b^2 \text{arccosh}\left(\frac{E}{2}\right)\right)~.
\end{align}
When written in this way, the density of states is manifestly positive in a vicinity of $E \sim 2$. Because of the sine, the density of states however becomes negative far away from $E=2$. The sign changes occur precisely at the location of the singular points \eqref{eq:singular points spectral curve}. This makes it possible to rescue the definition of the theory and make it non-perturbatively well-defined. This has no influence on perturbative quantities and we will postpone the discussion to \cite{paper3}.
\begin{figure}[ht]
    \centering
    \includegraphics[width=0.6\linewidth]{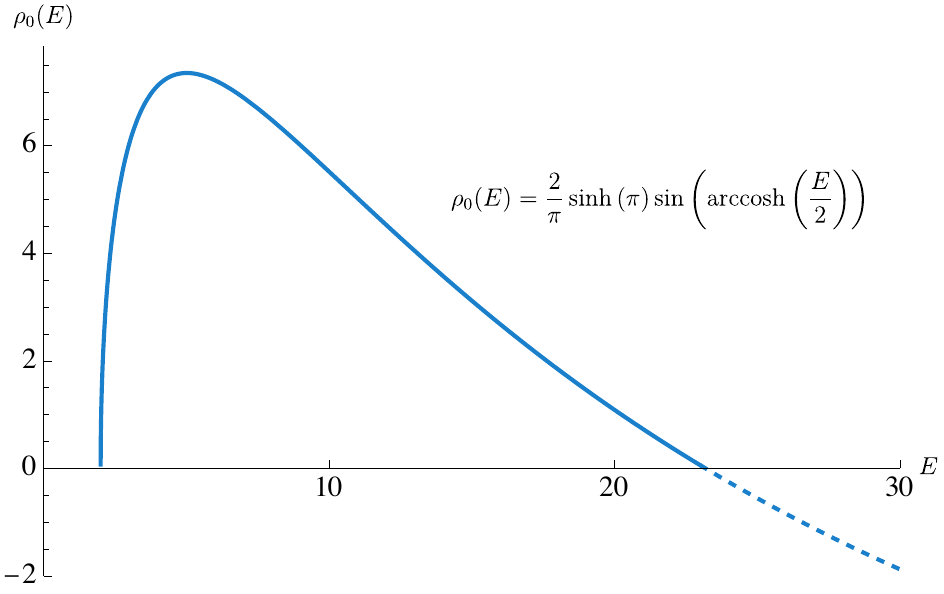}
    \caption{Density of states for the first matrix for $b^2=i$. Close to $E\approx 2$ the density behaves as $\rho_0(E)\approx \sqrt{E-2}$.}
    \label{fig:enter-label}
\end{figure}

One can similarly also compute the structure of the eigenvalues of the second matrix, also interpreted as a Hamiltonian. They are supported on the interval $(-\infty,-2]$ with density of states
\be 
\rho_0^{(2)}(E^{(2)})=\frac{2}{\pi} \sinh\big(\pi i b^{-2}\big) \sin \left( i b^{-2} \text{arccosh}\left(-\frac{E^{(2)}}{2}\right)\right)~,
\ee
which is also positive for $E^{(2)}$ close to the edge $-2$.

\paragraph{Topological recursion.} 
In this discussion it turns out to be most convenient to parameterize the spectral curve in terms of the $w$ coordinates so that 
\begin{equation}\label{eq:xy spectral curve w}
    \mathsf{x}(w) = -2\cos(\pi b^{-1}w), \quad \mathsf{y}(w) = 2\cos(\pi b w)\,,
\end{equation}
and 
\begin{subequations}
\begin{align}
    \omega^{(b)}_{0,1}(w) &= -\frac{4\pi \cos(\pi b w)\sin(\pi b^{-1} w)}{b}\, \d w ~ ,\\
    \omega^{(b)}_{0,2}(w_1,w_2) &= \left(\frac{1}{(w_1-w_2)^2} - \frac{1}{(w_1+w_2)^2}\right)\d w_1\, \d w_2\, .
\end{align}
\end{subequations}
In this parametrization the branch points of the spectral curve correspond to $w = \pm m b$ for $m\in\mathbb{Z}_{\geq 1}$, with the local Galois inversion given by $\sigma_m(w) = 2mb - w$. The higher resolvent differentials are then determined by the topological recursion (\ref{eq:topological recursion}) with the recursion kernel given by (\ref{eq:definition recursion kernel}):
\begin{equation}\label{eq:recursion kernel w basis}
    K_m(w_1,w) = -\frac{b w_1\left(\frac{1}{w_1^2-w^2}-\frac{1}{w_1^2-\sigma_m(w)^2}\right)}{4\pi\left[\sin(\pi b^{-1} \sigma_m(w))\cos(\pi b\sigma_m(w))+\sin(\pi b^{-1} w)\cos(\pi b w)\right]}\frac{\d w_1}{\d w}\, .
\end{equation}
The plus sign in the denominator arises because $\d \sigma_m(w) = -\d w$. 

As an example, we can then straightforwardly apply the topological recursion to obtain e.g. $\omega^{(b)}_{0,3}(w_1,w_2,w_3)$ 
\begin{align}\label{eq:omega03 from topo}
   &\omega^{(b)}_{0,3}(w_1,w_2,w_3)\nonumber\\
   &\qquad=\sum_{m=1}^\infty\Res_{w=w^*_m} K_m(w_1,w) \Big(\omega_{0,2}(w,w_2)\omega_{0,2}(\sigma_m(w),w_3) + (w_2 \leftrightarrow w_3)\Big) \nonumber\\
    &\qquad=-\sum_{m=1}^\infty\frac{16 m^3 b^4 (-1)^m\,w_1w_2w_3\, \d w_1 \, \d w_2\, \d w_3}{\pi^3 \sin(\pi m b^2) (w_1^2-(w_m^*)^2)^2(w_2^2-(w_m^*)^2)^2(w_3^2-(w_m^*)^2)^2}~.
\end{align}
where the overall minus sign again comes from $\d\sigma_m(w) = - \d w$. Similarly, $\omega^{(b)}_{1,1}(w_1)$ is given by
\begin{align}
    \omega^{(b)}_{1,1}(w_1) &= -\sum_{m=1}^\infty \frac{(-1)^m w_1 \, \d w_1}{48\pi^3 m\sin(\pi m b^2)(w_1^2-(w_m^*)^2)^4}\bigg(24b^4 m^4 + 12 b^2 m^2(w_1^2-(w_m^*)^2)\nonumber\\
    & \quad\quad+ (-6+m^2\pi^2(1+b^4))(w_1^2-(w_m^*)^2)^2\bigg)\, .
\end{align}

\paragraph{Comparison with the $(p,q)$ minimal string.} Let us compare this spectral curve to the spectral curve of the $(p,q)$ minimal string \cite{Seiberg:2004at}, which can be also parametrized analogously, 
\be 
\xx(w)=-2 \cos(\pi b^{-1} w)~ , \quad \yy(w)=2 \cos(\pi b w)~,\label{eq:spectral curve pq minimal string}
\ee
but with $b^2=\frac{p}{q} \in \QQ$. The coordinate $w$ is not a rational parametrization because 
\be 
\xx(\pm w+2 n \sqrt{pq})=\xx(w)~,\quad \yy(\pm w+2 n \sqrt{pq})=\yy(w)~, \label{eq:spectral curve pq minimal string ambiguity}
\ee
with $n \in \ZZ$. To pass to a rational parametrization, we set $w=\frac{1}{\pi} \sqrt{pq} \arccos(z)$ for a new coordinate $z$. The multi-valued structure of $\arccos$ precisely absorbs the ambiguity \eqref{eq:spectral curve pq minimal string ambiguity}. Thus in these coordinates, the spectral curve reads
\be 
\xx(z)=-2 \cos(q \arccos(z))=-2 T_q(z)~,\quad \yy(z)=2 \cos(p \arccos(z))=2 T_p(z)~,
\ee
with $T_m(z)$ the Chebyshev polynomials. This corresponds to the conformal background discussed in section~\ref{subsec:double scaling}.

Because of the additional invariance in \eqref{eq:spectral curve pq minimal string ambiguity}, there are only finitely many branch points located at 
\be 
z_m^*=\cos\left(\frac{\pi m}{q}\right)
\ee
with $m=1,\dots,q-1$, compare with \eqref{eq:branch points}. There are also only finitely many nodal singularities located at
\be 
z_{(r,s)}^{\pm}=\cos\left(\frac{\pi (r p\pm s q)}{pq}\right) ~.
\ee
with $r=1,\dots,q-1$ and $s=1,\dots,p-1$. Notice also that $z_{(r,s)}^\pm=z_{(q-r,p-s)}^\pm$ and thus there are exactly $\frac{1}{2}(p-1)(q-1)$ nodal singularities matching the general discussion of footnote \ref{footnote:nodal singularities minimal string}. They map to the Kac table of the Virasoro minimal model on the worldsheet.
\subsection{Relation between observables}
We claim that the dictionary to the bulk diagrams $\mathsf{A}_{g,n}^{(b)}$ is given by 
\begin{align} 
\mathsf{A}_{g,n}^{(b)}(p_1,\dots,p_n)&=\int_{\gamma}  \prod_{j=1}^n \frac{\mathrm{e}^{2\pi i p_j w_j}}{4\pi i p_j} \, \omega^{(b)}_{g,n}(w_1,\dots,w_n) \label{eq:Agn omegagn relation w}\\
&=\sum_{m_1,\dots,m_n=1}^\infty  \Res_{z_1=z_{m_1}^*} \cdots \Res_{z_n=z_{m_n}^*} \prod_{j=1}^n \frac{\cos(2 \pi p_j \sqrt{z_j})}{p_j}\,  \omega^{(b)}_{g,n}(z_1,\dots,z_n)~. \label{eq:Agn omegagn relation}
\end{align}
The first expression is in terms of the coordinate $w_j=\sqrt{z_j}$, in which $\omega^{(b)}_{g,n}$ has poles at $w_j=\pm m b$ for $m \in \ZZ_{\ge 1}$.
The contour $\gamma$ runs to the right of the series of singularities $\pm m b$ for each $w_j$. The first equation \eqref{eq:Agn omegagn relation w} is valid provided that $\Re(b p_j)>0$. It can be viewed as an inverse Laplace transformation of the $\omega^{(b)}_{g,n}$'s. We can then pull the contour over the singularities which picks up the residue at the poles $\pm m b$. For $\omega^{(b)}_{g,n}$, the two residues are identical and thus the residue becomes
\be 
\Res_{w_j=m b}\frac{1}{2} (\mathrm{e}^{2\pi i p_j w_j}+\mathrm{e}^{-2\pi i p_j w_j})\, \omega^{(b)}_{g,n}(w_1,\dots,w_n)=\Res_{w_j=m b} \cos(2\pi p_j w_j) \, \omega^{(b)}_{g,n}(w_1,\dots,w_n)~,
\ee
which becomes \eqref{eq:Agn omegagn relation} when written in terms of the variables $z_j$.
The second equation \eqref{eq:Agn omegagn relation} can be taken to be the defining relation for all values of $p_j$.

The inverse transform that expresses the resolvents in terms of the string amplitudes is given by
\begin{equation}\label{eq:omega in terms of A}
    \omega^{(b)}_{g,n}(w_1,\ldots,w_n) = (-2\pi)^n \int \prod_{j=1}^n\left(-2p_j \d p_j \sin(2\pi p_j w_j)\right)\mathsf{A}_{g,n}^{(b)}(p_1,\ldots,p_n) \prod_{j=1}^n \d w_j\, .
\end{equation}
The integrals over the Liouville momenta $p_j$ are to be computed in the following sense. By expanding the sines, we are integrating polynomials times exponentials of the form
\be\label{eq:p integral complex exponential}
    \int \d p \, \mathrm{e}^{2\pi i (w\pm m b) p} p^{2a+1}~,
\ee
for an integer $m$. The integral is then taken to run from $0$ to infinity in a direction of the complex plane such that the integral converges.

\subsection{Reducing to sums over stable graphs}
As reviewed in section~\ref{subsec:relation 2d gravity}, the differentials $\omega^{(b)}_{g,n}$ can be expressed as integrals over the moduli space of surfaces, see eq.~\eqref{eq:omegagn general intersection number formula}. When translating the relation to $\mathsf{A}_{g,n}^{(b)}$, this relation takes the form
\begin{align} 
\mathsf{A}_{g,n}^{(b)}(p_1,\dots,p_n) &=\sum_{\Gamma \in \mathcal{G}_{g,n}^\infty}\frac{1}{|\text{Aut}(\Gamma)|}\int' \prod_{e \in \mathcal{E}_\Gamma} (-2 p_e \, \d p_e)\prod_{v \in \mathcal{V}_\Gamma} \left(\frac{b(-1)^{m_v}}{\sqrt{2}\sin(\pi m_v b^2)}\right)^{2g_v-2+n_v}\nonumber\\
 &\quad\times  \prod_{j \in I_v} \sqrt{2} \sin(2\pi m_v b p_j) \mathsf{V}^{(b)}_{g_v,n_v}(i\boldsymbol{p}_v)~. \label{eq:Agn through quantum volumes}
\end{align}
Details on the derivation of this formula can be found in appendix \ref{app:intersection theory}.\footnote{Strictly speaking all the results in \cite{Eynard:2011ga, Dunin-Barkowski:2012kbi} were derived for a finite number of branch points, but from the presence of the inverse $\sin(\pi m b^2)$ factors, it is clear that all sums converge exponentially fast and thus convergence is not a problem.} Here $I_v$ is the set of momenta associated with the vertex $v$. There are two new ingredients in this formula.
The quantity $\mathsf{V}_{g,n}^{(b)}(i \boldsymbol{p}) \equiv \mathsf{V}^{(b)}_{g,n}(ip_1,\dots,ip_n)$ is the quantum volume defined in \cite{Collier:2023cyw}.\footnote{Note that we multiplying the arguments by an extra $i$ since we are parametrizing the Liouville momenta by $p_j=-i P_j$.} It is a polynomial in $\QQ[\frac{b^2+b^{-2}}{4},p_1^2,\dots,p_n^2]$ of order $3g-3+n$ and can be defined as a certain intersection number of moduli space or alternatively from a recursion relation analogous to Mirzakhani's recursion relation of the Weil-Petersson volumes \cite{Mirzakhani:2006fta}.
The primed integral $\int'$ means the following. By expanding the sines as in the discussion around (\ref{eq:omega in terms of A}) we encounter integrals of the form
\begin{equation}
    \int \d p\, \mathrm{e}^{2\pi i m b p}p^{2a+1}
\end{equation}
Here $m$ is either the sum or difference of neighboring colors. For $m\ne 0$ we take the integral to run from $0$ to infinity in a direction such that the integral converges. 
However, it can happen that $m=0$ if the colors of the two components we are connecting agrees. In this case, the integral clearly does not converge and we simply discard it, i.e.
\be\label{eq:primed integral}
    \int' \d p \, \mathrm{e}^{2\pi i m b p} p^{2a+1}=\begin{cases}
            \Gamma(2a+2)(-2\pi i b m)^{-2a-2}~,  & m\ne 0 \\
            0 ~, & m=0~.
    \end{cases}
\ee
The logic is that in this case, the integral is instead accounted for in the formula by the stable graph where the two components are merged.

Let us evaluate \eqref{eq:Agn through quantum volumes} for some simple examples. We group different terms according to the topology of the corresponding stable graphs.
\begin{table}
\begin{center}
\begin{tabular}{c|c}
     \begin{tikzpicture}[baseline={([yshift=-.5ex]current bounding box.center)},scale=.6]
    \node[shape=circle,draw=black, very thick] (A) at (0,0) {$0$};
    \draw[very thick] (A) to (60:1.3) node[right] {1};
    \draw[very thick] (A) to (180:1.3) node[left] {2};
    \draw[very thick] (A) to (-60:1.3) node[right] {3};
\end{tikzpicture} &  $\!\begin{aligned}\sum_{m=1}^\infty \frac{2b (-1)^m}{\sin(\pi m b^2)} \prod_{i=1}^3 \sin(2\pi m b p_i)\end{aligned}$ \\
\hline 
\hline
    \begin{tikzpicture}[baseline={([yshift=-.5ex]current bounding box.center)},scale=.6]
    \node[shape=circle,draw=black, very thick] (A) at (0,0) {$1$};
    \draw[very thick] (A) to (0:1.3) node[right] {1};
\end{tikzpicture} & \rule[-1.4em]{0pt}{3.3em}$\!\begin{aligned}\sum_{m=1}^\infty \frac{b  (-1)^m \sin(2\pi m b p_1) }{24\sin(\pi m b^2)}  \bigg(\frac{b^2+b^{-2}}{4}-p_1^2\bigg) \end{aligned}$\\
\hline
    \begin{tikzpicture}[baseline={([yshift=-.5ex]current bounding box.center)},scale=.6]
    \node[shape=circle,draw=black, very thick] (A) at (0,0) {$0$};
    \draw[very thick] (A) to (0:1.3) node[right] {1};
    \draw[very thick, out=145, in=-145, looseness=3] (A) to (A);
\end{tikzpicture} & \rule[-1.4em]{0pt}{3.3em}$\!\begin{aligned}-\sum_{m=1}^\infty \frac{b  (-1)^m\sin(2\pi m b p_1)}{16\pi^2 b^2 m^2\sin(\pi m b^2)}  \end{aligned}$\\
\hline
\hline
     \begin{tikzpicture}[baseline={([yshift=-.5ex]current bounding box.center)},scale=.6]
    \node[shape=circle,draw=black, very thick] (A) at (0,0) {$0$};
    \draw[very thick] (A) to (45:1.3) node[right] {1};
    \draw[very thick] (A) to (135:1.3) node[left] {2};
    \draw[very thick] (A) to (-135:1.3) node[left] {3};
    \draw[very thick] (A) to (-45:1.3) node[right] {4};
\end{tikzpicture} &  $\!\begin{aligned}\sum_{m=1}^\infty \bigg(\frac{\sqrt{2} b  (-1)^m}{\sin(\pi m b^2)}\bigg)^2 \bigg(\frac{b^2+b^{-2}}{4}-\sum_{j=1}^4 p_j^2 \bigg)\prod_{i=1}^4 \sin(2\pi m b p_i)\end{aligned}$  \\
\hline
\begin{tikzpicture}[baseline={([yshift=-.5ex]current bounding box.center)},scale=.6]
    \node[shape=circle,draw=black, very thick] (A) at (0,0) {$0$};
    \node[shape=circle,draw=black, very thick] (B) at (-1.5,0) {$0$};
    \draw[very thick] (A) to (B);
    \draw[very thick] (A) to (45:1.3) node[right] {1};
    \draw[very thick] (B) to ++(135:1.3) node[left] {2};
    \draw[very thick] (B) to ++(-135:1.3) node[left] {3};
    \draw[very thick] (A) to (-45:1.3) node[right] {4};
    \node at (-.75,-1.5) {$+ \text{2 perms}$};
\end{tikzpicture} & $\!\begin{aligned} &\sum_{m_1,m_2=1}^\infty \prod_{j=1}^2\frac{ (-1)^{m_j}}{\pi \sin(\pi m_j b^2)} \prod_{i=1,4}\sin(2\pi m_1 b p_i)\prod_{i=2,3}\sin(2\pi m_2 b p_i)\nonumber\\
&\quad\times\bigg(\frac{\delta_{m_1 \ne m_2}}{(m_1-m_2)^2}-\frac{1}{(m_1+m_2)^2} \bigg)+\text{2 perms}\end{aligned}$ 
\end{tabular}
\end{center}
\caption{The stable graphs contributing to $\mathsf{A}_{g,n}^{(b)}$ according to eq.~\eqref{eq:Agn through quantum volumes} for the first few cases.} \label{tab:stable graphs Agn 03, 11, 04}
\end{table}
This leads to Table~\ref{tab:stable graphs Agn 03, 11, 04}. Summing these contributions recovers in particular the equations we used in our previous paper \cite{paper1}.

\paragraph{String amplitudes from ``Feynman rules."}
In order to demystify the discussion of stable graphs presented in the last subsection, here we illustrate the structure of the string amplitudes (\ref{eq:Agn through quantum volumes}) by explicitly representing the stable graphs as specific degenerations of the worldsheet surface. We interpret (\ref{eq:Agn through quantum volumes}) as a sum over Feynman diagrams for the closed string field theory in a particular gauge, with specific Feynman rules associated to each degeneration of the worldsheet surface. In these rules, each component of the degenerated surface receives a factor proportional to the Virasoro minimal string quantum volume $\mathsf{V}_{g,n}^{(b)}$, which we interpet as an on-shell string vertex. We work through the three examples listed in table \ref{tab:stable graphs Agn 03, 11, 04} in turn.

The string amplitude of the three-punctured sphere $\mathsf{A}_{0,3}^{(b)}$, corresponding to the stable graph in the first line of table \ref{tab:stable graphs Agn 03, 11, 04} is represented by the single non-degenerate pair of pants in (\ref{eq:A03 Feynman rule}). Using that $\mathsf{V}_{0,3}^{(b)}(ip_1,ip_2,ip_3)=1$ we have
\begin{equation}\label{eq:A03 Feynman rule}
    \mathsf{A}_{0,3}^{(b)}(p_1,p_2,p_3) =
    \begin{tikzpicture}[baseline={([yshift=7pt]current bounding box.center)}]
    \def\XR{0.25}; 
    \def\YR{0.5};  

    \begin{scope}[shift={(-2,0)}]
    \draw[fill=blue, draw=blue, opacity=.2] (6,2.5) to[out=0, in = 180] (8.25,1.5) to [out=200, in =160, looseness=.75] (8.25,0.5) to[out=180,in = 0] (6,-.5) to [out=20, in = -20, looseness=.95] (6,0.5) to[out= 0, in = 0] (6,1.5) to [out=20, in = -20, looseness=.95] (6,2.5);
    
    \node[scale=1] (A) at (4.5,.9) {$\displaystyle\sum_{m_1\geq 1}$};
    \draw[very thick] (6,0) ellipse (0.25 and .5);
    
    \draw[very thick] (6,2) ellipse (0.25 and .5);
    
    \draw[very thick] (8.3,1) ellipse (0.25 and .5);

    \draw[thick] (6,.5) to[out=0, in =0] (6,1.5); 
    
    \draw[thick] (6,2.5) to[out=0, in = 180] (8.25,1.5);
    
    \draw[thick] (6,-.5) to[out=0, in = 180] (8.25,0.5);
    \node[scale= 1] (B) at (5.6,2.9) {$\sqrt{2}\sin(2\pi m_1 b p_1)$};
    \node[scale= 1] (B) at (5.6,-.8) {$\sqrt{2}\sin(2\pi m_1 b p_2)$};
    \node[scale= 1] (B) at (10.2,1.) {$\sqrt{2}\sin(2\pi m_1 b p_3)$};
    
    \node[scale=1.] (A) at (7.1,1) {$m_1$};
    
    \node[scale= 1.] (B) at (8.7,-1.5) {$\frac{b(-1)^{m_1}}{\sqrt{2}\sin(\pi m_1 b^2)}$};
    \draw[->,very thick] (8.5,-1) -- (7.4,.6);
    
    \end{scope}
    \end{tikzpicture}
    \, .
\end{equation}

Next up we have the once-punctured torus which is the sum over two stable graphs in (\ref{eq:Agn through quantum volumes}):
\begin{equation}\label{eq:A11 Feynamn diagrams}
    \mathsf{A}_{1,1}^{(b)}(p_1) = 
    \begin{tikzpicture}[baseline={([yshift=0pt]current bounding box.center)}]
    \node[scale=1] (A) at (-3.6,-.1) {$\displaystyle\sum_{m_1\geq 1}$};
    \begin{scope}[shift={(0,0)}]
    \node[scale= 1] (B) at (-1.9,.9) {$\sqrt{2}\sin(2\pi m_1 b p_1)$};
    
    \draw[fill=blue,draw=blue, opacity=.2] (-1,1/2) to[out=0, in = 180] (1/2,1) to[out=0, in = 0,looseness=2] (1/2,-1) to[out=180, in = 0] (-1,-1/2) to[out = 0, in = 0,looseness=.8] (-1,1/2);

    \draw[fill=white,draw=white] (1/4,-.1) to[out=60, in = 120] (3/4,-.1) to[out=200, in = -20,looseness=.8] (1/4,-.1);

    \draw[very thick] (-1,0) ellipse (1/4 and 1/2);
    \draw[very thick] (-1,1/2) to[out=0, in = 180] (1/2,1) to[out=0, in = 0,looseness=2] (1/2,-1) to[out=180, in = 0] (-1,-1/2);

    \draw[thick] (0,0) to[out=-30, in = 210] (1,0);
    \draw[thick] (1/4,-.1) to[out=60, in = 120] (3/4,-.1);

    \node at (1/2,-1/2) {$m_1$};
    \draw[->,very thick] (-.3,-1.7) -- (-.3,-.3);
    \node[scale= 1] (B) at (-.3,-2.2) {$\frac{b(-1)^{m_1}}{\sqrt{2}\sin(\pi m_1 b^2)} \mathsf{V}_{1,1}^{(b)}(ip_1)$};
        
    \end{scope}
    
    \begin{scope}[shift={(5.4,0)}]
    
    \node[scale= 1] (B) at (-1.9,.9) {$\sqrt{2}\sin(2\pi m_1 b p_1)$};
    \node[scale=1.] (A) at (-3.,0) {$+$ };
    \draw[very thick, <-] (1.7,0) to[bend right=30, looseness=1.] (1.8,1.7) ;
    \draw[->,very thick] (-.3,-1.7) -- (-.3,-.3);
    \node[scale= 1] (B) at (-.3,-2.2) {$\frac{b(-1)^{m_1}}{\sqrt{2}\sin(\pi m_1 b^2)}$};
    
    \node[scale=1] at (.5,2.) {$\displaystyle\int' (-2q\, \mathrm{d}q)\sin(2\pi m_1 b q)^2$};
    
        \draw[fill=blue, draw=blue, opacity=.2] (-1,1/2) to[out=0, in = 180] (1/2,1) to[out=0, in = 90] (3/2,0) to[out=-90, in = 0] (1/2,-1) to [out = 180, in = 0] (-1,-1/2) to[out= 0, in = 0,looseness=.8] (-1,1/2);
    
        \draw[fill=white, draw=white]  (3/2,0) to [out= 120, in = 0] (1/2, 1/2) to[out = 180, in = 180, looseness=1.75] (1/2,-1/2) to[out = 0, in = -120] (3/2,0);
    
        \draw[very thick] (-1,0) ellipse (1/4 and 1/2);
        \draw[very thick] (-1,1/2) to[out=0, in = 180] (1/2,1) to[out=0, in = 90] (3/2,0) to[out=-90, in = 0] (1/2,-1) to [out = 180, in = 0] (-1,-1/2);
        \draw[very thick] (3/2,0) to [out= 120, in = 0] (1/2, 1/2) to[out = 180, in = 180, looseness=1.75] (1/2,-1/2) to[out = 0, in = -120] (3/2,0);
    
        \node at (-3/8,0) {$m_1$};
    
        \fill (3/2,0) ellipse (0.1 and 0.1);

        \end{scope}
    \end{tikzpicture}
    \, .
\end{equation}
The first graph in (\ref{eq:A11 Feynamn diagrams}) corresponds to a non-degenerate once-punctured torus, while the second surface is a pair of pants glued together at two nodal points where the surface degenerates, an example of a non-separating degeneration.

Finally the last two stable graphs in table \ref{tab:stable graphs Agn 03, 11, 04} are the building blocks of $\mathsf{A}_{0,4}^{(b)}$ in (\ref{eq:Agn through quantum volumes}). We obtain a four-punctured sphere, as well as a surface with a nodal point that connects two three-punctured spheres. The two components are labelled by different color indices $m_1$ and $m_2$. Graphically these two cases are shown below:
\begin{equation}\label{eq:A04 Feynam diagrams}
\begin{aligned}
    &\mathsf{A}_{0,4}^{(b)}(p_1,p_2,p_3,p_4)= \\
    &
    \begin{tikzpicture}[baseline={([yshift=10pt]current bounding box.center)}]
        \def\XR{0.25}; 
        \def\YR{0.5};  
        \begin{scope}[shift={(-2,0)}]
        
        \draw[fill=blue,draw=blue,opacity=.2] (0,2.5) to[out=0, in=180] (1.5,1.75) to[out=0, in =180] (3,2.5) to[out=200,in=160,looseness=.95] (3,1.5) to[out=180, in = 180] (3,.5) to[out=200, in = 160,looseness=.95] (3,-.5) to[out=180, in = 0] (1.5,.25) to[out=180, in = 0] (0,-.5) to[out=20, in = -20,looseness=.95] (0,.5) to[out=0, in = 0] (0,1.5) to[out=20, in =-20,looseness=.95] (0,2.5);
        
        \draw[very thick] (0,0) ellipse (0.25 and .5);
        \draw[very thick] (0,2) ellipse (0.25 and .5);
        \draw[very thick] (3,0) ellipse (0.25 and .5);
        \draw[very thick] (3,2) ellipse (0.25 and .5);
        \draw[thick] (0,.5) to[out=0, in = 0] (0,1.5);
        \draw[thick] (3,.5) to[out=180, in = 180] (3,1.5);
        \draw[thick] (0,2.5) to[out=0, in = 180] (1.5,1.75) to[out=0, in = 180] (3,2.5);
        \draw[thick] (0,-.5) to[out=0, in = 180] (1.5,.25) to[out=0, in=180] (3,-.5);
        \node[scale=1.] (m1) at (1.5,1) {$m_1$};
        \node[scale=1] (A) at (-1.3,.9) {$\displaystyle\sum\limits_{m_1\geq 1}$};

        \node[scale= 1] (B) at (-.2,2.8) {$\sqrt{2}\sin(2\pi m_1 b p_1)$};
        \node[scale= 1] (B) at (3,2.8) {$\sqrt{2}\sin(2\pi m_1 b p_4)$};
        \node[scale= 1] (B) at (-.2,-.8) {$\sqrt{2}\sin(2\pi m_1 b p_2)$};
        \node[scale= 1] (B) at (3,-.8) {$\sqrt{2}\sin(2\pi m_1 b p_3)$};
        \draw[->,very thick] (1.5,-2) -- (m1);
        \node[scale= 1] (B) at (1.5,-2.4) {$\left(\frac{b(-1)^{m_1}}{\sqrt{2}\sin(\pi m_1 b^2)}\right)^2\mathsf{V}_{0,4}^{(b)}(ip_1,ip_2,ip_3,ip_4)$};
        \end{scope}
        
        \begin{scope}[shift={(-1.8,0)}]
        \node[scale=1.] (A) at (3.7,1) {$+$};
        \node[scale=1.] (A) at (4.7,.8) {$\displaystyle\sum\limits_{m_1,m_2\geq 1}$};
        \node[scale=1.] (A) at (11.1,1) {$+$ };
        \node[scale=1.] (A) at (12.1,1) {2 perm};

        \draw[fill=blue, draw=blue, opacity=.2] (6,2.5) to[out=0, in = 180] (8.25,1) to [out=180, in =0] (6,-.5) to [out=20, in = -20, looseness=.95] (6,0.5) to[out= 0, in = 0] (6,1.5) to [out=20, in = -20, looseness=.95] (6,2.5);
        
        \draw[fill=vert, draw=vert, opacity=.2] (10.5,2.5) to[out=180, in = 0] (8.25,1) to[out=0, in = 180] (10.5,-.5) to[out=160, in = 200, looseness=.95] (10.5,0.5) to[out=180, in = 180] (10.5,1.5) to[out=160, in = 200, looseness=.95] (10.5,2.5);
        
        \draw[very thick] (6,0) ellipse (0.25 and .5);
        \draw[very thick] (6,2) ellipse (0.25 and .5);
        
        \draw[very thick] (10.5,0) ellipse (0.25 and .5);
        
        \draw[very thick] (10.5,2) ellipse (0.25 and .5);lf
        \draw [domain=270:440,thick] plot ({10.5+\XR*cos(\x)}, {\YR*sin(\x)}); 
        
        \draw[thick] (6,.5) to[out=0, in =0] (6,1.5); 
        \draw[thick] (10.5,.5) to[out=180, in =180] (10.5,1.5); 
        \draw[thick] (6,2.5) to[out=0, in = 180] (8.25,1) to[out=0, in=180] (10.5,2.5);
        \draw[thick] (6,-.5) to[out=0, in = 180] (8.25,1) to[out=0, in = 180] (10.5,-.5);

        \node[scale=1., inner sep = 1.5pt] (m1right) at (6.9,1) {$m_1$};
        \node[scale=1., inner sep = 1.5pt] (m2right) at (9.6,1) {$m_2$};
        \fill (8.25,1.) ellipse (0.1 and 0.1);
        
        \draw[very thick, <-] (m1right) to[bend left=25, looseness=1.] (7.7,-1.8);
        \node[scale= 1.] (B) at (7.1,-2.4) {$\frac{b(-1)^{m_1}}{\sqrt{2}\sin(\pi m_1 b^2)}$};
        \node[scale= 1.] (B) at (9.7,-2.4) {$\frac{b(-1)^{m_2}}{\sqrt{2}\sin(\pi m_2 b^2)}$};
        \draw[very thick, <-] (m2right) to[bend right=25, looseness=1.] (8.9,-1.8);
        \node[scale= 1] (B) at (6.2,2.8) {$\sqrt{2}\sin(2\pi m_1 b p_1)$};
        \node[scale= 1] (B) at (6.2,-.8) {$\sqrt{2}\sin(2\pi m_1 b p_2)$};
        \node[scale= 1] (B) at (10.5,2.8) {$\sqrt{2}\sin(2\pi m_2 b p_4)$};
        \node[scale= 1] (B) at (10.5,-.8) {$\sqrt{2}\sin(2\pi m_2 b p_3)$};
        \draw[<-,very thick] (8.25,1.2) -- (8.25,3.3);
        \node[scale= 1] (B) at (8.25,3.6) {$\displaystyle\int' (-2q\mathrm{d}q)\sin(2\pi m_1  b q)\sin(2\pi m_2  b^{-1} q)$};

        \end{scope}
    \end{tikzpicture}
     .
\end{aligned}
\end{equation}
The general string amplitude $\mathsf{A}_{g,n}^{(b)}$ in (\ref{eq:Agn through quantum volumes}) may similarly be obtained by repeated application of these Feynman rules. However note that the number of stable graphs (Feynman diagrams) grows very quickly with the genus of the surface and the number of boundary insertions.

\subsection{A semiclassical limit of the string amplitudes}
In the Virasoro minimal string, the string amplitudes reduce precisely to the Weil-Petersson volumes in the limit in which the worldsheet central charge is taken to infinity, in accordance with the fact that the worldsheet theory reduces to JT gravity in this semiclassical limit \cite{Collier:2023cyw}. One might wonder whether the string amplitudes of the complex Liouville string exhibit a similar simplification in an analogous semiclassical limit in which the imaginary part of the worldsheet central charge is taken to infinity; after all, in this limit the sine dilaton gravity theory that describes the worldsheet theory reduces to de Sitter JT gravity \cite{paper4}. Here we will see that a similar simplification occurs at the level of the semiclassical limit of the string amplitudes.

We will take the $\im c\to \infty$ limit as (recall that $-ib^2\in\mathbb{R}_+$)
\begin{equation}\label{eq:JT limit resolvents}
-ib^2 \to \infty ~.
\end{equation}
In this limit, we scale the Liouville momenta with $b$ so that
\begin{equation}
    p \sim -\frac{i \ell b}{4\pi} ~.
\end{equation}
Here $\ell$ is held fixed. In the complex Liouville string it is natural for the Liouville momenta $p$ to have either the opposite or same $\mathrm{e}^{\frac{\pi i}{4}}$ phase as $b$ (the two situations are related by the duality symmetry), corresponding to either real or purely imaginary $\ell$, respectively. In the semiclassical limit $\ell$ will be identified with a geodesic length. 

The behavior of the string amplitudes in the semiclassical limit is most transparent in the representation (\ref{eq:Agn through quantum volumes}) involving the sum over stable graphs corresponding to degenerations of the worldsheet surface. Associated with each vertex of the stable graph is a factor of the quantum volume $\mathsf{V}_{g_v,n_v}^{(b)}$. In this semiclassical limit, the quantum volumes simply reduce to the corresponding Weil-Petersson volumes $V_{g_v,n_v}$ \cite{Collier:2023cyw}
\begin{equation}\label{eq:VMS to WP volumes}
    \mathsf{V}_{g,n}^{(b)}(i\boldsymbol{p}) \sim \left(\frac{b^2}{8\pi^2}\right)^{3g-3+n}V_{g,n}(\bm{\ell})\, .
\end{equation}
The corrections are suppressed in powers of $1/b^2$. In the sum over stable graphs, we see that the leading contribution comes from the $m_v=1$ terms in the sums over colors; the contributions of higher colors are exponentially suppressed at large $b^2$. Each stable graph with all colors set to one hence has the same exponential scaling at large $b^2$. However, each integration over internal momenta is further suppressed by a factor of $1/b^4$, one factor of $b^{-2}$ from integration over an internal edge \eqref{eq:primed integral} and another from the sub-volumes \eqref{eq:VMS to WP volumes} comprising the degenerated surface. Therefore, we conclude that the leading contribution comes solely from the trivial stable graph; the contributions from degenerated surfaces are all subleading in the semiclassical limit. We thus find
\begin{equation}\label{eq:semiclassical limit of Agn}
    \mathsf{A}_{g,n}^{(b)}(\boldsymbol{p}) \sim \left(\frac{i b^4}{16\pi^3}\e^{\pi i b^2}\right)^{2g-2+n}\prod_{j=1}^n \left(\frac{4\pi}{b}\sin(-\tfrac{i\ell_j b^2}{2})\right)V_{g,n}(\bm{\ell})\, .
\end{equation}
Hence the semiclassical limit of the string amplitudes reduces to the corresponding Weil-Petersson volume, up to a renormalization of the vertex operators and of the string coupling constant. Notably, the renormalization of the string coupling constant that appears above is \emph{purely imaginary}, leading to oscillations in the sum over genera --- indeed, we will see in \cite{paper3} that the effective string coupling deduced from the large-genus asymptotics of the string amplitudes is imaginary. We take this as an indication that the semiclassical limit of the complex Liouville string corresponds to \emph{de Sitter} JT gravity \cite{Maldacena:2019cbz, Cotler:2019nbi,Cotler:2024xzz}.\footnote{In this context, imaginary $\ell$ is more natural \cite{Cotler:2019nbi,Cotler:2024xzz}. This corresponds to the case where the Liouville momenta $p$ have the same rather than opposite phase as $b$.}

It is interesting to compare this to the semiclassical limit of the spectral curve itself. The semiclassical limit of the string amplitudes led to a projection to the $m=1$ term in the sum over colors (\ref{eq:Agn through quantum volumes}), so we expand the spectral curve (\ref{eq:xy spectral curve w}) around the $m=1$ branch point by writing\footnote{In principle we could expand the spectral curve around any of the other branch points, but these would lead to string amplitudes that are non-perturbatively suppressed compared to (\ref{eq:semiclassical limit of Agn}) in the semiclassical limit.}
\begin{equation}\label{eq:zoom near first branch pt}
    w \sim b + \frac{2u}{b}\, .
\end{equation}
This expansion of the spectral curve yields
\begin{align}
    \mathsf{x}(u) &\sim 2 - \frac{4\pi^2}{b^4} u^2\\
    \mathsf{y}(u) &\sim 2 \cos(\pi b^2 +2\pi u)\nonumber\\
    &= 2(\cos(\pi b^2)\cos(2\pi u) - \sin(\pi b^2)\sin(2\pi u))\, .
\end{align}
In this expansion the spectral curve now has just a single branch point corresponding to $u=0$. The input to topological recursion then becomes
\begin{subequations}
\begin{align}
    \omega_{0,1}^{(b)}(u) &= \frac{16\pi^2 u}{b^4}\left(\cos(\pi b^2)\cos(2\pi u) - \sin(\pi b^2)\sin(2\pi u)\right)\d u\label{eq:omega01 semiclassical}\\
    \omega_{0,2}^{(b)}(u_1,u_2) &= \left(\frac{1}{(u_1-u_2)^2} - \frac{1}{(u_1+u_2+b^2)^2}\right)\d u_1\, \d u_2\nonumber\\
    &\sim \frac{\d u_1\, \d u_2}{(u_1-u_2)^2}\, .
\end{align}
\end{subequations}
The first term involving $\cos(2\pi u)$ in (\ref{eq:omega01 semiclassical}) may appear unfamiliar, but it actually does not give any contribution to the topological recursion (\ref{eq:topological recursion}), because it is continuous around the branch point $u=0$ (in other words, it is projected out by the combination $\omega_{0,1}^{(b)}(u)-\omega_{0,1}^{(b)}(-u)$ that appears in the recursion kernel). In the semiclassical limit we may then take $\omega_{0,1}^{(b)}$ to be given by
\begin{align}\label{eq:reduce to JT gravity spectral curve}
    \omega_{0,1}^{(b)}(u) &\sim \frac{8\pi^2 \mathrm{e}^{-\pi i b^2}}{ib^4}u \sin(2\pi u) \d u \nonumber\\
    & = \left(\frac{i b^4}{16\pi^3}\mathrm{e}^{\pi i b^2}\right)^{-1}\omega_{0,1}^{(\text{JT})}(u)\, .
\end{align}
This is proportional to the input of JT gravity to topological recursion, $\omega_{0,1}^{(\text{JT})}(u) = \tfrac{u}{2\pi}\sin(2\pi u)\d u$ \cite{Saad:2019lba}, with the constant of proportionality precisely reproducing the renormalization of the string coupling that we observe in the semiclassical limit of the string amplitudes (\ref{eq:semiclassical limit of Agn}). The remaining normalization factors in (\ref{eq:semiclassical limit of Agn}) are produced by the semiclassical limit of the map between the string amplitudes and the resolvent differentials (\ref{eq:Agn omegagn relation}). 
We can similarly zoom into any of the other branch points of the spectral curve \eqref{eq:spectral curve} and find that, once again, it reduces to the JT gravity spectral curve; in particular, we also observe an imaginary renormalization of the string coupling, as in \eqref{eq:reduce to JT gravity spectral curve}. 
More generally, the resolvent differentials of the complex Liouville string reduce to those of JT gravity in the limit \eqref{eq:JT limit resolvents} near the $m=1$ branch  point \eqref{eq:zoom near first branch pt}. In the conventions of this paper, we have 
\begin{align}\label{eq:resolvents reduction to JT}
\omega^{(b)}_{g,n}(w_1,\ldots,w_n) \longrightarrow \left(\frac{i b^4}{16\pi^3}\mathrm{e}^{\pi i b^2}\right)^{2g-2+n} \omega^{(\mathrm{JT})}_{g,n}(u_1,\dots,u_n) ~.
\end{align}

\subsection{Recursion relation} \label{subsec:recursion relation}

Having established the topological recursion for the matrix integral and the relation between the resolvent differentials and the string amplitudes, we are now in a position to write down a recursion relation for the string amplitudes themselves.

\begin{figure}[ht]
    \centering
    \begin{tikzpicture}[baseline={([yshift=-.5ex]current bounding box.center)},scale=1]
        
        \draw[thick,blue] (1,1.32) to[out=0, in = 0,looseness=.8] (1,1/16);
        \draw[thick, blue, dashed] (1,1.32) to[out=180, in = 180, looseness=.8] (1,1/16);

        \draw[thick,blue] (1-2/8,-1/16) to[out=-90,in=-90] (.3,-1/16);
        \draw[thick,blue,dashed] (1-2/8,-1/16) to[out=90,in=90] (.3,-1/16);

        \draw[fill=blue,draw=blue,opacity=.4] ({0+1/4*cos(60)},{1+1/2*sin(60)}) to[out=-30, in = 180] (1,1.32) to[out=0, in = 0,looseness=.8] (1,1/16) to[out=0, in = 90] (1-2/8,-1/16) to[out = -90, in = -90] (.3,-1/16) to[out=90, in = 0] (0,1/2) to[out=0, in = -30,looseness=.75] ({{0+1/4*cos(60)}},{1+1/2*sin(60)});

        \draw[very thick] (0,1) ellipse (1/4 and 1/2);
        \draw[very thick] (0,-1) ellipse (1/4 and 1/2);
        \draw[very thick] (0,1/2) to[out=0, in = 0] (0,-1/2);
        \draw[very thick] ({0+1/4*cos(60)},{1+1/2*sin(60)}) to[out= -30, in = 180] (2,3/2) to [out=0, in=90] (3,0) to [out=-90, in= 0] (2,-3/2) to[out=180, in = 30] ({{0+1/4*cos(-60)}},{-1+1/2*sin(-60)}); 

        \node at (0,1) {$p_1$};
        \node at (0,-1) {$p_2$};

        \draw[thick] (5/8,0) to[out=-30, in = 210] (11/8,0);
        \draw[thick] (1-2/8,-1/16) to[out=60, in = 120] (1+2/8,-1/16);

        \draw[thick] (14/8+1/8,0) to [out=-30, in=210] (20/8+1/8,0);
        \draw[thick] (17/8-2/8+1/8,-1/16) to[out=60, in = 120] (17/8+2/8+1/8,-1/16);

        \node[scale = 1] at (1.75,-3/4) {$\mathsf{A}_{g-1,n+1}^{(b)}$};

        \begin{scope}[shift={(10,0)}]
            \draw[thick,blue] (1/2, 1.3) to[out=0, in =0,looseness=.2] (1/2,-1.3);
            \draw[thick,blue,dashed] (1/2, 1.3) to[out=180, in =180,looseness=.2] (1/2,-1.3);
            \draw[draw=blue, fill=blue, opacity=.4] ({0+1/4*cos(60)},{1+1/2*sin(60)}) to[out=-30, in = 180] (1/2,1.3) to[out=0, in = 0,looseness=.2] (1/2,-1.3) to[out=180, in = 30] ({{0+1/4*cos(-60)}},{-1+1/2*sin(-60)}) to[out=30, in =0, looseness=.75] (0,-1/2) to[out=0, in=0] (0,1/2) to[out=0, in = -30,looseness=.75] ({{0+1/4*cos(60)}},{1+1/2*sin(60)});
    
            \draw[very thick] (0,1) ellipse (1/4 and 1/2);
            \draw[very thick] (0,-1) ellipse (1/4 and 1/2);
            \draw[very thick] (0,1/2) to[out=0, in = 0] (0,-1/2);
            \draw[very thick] ({0+1/4*cos(60)},{1+1/2*sin(60)}) to[out= -30, in = 180] (2,3/2) to [out=0, in=90] (3,0) to [out=-90, in= 0] (2,-3/2) to[out=180, in = 30] ({{0+1/4*cos(-60)}},{-1+1/2*sin(-60)}); 
    
            \node at (0,1) {$p_1$};
            \node at (0,-1) {$p_2$};
    
            \draw[thick] (5/8,0) to[out=-30, in = 210] (11/8,0);
            \draw[thick] (1-2/8,-1/16) to[out=60, in = 120] (1+2/8,-1/16);
    
            \draw[thick] (14/8+1/8,0) to [out=-30, in=210] (20/8+1/8,0);
            \draw[thick] (17/8-2/8+1/8,-1/16) to[out=60, in = 120] (17/8+2/8+1/8,-1/16);
    
            \node[scale = 1] at (1.75,-3/4) {$\mathsf{A}_{g,n-1}^{(b)}$};
        \end{scope}

        \begin{scope}[shift={(5,0)}]
            \draw[thick, blue] (1.7,1.48) to[out=0,in=0,looseness=.15] (1.7,-1.48);
            \draw[thick, blue, dashed] (1.7,1.48) to[out=180,in=180,looseness=.15] (1.7,-1.48);
    
            \draw[thick, blue] (1/4+.05,0) to[out=0, in =180] (1,1/4) to[out=0, in = 90] (3/2,0) to[out= -90, in =60,looseness=.4] (3/2,-1.45);
            \draw[thick, blue, dashed] (3/2,-1.45) to [out=150, in =-90,looseness=.4] (3/2-.05,0) to[out=90, in = 0] (1,1/4-0.05) to[out=180, in = 0] (1/4+.05,-.05);
    
            \draw[fill=blue, draw=blue, opacity=.4] ({0+1/4*cos(60)},{1+1/2*sin(60)}) to[out= -30, in=180] (3/4,1.28) to[out=0, in = 180,looseness=.9] (1.7,1.48) to[out=0,in=0,looseness=.15] (1.7,-1.48) to[out=180, in = -10] (3/2,-1.45) to[out=60, in = -90,looseness=.4] (3/2,0) to[out = 90, in = 0] (1,1/4) to[out=180, in = 0] (1/4+.05,0) to[out=90, in = 0] (0,1/2) to[out= 0, in = -30,looseness=.75] ({{0+1/4*cos(60)}},{1+1/2*sin(60)});
    
            \draw[fill=blue, draw=blue, opacity=.25] (3/2,-1.45) to[out=150, in =-90,looseness=.4] (3/2-.05,0) to[out=90, in = 0] (1,1/4-0.05) to[out=180, in = 0] (1/4+.05,-.05) to (1/4+.05,0) to[out=0, in =180] (1,1/4) to[out=0, in = 90] (3/2,0) to[out= -90, in =60,looseness=.4] (3/2,-1.45);
    
            \draw[very thick] (0,1) ellipse (1/4 and 1/2);
            \draw[very thick] (0,-1) ellipse (1/4 and 1/2);
            \draw[very thick] (0,1/2) to[out=0, in = 0] (0,-1/2);
            \draw[very thick] ({0+1/4*cos(60)},{1+1/2*sin(60)}) to[out= -30, in = 180] (2,3/2) to [out=0, in=90] (3,0) to [out=-90, in= 0] (2,-3/2) to[out=180, in = 30] ({{0+1/4*cos(-60)}},{-1+1/2*sin(-60)}); 
    
            \node at (0,1) {$p_1$};
            \node at (0,-1) {$p_2$};
    
            \draw[thick] (5/8,0) to[out=-30, in = 210] (11/8,0);
            \draw[thick] (1-2/8,-1/16) to[out=60, in = 120] (1+2/8,-1/16);
    
            \draw[thick] (14/8+1/8,0) to [out=-30, in=210] (20/8+1/8,0);
            \draw[thick] (17/8-2/8+1/8,-1/16) to[out=60, in = 120] (17/8+2/8+1/8,-1/16);
    
            \node[scale = 1] at (.8,-1.85) {$\mathsf{A}_{h,1+|\mathcal{I}|}^{(b)}$};
            \node[scale = 1] at (2.5,-1.85) {$\mathsf{A}_{g-h,1+|\mathcal{J}|}^{(b)}$};
        \end{scope}

\end{tikzpicture}
\caption{The three distinct ways of embedding a pair of pants with a distinguished external cuff (labelled by $p_1$ above) into a surface. These correspond to the three classes of terms in (\ref{eq:Agn simpler residue recursion}) and (\ref{eq:Agn recursion relation}). There is a factor of the sphere three-point amplitude $\mathsf{A}_{0,3}^{(b)}$ corresponding to this pair of pants for each term in the recursion.}
\label{fig:recursion pants}
\end{figure}
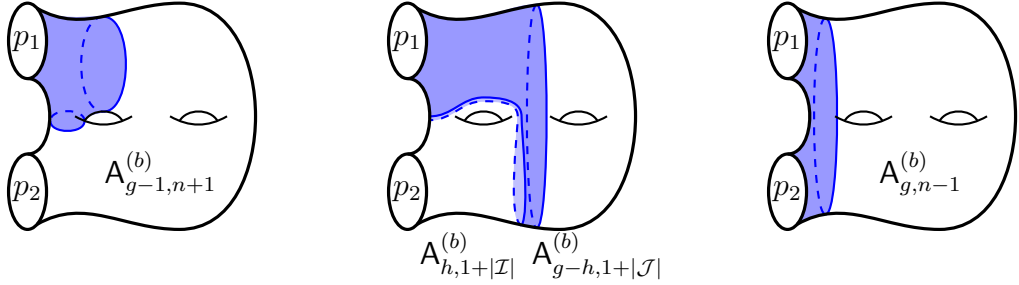

Translating the topological recursion (\ref{eq:topological recursion}) to the string amplitudes via (\ref{eq:Agn omegagn relation}), we arrive at the following recursive representation
\begin{align}
        &\quad p_1 \mathsf{A}_{g,n}^{(b)}(p_1,\boldsymbol{p})\nonumber\\
        &= \frac{\pi}{2} \sum_{m=1}^\infty \frac{b(-1)^m\sin(2\pi m b p_1)}{\sin(\pi m b^2)}\Res_{u=0}\bigg\{\frac{\sin(4\pi u p_1)}{\sin(2\pi b u)\sin(2\pi b^{-1}u)}\bigg[\int 2q \d q\, 2q' \d q'\nonumber\\
        &\quad\quad\times \bigg(\sum_{\pm}\pm \cos(4\pi u(q\pm q'))\cos(2\pi m b(q\mp q'))\bigg)\nonumber\\
        &\quad\quad\times\bigg(\mathsf{A}_{g-1,n+1}^{(b)}(q,q',\boldsymbol{p}) + \sum_{h=0}^g \sideset{}{'}\sum_{\mathcal{I},\mathcal{J}}\mathsf{A}_{h,1+|\mathcal{I}|}^{(b)}(q,\boldsymbol{p}_{\mathcal{I}})\mathsf{A}_{g-h,1+|\mathcal{J}|}^{(b)}(q',\boldsymbol{p}_{\mathcal{J}})\bigg)\nonumber\\
        &\quad - 2\sum_{j=2}^n\int 2q \d q \bigg(\sum_{\pm} \pm \cos(4\pi u(q\pm p_j))\cos(2\pi m b(q\mp p_j))\bigg)\mathsf{A}_{g,n-1}^{(b)}(q,\boldsymbol{p}\setminus  p_j)\bigg]\bigg\}\, .\label{eq:Agn recursion first pass}
\end{align}
Here $\boldsymbol{p} = \{p_2,\ldots, p_n\}$, and the sum in the third line runs over all subsets $\mathcal{I}\cup \mathcal{J} = \{p_2,\ldots,p_n\}$ excluding $(h,\mathcal{I}) = (0,\emptyset)$ and $(h,\mathcal{J}) = (g,\emptyset)$. The integrals over $q,q'$ are defined as in the first case of (\ref{eq:primed integral}). In practice we expand the string amplitudes into sums of terms involving complex exponentials $\mathrm{e}^{2\pi i m b q}$, and hence the integrand does not exhibit poles in $q$ term-by-term and we may freely deform the $q$ contour in order to apply (\ref{eq:primed integral}).
We can make use of the symmetry properties of the string amplitudes to simplify this recursive representation somewhat\footnote{There is an exception. In writing (\ref{eq:Agn simpler residue recursion}) we have used the fact that the string amplitudes depend on a particular momentum $q$ via a sum of terms involving even polynomials in $q$ times factors of $\sin(2\pi m b q)$, for $m$ an integer. The recursion as written below doesn't apply for $\mathsf{A}_{1,1}^{(b)}$ because $\mathsf{A}_{0,2}^{(b)}$, which appears in the recursion, does not take this form. Nevertheless one can verify that the final form of the recursion relation given in equation (\ref{eq:Agn recursion relation}) holds in this case (up to a factor of $\frac{1}{2}$ due to a symmetry factor of the configuration).}
\begin{align}
        &\quad p_1 \mathsf{A}^{(b)}_{g,n}(p_1,\boldsymbol{p}) \nonumber\\
        &= \frac{\pi}{2}\Res_{u=0}\bigg\{ \frac{\sin(4\pi u p_1)}{\sin(2\pi b u)\sin(2\pi b^{-1}u)}\bigg[\int 2q \d q \, 2q' \d q' \cos(4\pi u q)\cos(4\pi u q')\mathsf{A}_{0,3}^{(b)}(q,q',p_1) \nonumber\\
        &\quad\quad\times\bigg(\mathsf{A}_{g-1,n+1}^{(b)}(q,q',\boldsymbol{p}) + \sum_{h=0}^g \sideset{}{'}\sum_{\mathcal{I},\mathcal{J}}\mathsf{A}_{h,1+|\mathcal{I}|}^{(b)}(q,\boldsymbol{p}_{\mathcal{I}})\mathsf{A}_{g-h,1+|\mathcal{J}|}^{(b)}(q',\boldsymbol{p}_{\mathcal{J}})\bigg)\nonumber\\
        &\quad - 2\sum_{j=2}^n\int 2q \d q\, \cos(4\pi u q)\cos(4\pi u p_j)\mathsf{A}_{0,3}^{(b)}(p_1,p_j,q)\mathsf{A}_{g,n-1}^{(b)}(q,\boldsymbol{p}\setminus p_j)\bigg]\bigg\}\, . \label{eq:Agn simpler residue recursion}
\end{align}  
The three different terms in the sum correspond to the three topologically distinct ways of embedding a pair of pants with a distinguished external leg $p_1$ into the surface $\Sigma_{g,n}$, as shown in figure \ref{fig:recursion pants}. Indeed there is a factor of $\mathsf{A}_{0,3}^{(b)}$ corresponding to this distinguished pair of pants for each term in the recursion.

\begin{figure}[ht]
    \centering
    \begin{tikzpicture}[baseline={([yshift=-.5ex]current bounding box.center)},scale=1]
        \draw[thick, <->] (-3.5,0) -- (3.5,0);
        \draw[thick, <->] (0,-2.3) -- (0,2.3);
        \foreach \x in {-2.12132, -1.59099, -1.06066, -0.53033, 0.53033, 1.06066, 1.59099, 2.12132}{
            \draw ({\x},{\x}) node[cross, very thick] {};
        }
    
        \foreach \x in {-2.12132, -1.59099, -1.06066, -0.53033, 0.53033, 1.06066, 1.59099, 2.12132}{
            \node[thick, cross out, scale=.9, draw=black] at ({\x},{\x}) {};
        }
    
        \foreach \x in {-1.88562, -0.942809, 0.942809, 1.88562}{
            \node[thick, cross out, scale=.9, draw=black] at({\x},{-\x}) {};
        }
    
        \draw[very thick, red, decoration={markings, mark=at position 0.40 with {\arrow{<}}}, postaction={decorate}] (2,2.4) to[out=225, in =90, looseness=.75] (0,0) to[out= -90, in=45,looseness=.75] (-2,-2.4);
    
        \node[above] at (2,2.4)  {$\Gamma$};
        \node(n)[inner sep=2pt] at (3.5,2) {$u$};
        \draw[line cap=round](n.north west)--(n.south west)--(n.south east);
\end{tikzpicture}
\caption{The contour of integration that defines the recursion kernel (\ref{eq:recursion kernel}).}\label{fig:recursion kernel contour}
\end{figure}
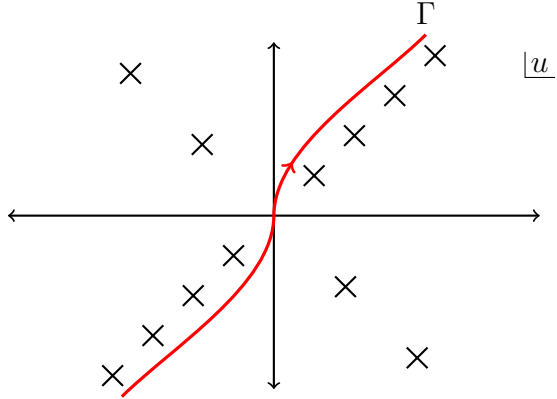

We can further massage the representation of the residue in (\ref{eq:Agn simpler residue recursion}) to write the recursion for the string amplitudes in a more conventional form. At the end of the day we find the following more familiar representation for the recursion relation satisfied by the string amplitudes
\begin{align}
        p_1 \mathsf{A}_{g,n}^{(b)}(p_1,\boldsymbol{p}) &= \int 2q \d q \, 2q' \d q' \, \mathsf{H}_b(q+q',p_1)\mathsf{A}_{0,3}^{(b)}(p_1,q,q')\nonumber\\
        &\quad\quad\times\left(\mathsf{A}_{g-1,n+1}^{(b)}(q,q',\boldsymbol{p}) + \sum_{h=0}^g \sideset{}{'}\sum_{\mathcal{I},\mathcal{J}}\mathsf{A}_{h,1+|\mathcal{I}|}^{(b)}(q,\boldsymbol{p}_{\mathcal{I}})\mathsf{A}_{g-h,1+|\mathcal{J}|}^{(b)}(q',\boldsymbol{p}_{\mathcal{J}})\right)\nonumber\\
        & \quad - \sum_{j=2}^n\int 2q \d q \, \sum_{\pm} \mathsf{H}_b(q,p_1\pm p_j)\mathsf{A}_{0,3}^{(b)}(p_1,p_j,q)\mathsf{A}_{g,n-1}^{(b)}(q,\boldsymbol{p}\setminus p_j)\, .\! \label{eq:Agn recursion relation}
\end{align}
Here the recursion kernel $\mathsf{H}_b$ is essentially identical to that which recently appeared in the recursion relations satisfied by the quantum volumes of the Virasoro minimal string \cite{Collier:2023cyw} 
\begin{equation}\label{eq:recursion kernel}
    \mathsf{H}_b(x,y) \coloneqq \frac{y}{2} - \frac{1}{2} \int_{\Gamma}\d u\, \frac{\sin(4\pi u x)\sin(4\pi u y)}{\sin(2\pi b u)\sin(2\pi b^{-1} u)}\, .
\end{equation}
The contour of integration $\Gamma$ is shown in figure \ref{fig:recursion kernel contour}.
The recursion kernel also admits the following infinite sum representation 
\begin{align}\nonumber
    \mathsf{H}_b(x,y) &= -\frac{1}{4\pi}\partial_y \log\bigg(\mathrm{e}^{\pi i y^2}\prod_{\pm }S_b\left(\frac{Q}{2}-x \pm y\right)\bigg) \\ \label{eq:recursion kernel double sine}
    &= y - \frac{1}{2}\sum_{m=0}^\infty \sum_{\sigma = \pm}\left[\frac{b^{-1}\sigma}{1+\mathrm{e}^{2\pi ib^{-1}((m+\frac{1}{2})b^{-1}-x-\sigma y)}} - \frac{b\sigma}{1+\mathrm{e}^{-2\pi i b((m+\frac{1}{2})b +x+\sigma y)}}\right]\, ,
\end{align}
where $S_b(x) = \frac{\Gamma_b(x)}{\Gamma_b(Q-x)}$ is the double-sine function. We elaborate on some details of the derivation of this recursive representation in appendix \ref{app:recursion derivation}.

A novel feature compared to recursion relations satisfied by the quantum volumes of the Virasoro minimal string is the presence of the non-trivial sphere three-point amplitude $\mathsf{A}_{0,3}^{(b)}$ for each topologically distinct term in the recursion corresponding to the pair of pants involving $p_1$.\footnote{Recall that in the Virasoro minimal string $\mathsf{V}_{0,3}^{(b)}(ip_1,ip_2,ip_3) = 1$.}

In order to efficiently implement the recursion, it will be useful to note some regularly appearing integral formulas involving the recursion kernel. We define for instance
\be
     F_{k;m,n}(y) \coloneqq \int 2x \d x \, x^{2k} \mathsf{H}_b(x,y)\sin(2\pi m b x)\sin(2\pi n b x)\,,
\ee
for $k\in\mathbb{Z}_{\geq 0}$ and $m,n\in\mathbb{Z}_{\geq 1}$. These integrals are simplest to evaluate in the situation that none of the arguments of the sines (in other words, none of the colors) coincide. In these situations we simply have
\begin{align}\nonumber
        F_{k;m,n}(y) &= \frac{y}{2}\int 2x \d x\, x^{2k}\sin(2\pi m b x)\sin(2\pi n b x)\\
        &=\frac{(-1)^{k}(2k+1)!}{2 (2\pi b)^{2k+2}}\left(\frac{1}{(m+n)^{2k+2}}-\frac{1}{(m-n)^{2k+2}}\right)y
\end{align}
for $m\ne n$. The integral formulas get more complicated when some of the colors coincide. In this situation the recursion kernel regulates the integral in essentially the same way as in the Virasoro minimal string \cite{Collier:2023cyw}. In this case we have
\begin{multline}
    F_{k;m,m}(y) = \frac{(-1)^k (2k+1)! }{2 (4\pi b m)^{2k+2}}y\\
     + \sum_{0\leq \ell+j\leq k+1}\frac{B_{2\ell}B_{2j}(1-2^{1-2\ell})(1-2^{1-2j})(2k+1)! b^{2\ell-2j}}{2(2\ell)!(2j)!(2k+3-2\ell-2j)!}y^{2k+3-2\ell-2j}\, ,
\end{multline}
where $B_{2\ell}$ are the Bernoulli numbers. The latter term above generates the Virasoro minimal string quantum volumes $\mathsf{V}_{g,n}^{(b)}$ that appear in the string amplitudes as in (\ref{eq:Agn through quantum volumes}). 

\subsection{Cohomological Field Theory and \texorpdfstring{$\mathrm{SU}(2)_q$}{SU(2)q} Yang-Mills theory} \label{subsec:CohFT}
We now refine the discussion and consider the cohomology classes in $\mathrm{H}^\bullet(\bM_{g,n},\CC)$ that appear in the intersection number formula for $\mathsf{A}_{g,n}^{(b)}$ \eqref{eq:Agn through quantum volumes}. They define a cohomological field theory (CohFT).\footnote{We thank Alessandro Giacchetto and Nikita Nekrasov for discussions about this.}

\paragraph{Definition.}
Let us recall the definition of a CohFT \cite{Kontsevich:1994qz}. Let $\mathcal{H}$ be a Hilbert space over $\CC$. Then a CohFT over $\CC$ is a collection of maps
\be 
\Omega_{g,n}: \mathcal{H}^{\otimes n} \longrightarrow \mathrm{H}^\bullet(\bM_{g,n},\CC) \label{eq:Omegagn maps}
\ee
assigning cohomology classes to collection of vectors. Given a CohFT, one can also define correlators of gravitational descendants,
\be 
\big \langle \tau_{k_1}(v_1) \cdots \tau_{k_n}(v_n) \big \rangle_g=\int_{\bM_{g,n}} \Omega_{g,n}(v_1,\dots,v_n) \prod_{i=1}^n \psi_i^{k_i}~,
\ee
where $\psi_i$ are the standard psi-classes.
These will be related to the string amplitudes. 
The maps \eqref{eq:Omegagn maps} satisfy two axioms:
\begin{enumerate}
    \item Symmetry: $\Omega_{g,n}$ is invariant under simultaneous permutation of its arguments and the marked points of $\bM_{g,n}$.
    \item Factorization: Let $\iota_{h,I} :\mathscr{D}_{h,I} \cong \bM_{h,|I|+1} \times \bM_{g-h,|I^c|+1}\longrightarrow \bM_{g,n}$ and $\iota_\text{irr}: \mathscr{D}_{\text{irr}} \cong \bM_{g-1,n+2} \longrightarrow \bM_{g,n}$ be the embedding maps of the boundary divisors. Then
    \be 
        \iota^*_{h,I}\Omega_{g,n}(\boldsymbol{v})=\sum_m \Omega_{h,|I|+1}(\boldsymbol{v}_I,e_m)   \Omega_{g-h,|I^c|+1}(\boldsymbol{v}_{I^c},e_m)~, \label{eq:cohft factorization axiom}
    \ee
    where $\{e_m\}_{m=1,\dots,\dim(\mathcal{H})}$ is a complete orthonormal basis. A similar statement holds for $\iota_\text{irr}$. In other words, when we restrict the cohomology class to the boundary of moduli space where the surface separates into two parts, it is given by the product of the classes on the two parts with a complete set of states inserted at the node.
\end{enumerate}
Sometimes a third axiom of a flat unit is added. It does not hold in the case of interest and we omit it.

\paragraph{Examples.} Let us mention two very simple examples. In both cases the Hilbert space is one-dimensional and we omit the basis vector.
\begin{enumerate}
    \item JT-gravity: $\Omega_{g,n}=\exp(2\pi^2 \kappa_1)$, where $2\pi^2 \kappa_1=[\omega_\text{WP}]$ is the cohomology class of the Weil-Petersson form. Since the Weil-Petersson form restricts to the direct sum of the form on both factors, the factorization axiom \eqref{eq:cohft factorization axiom} clearly holds. One can recover the Weil-Petersson volumes out of the gravitational descendant correlators.
    \item Virasoro minimal string: 
    $\Omega_{g,n}=\exp\big(\frac{c-13}{24}\kappa_1-\sum_{m \ge 1} \frac{B_{2m}}{2m(2m)!} \kappa_{2m}\big)$. The higher kappa-classes again restrict to their direct sum on both factors and the factorization axiom holds.
\end{enumerate}
There are many other examples of CohFTs in the literature such as the Hodge class $\Lambda=c(\mathbb{E})$ with $\mathbb{E}$ the Hodge bundle, Norbury's Theta-class $\Theta_{g,n}$ needed for supersymmetric JT-gravity \cite{Norbury:2017eih}, Witten's $r$-spin class \cite{witten1993algebraic} ($\dim \mathcal{H}=r-1$), the Chern-character of the Verlinde bundle \cite{Marian:2015ezv} ($\dim \mathcal{H}$ is the number of representations of the current algebra $\mathfrak{g}_k$), and the pushforward of Gromov-Witten with target a projective variety $X$ classes to $\bM_{g,n}$ \cite{Kontsevich:1994qz}.

\paragraph{CohFTs and topological recursion.}
Cohomological field theories are closely related to topological recursion. In fact, every semi-simple cohomological field theory produces a spectral curve such that the differentials $\omega_{g,n}$ as computed from topological recursion are related to the descendant correlators as
\be 
\omega_{g,n}(z_1,\dots,z_n)=\sum_{(m_1,k_1),\dots,(m_n,k_n)} \big \langle \tau_{k_1}(e_{m_1}) \cdots \tau_{k_n}(e_{m_n}) \big \rangle_g \prod_{i=1}^n \d \eta_{k_i}(z_i)~,
\ee
for some set of differentials $\d \eta_{k_i}(z_i)$. See \cite{Dunin-Barkowski:2012kbi} for the precise formula. Here $m_i$ runs over the set of branch points and $k_i$ runs over $\ZZ_{\ge 0}$. This is very similar to what we explained in appendix~\ref{app:intersection theory}, but it produces a spectral curve with a special choice of coordinates.

The reverse also holds for a global spectral curve under certain conditions. In particular, it does hold for a compact spectral curve with holomorphic differentials $\d \xx(z)$ and $\d \yy(z)$ \cite{Dunin-Barkowski:2015caa}. The spectral curve of interest \eqref{eq:spectral curve} is not compact, but $\d \xx(z)$ and $\d \yy(z)$ are holomorphic. Furthermore, since the sum over the branch points converges absolutely, one can check that the proofs given in \cite{Dunin-Barkowski:2015caa, Dunin-Barkowski:2016dec} continue to go through. Thus we can indeed uniquely define a CohFT out of the complex Liouville string. We will not write down explicit formulas since they become rather complicated.

\paragraph{The topological field theory.}
Out of a cohomological field theory, we can always define a topological field theory by taking out the degree 0 piece of the cohomology and identifying canonically $\mathrm{H}^0(\bM_{g,n},\CC) \equiv \CC$. Thus we get maps
\be 
\Omega_{g,n}^\text{TQFT}: \mathcal{H}^{\otimes n} \longrightarrow \CC~.
\ee
This can be directly extracted from \eqref{eq:Agn through quantum volumes} by taking the degree zero piece of the integrand. 
We first notice that in cohomology, any non-trivial graph corresponds to an intersection number in a lower-dimensional moduli space and does not contribute to the degree zero piece.\footnote{More formally, we can write the contribution of a graph $\Gamma$ as an integral over the moduli space $\bM_\Gamma$ as in appendix~\ref{app:intersection theory}. The pushforward $(\iota_\Gamma)_*$ from the inclusion $\iota_\Gamma:\bM_\Gamma \longrightarrow \bM_{g,n}$ shifts the degree of the cohomology classes upward by the codimension of $\bM_\Gamma$. Thus only $\bM_\Gamma=\bM_{g,n}$ can contribute.}\footnote{It is a general feature of CohFTs that the cohomological classes can be written as sums over stable graphs. This corresponds to the action of the Givental R-matrix \cite{givental2001gromov}.}
It remains to pick the degree zero piece of the integrand of the quantum volumes, which is the second example discussed above. The integrand is an exponential and thus its degree zero piece is 1. Thus we get the degree zero piece from \eqref{eq:Agn through quantum volumes} by restricting to the trivial graph and replacing the quantum volume with $1$. There are no internal edges and hence we simply get
\be 
\mathsf{A}_{g,n}^{(b),\text{TQFT}}(\boldsymbol{p})=\sum_{m=1}^\infty \left(\frac{b  (-1)^{m}}{\sqrt{2} \sin(\pi m b^2)}\right)^{2g-2+n} \prod_{i=1}^n \sqrt{2} \sin(2\pi m b p_i)~. \label{eq:TQFT}
\ee
This is the correlation function of $\mathrm{SU}(2)_q$ Yang-Mills theory and up to normalization computes the Schur index of $\mathrm{SU}(2)$ $\mathcal{N}=2$ theories in four dimensions of class $\mathcal{S}$ \cite{Gadde:2011ik}. 
We also remark that we would have obtained the trivial TQFT if we had performed this procedure for JT-gravity or the Virasoro minimal string.
\paragraph{$\mathrm{SU}(2)_q$ Yang-Mills and the Schur index.}
$\mathrm{SU}(2)_q$ Yang-Mills theory (in the zero-area limit) is the topological field theory associated to the quantum group $\SU(2)_q$. Its representations are labelled by the dimension $m \in \ZZ_{\ge 1}$. They have character and quantum dimension
\be 
\text{ch}_q(m,a)=\frac{a^{m}-a^{-m}}{a-a^{-1}}~, \quad \dim_q(m)=\frac{q^{\frac{m}{2}}-q^{-\frac{m}{2}}}{q^{\frac{1}{2}}-q^{-\frac{1}{2}}}~.
\ee
Here, $a \in \mathrm{S}^1 \subset \CC$ parametrizes the Cartan torus of $\SU(2)$. The TQFT correlators are
\be 
\mathsf{Z}_{g,n}^{\mathrm{SU}(2)_q}(\boldsymbol{a})=\sum_{m=1}^\infty \frac{\prod_{i=1}^n \text{ch}_q(m,a_i)}{\dim_q(m)^{2g-2+n}}~.
\ee
$\mathrm{SU}(2)_q$ admits a hermitian dagger when $|q|=1$ or when $q>0$. We are interested in the case $q>0$. By using the Weyl group symmetry we can assume that $0<q<1$.
When we identify $q=\mathrm{e}^{2\pi i b^2}$ and $a_j=-\mathrm{e}^{2\pi i b p_j}$, we have\footnote{Recall that $b p_j \in \RR$ and thus we indeed have $|a_j|=1$.}
\be 
\mathsf{A}_{g,n}^{(b),\text{TQFT}}(\boldsymbol{p})=\left(-\frac{b}{\sqrt{2}\sin(\pi b^2)}\right)^{2g-2+n} \prod_{j=1}^n \sqrt{2} \sin(2\pi b p_j) \, \mathsf{Z}_{g,n}^{\mathrm{SU}(2)_q}(\boldsymbol{a})~.
\ee
Thus, after changing the normalization of the punctures and the Euler term the two theories agree.\footnote{$\mathrm{SU}(2)_q$ Yang-Mills theory is a semisimple TQFT and thus the corresponding CohFT is also semisimple as predicted by the correspondence between spectral curves and CohFTs.} 

We can also further relate this to the Schur index of four-dimensional gauge theories. The relation is well-known and arises by putting 6d $\mathcal{N}=(2,0)$ theory of type $\mathrm{A}_1$ on $\mathrm{S}^3 \times \mathrm{S}^1_\beta \times \Sigma_{g,n}$ with a suitable partial topological twist \cite{Gaiotto:2009hg, Gadde:2011ik}. Compactifying on $\Sigma_{g,n}$ leads to the supersymmetric index of the corresponding class $\mathcal{S}$ theory in four-dimensions, while compactifying on $\mathrm{S}^3 \times \mathrm{S}^1_\beta$ leads to 2d $\mathrm{SU}(2)_q$ Yang-Mills theory on $\Sigma_{g,n}$. The index obtained in this way is the Schur index, which is a degeneration of the more general superconformal index. For a class $\mathcal{S}$ theory on $\Sigma_{g,n}$, this index takes the form
\be 
\mathsf{Z}_{g,n}^\text{Schur}(\boldsymbol{a})=\frac{\prod_{j=1}^n \mathcal{N}(a_j)}{\mathcal{N}_0^{2g-2+n}} \, \mathsf{Z}_{g,n}^{\mathrm{SU}(2)_q}(\boldsymbol{a})~,
\ee
where
\be 
\mathcal{N}(a)=\frac{1}{\prod_{m=1}^\infty (1-q^m)(1-a^2 q^m)(1-a^{-2} q^m)}~, \quad \mathcal{N}_0=\frac{1}{\prod_{m=2}^\infty(1-q^m)}~.
\ee
Here $q=\mathrm{e}^{-\beta}$ is also real. These infinite products correspond to passing to characters of an affine algebra. Identifying $\beta = -2\pi i b^2$, we can thus write\footnote{The additional minus sign in the relation between $a_j$ and $p_j$ corresponds to not inserting $(-1)^\mathrm{F}$ in the definition of the Schur index.} 
\be 
\mathsf{A}_{g,n}^{(b),\text{TQFT}}(\boldsymbol{p}) = \left(\frac{i\sqrt{2}b\, q^{\frac{13}{24}}}{\eta(b^2)}\right)^{2g-2+n}\prod_{j=1}^n\frac{\vartheta_1(2bp_j|b^2)}{\sqrt{2}q^{\frac{1}{8}}}\mathsf{Z}_{g,n}^\text{Schur}(\boldsymbol{a})\, .
\ee
This also relates $\mathsf{A}_{g,n}^{(b),\text{TQFT}}$ to $\SL(2,\CC)$ Chern-Simons theory through Schur quantization as explained recently in \cite{Gaiotto:2024osr}.

\section{Checks} \label{sec:checks}
In this section, we will demonstrate that the topological recursion based on the spectral curve \eqref{eq:spectral curve} reproduces all the properties of the string diagrams $\mathsf{A}_{g,n}^{(b)}$ that we derived from the worldsheet in our previous paper \cite{paper1}. Sometimes it will be convenient to use the form as coming from the topological recursion and sometimes the recursion relation derived in section~\ref{subsec:recursion relation}.
\subsection{Simple properties}
Let us first notice some properties that are obvious from eq.~\eqref{eq:Agn through quantum volumes}. 
\paragraph{Oddness and one series of trivial zeros.}
The quantum volumes $\mathsf{V}_{g,n}^{(b)}$ are even functions of their arguments. Every external momentum appears additionally in one factor $\sqrt{2} \sin(2\pi m_{v_j} b p_j)$, which shows that $\mathsf{A}_{g,n}^{(b)}$ is an odd function of its arguments. The oddness of the string amplitudes is required from the worldsheet definition due to a property of the leg factors. The presence of the sine shows also that $\mathsf{A}_{g,n}^{(b)}$ vanishes when $p_j=\frac{m}{2b}$ for $m \in \ZZ$ and for any $j$. In \cite{paper1}, we referred to these zeros as the trivial zeros from the worldsheet since they are a consequence of the chosen leg factors. There is a second series of trivial zeros located at $p_j=\frac{mb}{2}$. These are not readily visible in the formula \eqref{eq:Agn through quantum volumes}, but follow from the duality symmetry discussed in section~\ref{subsec:duality symmetry} below.

\paragraph{$b \to -b$ symmetry.} The string diagram $\mathsf{A}_{g,n}^{(b)}$ is invariant under $b \to -b$. Under this replacement, \eqref{eq:Agn through quantum volumes} receives a factor
\be 
(-1)^n (-1)^{\sum_v (2g_v-2+n_v)}=(-1)^{n+(2g-2+n)}=1~,
\ee
as required from the worldsheet representation of the string amplitudes. Here we used additivity of the Euler characteristic of the stable graph.
\paragraph{Swap symmetry.} A more interesting symmetry is the swap symmetry that sends $b \to -i b$ and $p_j \to i p_j$ simultaneously. This corresponds to swapping the two Liouville CFTs on the worldsheet. This operation leads to the overall factor
\begin{align}\label{eq:Agn swap symmetry}
    i^{\sum_v (2g_v-2+n_v)} (-1)^{\# \text{edges}} (-1)^{\sum_v (3g_v-3+n_v)}=i^{2g-2+n} (-1)^{3g-3+n}=(-i)^n~.
\end{align}
Here the first of the three factors come from the term raised to the Euler characteristic in \eqref{eq:Agn through quantum volumes}. For the second, we also rotate $p_e \to i p_e$ for all internal edges, which leads to a Jacobian of $-1$ for every edge. The third factor comes from the corresponding property of the quantum volume which was discussed in \cite{Collier:2023cyw}. We then use that the number of edges is $3g-3+n-\sum_v(3g_v-3+n_v)$. Thus we have
\be 
\mathsf{A}_{g,n}^{(-ib)}(i \boldsymbol{p})=(-i)^n\, \mathsf{A}_{g,n}^{(b)}(\boldsymbol{p})~,
\ee
as required from the worldsheet.
\paragraph{Special case of $(g,n)=(0,4)$ and $(1,1)$.} Finally, we notice that \eqref{eq:Agn through quantum volumes} can be straightforwardly evaluated in the special cases $(g,n)=(0,3)$, $(0,4)$ and $(1,1)$, see Table~\ref{tab:stable graphs Agn 03, 11, 04}. We investigated those cases in detail in our previous paper \cite{paper1} and provided overwhelming evidence for the correctness of \eqref{eq:Agn through quantum volumes} in these cases.
\subsection{Duality symmetry}\label{subsec:duality symmetry}
We claim that $\mathsf{A}_{g,n}^{(b)}$ as computed by \eqref{eq:Agn through quantum volumes} satisfy
\be 
\mathsf{A}^{(b^{-1})}_{g,n}(\boldsymbol{p})=(-1)^n \, \mathsf{A}^{(b)}_{g,n}(\boldsymbol{p})~, \label{eq:duality symmetry}
\ee
in accordance with the corresponding symmetry on the worldsheet. On the worldsheet, this property is completely manifest from the bootstrap definition of Liouville theory. The sign $(-1)^n$ comes from the transformation property of the leg factors.
This property is non-trivial from the matrix integral side. Notice that from the matrix integral point of view, duality exchanges $\xx(z)$ and $\yy(z)$ in the spectral curve and thus corresponds to exchanging the two matrices in the two-matrix integral. Let us note that this property is very constraining. In particular, we could have assumed that the quantum volumes $\mathsf{V}_{g,n}^{(b)}$ appearing in \eqref{eq:Agn through quantum volumes} are some arbitrary polynomials of $\frac{b^2+b^{-2}}{4}$ and $p_j^2$ of degree $3g-3+n$. Imposing duality symmetry recursively fixes them all.

\paragraph{$x$-$y$ symmetry of topological recursion.} For the partition functions $\mathsf{A}_{g,0}^{(b)}=\omega_{g,0}$ (also denoted by $F_g$ in the literature), this is a consequence of the $x$-$y$ symmetry discussed in section~\ref{subsec:top rec review}.

\paragraph{Direct proofs for low $g$ and $n$.} For $(g,n)=(0,3),\, (0,4),\, (1,1)$, we gave direct proofs of duality symmetry in our previous paper \cite{paper1}. One can in principle push these to higher $(g,n)$, but it becomes more and more cumbersome.

\paragraph{Consequence of the recursion relation.} Instead, one can deduce the duality relation inductively from the recursion relation \eqref{eq:Agn recursion relation}. The key observation is that the recursion kernel is invariant under the duality symmetry
\begin{equation}
    \mathsf{H}_{b^{-1}}(x,y) = \mathsf{H}_b(x,y)\,,
\end{equation}
while the sphere three-point amplitude is odd
\begin{equation}
    \mathsf{A}_{0,3}^{(b^{-1})}(p_1,p_2,p_3) = -\mathsf{A}_{0,3}^{(b)}(p_1,p_2,p_3)\, .
\end{equation}
It is simplest to see the former from the rewriting of the recursion kernel in terms of the double-sine function as in (\ref{eq:recursion kernel double sine}). We can then proceed inductively, starting with $\mathsf{A}_{0,4}^{(b)}$, $\mathsf{A}_{1,1}^{(b)}$ and so on. Applying the recursion relation, we find
\begin{align}
        &\quad p_1\mathsf{A}_{g,n}^{(b^{-1})}(p_1,\boldsymbol{p})\nonumber\\
        &= \int 2q \d q \, 2q' \d q' \, \mathsf{H}_b(q+q',p_1)(-1)\mathsf{A}_{0,3}^{(b)}(p_1,q,q')\nonumber\\
        &\quad\quad\times(-1)^{n+1}\left(\mathsf{A}_{g-1,n+1}^{(b)}(q,q',\boldsymbol{p}) + \sideset{}{'}\sum_{h=0}^g \sum_{\mathcal{I},\mathcal{J}}\mathsf{A}_{h,1+|\mathcal{I}|}^{(b)}(q,\boldsymbol{p}_{\mathcal{I}})\mathsf{A}_{g-h,1+|\mathcal{J}|}^{(b)}(q',\boldsymbol{p}_{\mathcal{J}})\right) \nonumber\\ \nonumber
        & \quad - \sum_{j=2}^n\int 2q \d q \, \sum_{\pm}\mathsf{H}_b(q,p_1\pm p_j)(-1)\mathsf{A}_{0,3}^{(b)}(p_1,p_j,q)(-1)^{n-1}\mathsf{A}_{g,n-1}^{(b)}(q,\boldsymbol{p}\setminus p_j)\\
        &= (-1)^n p_1\mathsf{A}_{g,n}^{(b)}(p_1,\boldsymbol{p})\, ,
\end{align}
as expected from the worldsheet. The only subtlety has to do with the contour of integration in the $q,q'$ integrals that appear in the recursive representation. In practice we compute these by expanding the string amplitudes into linear combinations of terms proportional to $q^{2k}\mathrm{e}^{\pm 2\pi i m b q}\mathrm{e}^{\pm 2\pi i n b q}$ and apply (\ref{eq:primed integral}). This procedure is unaffected by the duality transformation, so the above discussion is not modified. 

Similarly, it is straightforward to show that the recursion kernel satisfies
\begin{equation}
    \mathsf{H}_{-ib}(ix, iy) = i\, \mathsf{H}_b(x,y)\, ,
\end{equation}
which may be used to demonstrate the swap symmetry (\ref{eq:Agn swap symmetry}) from the recursive representation of the string amplitudes.

\subsection{Analytic structure}
We next discuss the analytic structure of \eqref{eq:Agn through quantum volumes} in more detail. We start by noticing that the formula \eqref{eq:Agn through quantum volumes} converges on the physical spectrum where $b p_j \in \RR$ thanks to the exponential suppression of the factors $\sin(\pi m b^2)^{-2g_v+n-n_v}$ for large $m$. Convergence persists in a neighborhood of the physical spectrum, but not for arbitrary choices of $p_j$. The corresponding divergences lead to the rich analytic structure of the string amplitudes that we discussed in our previous paper \cite{paper1}. We will now see how to recover that analytic structure. 

\paragraph{Analytic continuation.} Let us first show that \eqref{eq:Agn through quantum volumes} can be analytically continued to complex momenta $\boldsymbol{p}$.
For this, we exchange the sum over the colors in \eqref{eq:Agn through quantum volumes} with the integral of $p_e$ and resum them in a different way. We write
\begin{align}\nonumber
    \sum_{m=1}^\infty\frac{(-1)^{mN}\prod_{j=1}^n \sin(2\pi m b p_j)}{\sin(\pi m b^2)^N}&=(2i)^{N-n} \sum_{m=1}^\infty (-1)^{mN} \frac{\prod_{j=1}^n (\mathrm{e}^{\pi i m b p_j}-\mathrm{e}^{-\pi i m b p_j})}{(\mathrm{e}^{\pi i m b^2}-\mathrm{e}^{-\pi i m b^2})^N} \\  
    &=(-1)^N(2i)^{N-n}\sum_{m=1}^\infty \sum_{\sigma_1,\dots,\sigma_n=\pm} \sigma_1 \cdots \sigma_n \nonumber\\  \nonumber
    &\quad\times \sum_{k=0}^\infty \binom{N+k-1}{N-1} (-1)^{mN}\mathrm{e}^{2\pi i m b(\sum_j \sigma_j p_j+(k+\frac{N}{2})b)} \\ \label{eq:Rewriting component of stable graph}
    &=\sum_{k=0}^\infty\sum_{\sigma_1,\dots,\sigma_n=\pm} \frac{(-1)^N(2i)^{N-n}\sigma_1 \cdots \sigma_n  \binom{N+k-1}{N-1}}{(-1)^N\mathrm{e}^{-2\pi i m b(\sum_j \sigma_j p_j+(k+\frac{N}{2})b)}-1}~.
\end{align}
Here $N = 2g-2+n$ corresponds to the component of the stable graph under consideration. These steps are all valid for small enough $\sum_j \sigma_j bp_j$, but the infinite sum in the last expression always converges and thus defines the analytic continuation of the expression to arbitrary momenta.

We can use this rewriting for every vertex in the stable graph. This leads naturally to a sum over a set of graphs that we denote by $\mathcal{G}_{g,n}^{\infty,\pm}$. For a graph in $\mathcal{G}_{g,n}^{\infty,\pm}$, we associate a color to every vertex that we call $k_v$ to distinguish it from $m_v$. We also associate a sign $\sigma$ to every half-edge (i.e.\ both ends of each internal edge have a sign and every external edge has a sign). The automorphism group is the autormorphism group without decorations. We get in this way
\begin{multline}
    \mathsf{A}_{g,n}^{(b)}(\boldsymbol{p})=(\sqrt{2}b i)^{2g-2+n}\sum_{\Gamma \in \mathcal{G}_{g,n}^{\infty,\pm}} \frac{1}{|\text{Aut}(\Gamma)|}\\
    \times \int' \prod_{e \in \mathcal{E}_\Gamma} (-2 p_e \, \d p_e) \frac{\big(\prod_{j \in I_v} \frac{i \sigma_j}{\sqrt{2}}\big) \binom{2g_v-3+n_v+k_v}{2g_v-3+n_v}\mathsf{V}^{(b)}_{g_v,n_v}(i\boldsymbol{p}_v)}{(-1)^{n_v}\mathrm{e}^{2 \pi i b (\sum_{j \in I_v} \sigma_j p_j-(2g_v-2+n_v+2k_v)\frac{b}{2})}-1}~. \label{eq:Agn second stable graph representation}
\end{multline}
Let us also give a more invariant definition of the primed integral. We can regularize the integral by inserting a factor $\mathrm{e}^{-\varepsilon p_e}$. Provided we chose the phase of $\varepsilon$ appropriately, this makes the integral convergent, even for the zero mode that we want to project out. We can thus define
\be 
\int' \d p \, f(p)\equiv \lim_{\varepsilon \to 0}\int \d p \, \mathrm{e}^{- \varepsilon p} \, f(p)~.
\ee
In the limit, we by definition pick out the regular term and discard all divergent terms. This precisely implements the prescription \eqref{eq:primed integral}. The upshot of this is that we may treat the primed integral as an ordinary contour integral and perform contour deformations etc. In particular, since the integrand is an analytic function, this defines a (possibly multivalued) analytic continuation of \eqref{eq:Agn through quantum volumes} to all values of complex momenta.

\paragraph{Analogy with Feynman diagrams.} Let us next explain how discontinuities are generated from this integral representation. Discontinuities come from the integrals over the internal momenta. This is precisely in analogy with Feynman diagrams, where discontinuities come from loop momentum integrations.\footnote{This analogy can presumably be made more precise, since we expect that one can identify the stable graphs with the Feynman diagrams of closed string field theory on this background in a particular gauge.} 
For a single integral, a discontinuity is generated whenever the poles and/or endpoints of the integrand undergo a monodromy that drags the integration contour along. The new integration contour is a linear combination of the old contour and a new contribution which captures the discontinuity of the integral. For higher-dimensional integrals, this is mathematically described by the Picard-Lefschetz theorem.

We will not need to go into the details of this, but simply need to recall that in QFT there is a simple set of cutting rules that captures the imaginary part of a Feynman diagram in terms of simpler diagrams obtained by cutting the original diagram and putting the momentum on the cut propagator on-shell. The most well-known form of such cutting rules are the Cutkosky rules \cite{Cutkosky:1960sp}, but it is actually more convenient to use the so-called holomorphic cutting rules introduced in \cite{Hannesdottir:2022bmo}, which express the imaginary part of the amplitude as a sum over all possible cuttings.\footnote{In the Cutkosky cutting rules, one only sums over simple cuttings, but this necessitates complex conjugation of one part of the diagram.}
In favorable cases such as the computation of the lowest threshold discontinuity, one can also show that the imaginary part equals the discontinuity as a consequence of the Schwarz reflection principle.

The logic here is the same, except for two differences: (i) There is no momentum conservation and thus we integrate over all internal momenta, even at tree level and (ii) putting a particle on-shell means that we are taking the residue of the integrand at a pole.

\paragraph{Cutting rules.}
Suppose we want to compute the discontinuity of $\mathsf{A}_{g,n}^{(b)}$ around a given $p_*$. Then we have to cut the internal lines of the appearing graphs in all possible ways. This means that we cut lines which can go `on-shell' meaning that the integrand develops a pole.

If we denote the contribution of a graph $\Gamma$ to $\mathsf{A}_{g,n}^{(b)}(p_1,\dots,p_n)$ by the graph itself, we have for example
\begin{align}
    \Disc_{p_*=p_1+p_2+\frac{Q}{2}=0} \begin{tikzpicture}[baseline={([yshift=-.5ex]current bounding box.center)},scale=.6]
    \node[shape=circle,draw=black, very thick] (A) at (0,0) {$0$};
    \node[shape=circle,draw=black, very thick] (B) at (-2,0) {$0$};
    \draw[very thick] (A) to (1,1) node[right] {$p_3$};
    \draw[very thick] (A) to (1,-1) node[right] {$p_4$};
    \draw[very thick] (A) to (B);
    \draw[very thick] (B) to (-3,1) node[left] {$p_1$};
    \draw[very thick] (B) to (-3,-1) node[left] {$p_2$};
\end{tikzpicture}= \begin{tikzpicture}[baseline={([yshift=-.5ex]current bounding box.center)},scale=.6]
    \node[shape=circle,draw=black, very thick] (A) at (0,0) {$0$};
    \node[shape=circle,draw=black, very thick] (B) at (-2,0) {$0$};
    \draw[very thick] (A) to (1,1) node[right] {$p_3$};
    \draw[very thick] (A) to (1,-1) node[right] {$p_4$};
    \draw[very thick] (A) to (B);
    \draw[very thick] (B) to (-3,1) node[left] {$p_1$};
    \draw[very thick] (B) to (-3,-1) node[left] {$p_2$};
    \draw[very thick, red] (-1,-.5) to (-1,.5);
\end{tikzpicture}~,
\end{align}
since there is a pole when the internal momentum is $p_e=p_*$ or $p_e=-p_*$. This creates a discontinuity, since these two poles undergo a monodromy around $p_e=0$, the end-point of the integral. Hence we get
\begin{align} 
\begin{tikzpicture}[baseline={([yshift=-.5ex]current bounding box.center)},scale=.6]
    \node[shape=circle,draw=black, very thick] (A) at (0,0) {$0$};
    \node[shape=circle,draw=black, very thick] (B) at (-2,0) {$0$};
    \draw[very thick] (A) to (1,1) node[right] {$p_3$};
    \draw[very thick] (A) to (1,-1) node[right] {$p_4$};
    \draw[very thick] (A) to (B);
    \draw[very thick] (B) to (-3,1) node[left] {$p_1$};
    \draw[very thick] (B) to (-3,-1) node[left] {$p_2$};
    \draw[very thick, red] (-1,-.5) to (-1,.5);
\end{tikzpicture}&=-2\pi i  \Res_{p=p_*} \, (-2p) \begin{tikzpicture}[baseline={([yshift=-.5ex]current bounding box.center)},scale=.6]
    \node[shape=circle,draw=black, very thick] (A) at (0,0) {$0$};
    \node[shape=circle,draw=black, very thick] (B) at (-3.5,0) {$0$};
    \draw[very thick] (A) to (1,1) node[right] {$p_3$};
    \draw[very thick] (A) to (1,-1) node[right] {$p_4$};
    \draw[very thick] (A) to (-1.3,0) node[above] {$p$};
    \draw[very thick] (B) to (-2.2,0) node[above] {$p$};
    \draw[very thick] (B) to (-4.5,1) node[left] {$p_1$};
    \draw[very thick] (B) to (-4.5,-1) node[left] {$p_2$};
\end{tikzpicture}\nonumber\\ \nonumber
&\quad -2\pi i  \Res_{p=-p_*} \, (-2p) \begin{tikzpicture}[baseline={([yshift=-.5ex]current bounding box.center)},scale=.6]
    \node[shape=circle,draw=black, very thick] (A) at (0,0) {$0$};
    \node[shape=circle,draw=black, very thick] (B) at (-3.5,0) {$0$};
    \draw[very thick] (A) to (1,1) node[right] {$p_3$};
    \draw[very thick] (A) to (1,-1) node[right] {$p_4$};
    \draw[very thick] (A) to (-1.3,0) node[above] {$p$};
    \draw[very thick] (B) to (-2.2,0) node[above] {$p$};
    \draw[very thick] (B) to (-4.5,1) node[left] {$p_1$};
    \draw[very thick] (B) to (-4.5,-1) node[left] {$p_2$};
\end{tikzpicture}\\
&=8\pi i p_* \Res_{p=p_*} \mathsf{A}^{(b)}_{0,3}(p_1,p_2,p)\mathsf{A}^{(b)}_{0,3}(p_3,p_4,p)~.
\end{align}
The overall sign can be deduced by carefully tracking in which sense the contour is dragged from the monodromy.
A similar logic generalizes to more complicated diagrams. For each diagram, we have to sum over all possible ways to cut the internal lines. There can be more than one choice, for example
\begin{multline} 
\Disc_{p_1+p_2+\frac{Q}{2}=0} \begin{tikzpicture}[baseline={([yshift=-.5ex]current bounding box.center)},scale=.6]
    \node[shape=circle,draw=black, very thick] (A) at (0,0) {$0$};
    \node[shape=circle,draw=black, very thick] (C) at (-2,0) {$1$};
    \node[shape=circle,draw=black, very thick] (B) at (-4,0) {$0$};
    \draw[very thick] (A) to (1,1) node[right] {$p_3$};
    \draw[very thick] (A) to (1,-1) node[right] {$p_4$};
    \draw[very thick] (A) to (C);
    \draw[very thick] (B) to (C);
    \draw[very thick] (B) to (-5,1) node[left] {$p_1$};
    \draw[very thick] (B) to (-5,-1) node[left] {$p_2$};
\end{tikzpicture} \\ 
= \begin{tikzpicture}[baseline={([yshift=-.5ex]current bounding box.center)},scale=.6]
    \node[shape=circle,draw=black, very thick] (A) at (0,0) {$0$};
    \node[shape=circle,draw=black, very thick] (C) at (-2,0) {$1$};
    \node[shape=circle,draw=black, very thick] (B) at (-4,0) {$0$};
    \draw[very thick] (A) to (1,1) node[right] {$p_3$};
    \draw[very thick] (A) to (1,-1) node[right] {$p_4$};
    \draw[very thick] (A) to (C);
    \draw[very thick] (B) to (C);
    \draw[very thick] (B) to (-5,1) node[left] {$p_1$};
    \draw[very thick] (B) to (-5,-1) node[left] {$p_2$};
    \draw[very thick, red] (-3,-.5) to (-3,.5);
\end{tikzpicture}+\begin{tikzpicture}[baseline={([yshift=-.5ex]current bounding box.center)},scale=.6]
    \node[shape=circle,draw=black, very thick] (A) at (0,0) {$0$};
    \node[shape=circle,draw=black, very thick] (C) at (-2,0) {$1$};
    \node[shape=circle,draw=black, very thick] (B) at (-4,0) {$0$};
    \draw[very thick] (A) to (1,1) node[right] {$p_3$};
    \draw[very thick] (A) to (1,-1) node[right] {$p_4$};
    \draw[very thick] (A) to (C);
    \draw[very thick] (B) to (C);
    \draw[very thick] (B) to (-5,1) node[left] {$p_1$};
    \draw[very thick] (B) to (-5,-1) node[left] {$p_2$};
    \draw[very thick, red] (-1,-.5) to (-1,.5);
\end{tikzpicture}~.
\end{multline}
The cut then divides the diagram into either two disconnected pieces or --- when cutting a loop --- gives a connected graph of genus $g-1$. We can then reorganize the sum over stable graphs as a sum over stable graphs of the pieces $\mathcal{G}_{h,|I|+1}^\infty \times \mathcal{G}_{g-h,|I^c|+1}^\infty$ or $\mathcal{G}_{g-1,n+2}^\infty$. The sum over these stable graphs then precisely reconstructs $\mathsf{A}_{h,m+1}^{(b)}(\boldsymbol{p}_I,p)\mathsf{A}_{g-h,m+1}^{(b)}(\boldsymbol{p}_{I^c},p)$ and $\mathsf{A}_{g-1,n+2}^{(b)}(\boldsymbol{p},p,p)$, respectively. In the latter case, we get an additional factor of $\frac{1}{2}$, since we lose a $\ZZ_2$ factor in the automorphism group of the graph that flips the cutted edge. Furthermore, the poles that undergo the monodromy are in this case at $p=\pm \frac{1}{2} p_*$. For example, we have
\begin{align} \nonumber
\Disc_{p_*=p_1+\frac{Q}{2}=0} \begin{tikzpicture}[baseline={([yshift=-.5ex]current bounding box.center)},scale=.6]
    \node[shape=circle,draw=black, very thick] (A) at (0,0) {$0$};
    \draw[very thick] (A) to (1,0) node[right] {$p_1$};
    \draw[very thick, looseness=3] (A) to[out=135, in=-135] (A);
\end{tikzpicture} &= \begin{tikzpicture}[baseline={([yshift=-.5ex]current bounding box.center)},scale=.6]
    \node[shape=circle,draw=black, very thick] (A) at (0,0) {$0$};
    \draw[very thick] (A) to (1,0) node[right] {$p_1$};
    \draw[very thick, looseness=3] (A) to[out=135, in=-135] (A);
    \draw[red, very thick] (-1.3,0) to (-.7,0);
\end{tikzpicture}\\ \nonumber
&= -\frac{1}{2} \times 2\pi i \Big[\Res_{p=\frac{1}{2}p_*}+\Res_{p=-\frac{1}{2}p_*} \Big]\, (-2p) \mathsf{A}_{0,3}^{(b)}(p_1,p,p) \\
&=2\pi p_* \Res_{p=\frac{1}{2}p_*}  \mathsf{A}_{0,3}^{(b)}(p_1,p,p)
\end{align}
The general result can hence be stated as
\begin{multline}
    \Disc_{p_*=0} \mathsf{A}_{g,n}^{(b)}(\boldsymbol{p})=2\pi i p_* \Res_{p=\frac{1}{2}p_*} \mathsf{A}_{g-1,n+2}^{(b)}(\boldsymbol{p},p,p)\\
    +4\pi i p_* \Res_{p=p_*}\sum_{\begin{subarray}{c} 0 \le h \le g \\ I \subseteq \{1,\dots,n\} \\ \text{stable} \end{subarray}} \mathsf{A}_{h,|I|+1}^{(b)}(\boldsymbol{p}_I,p)\mathsf{A}_{g-h,|I^c|+1}^{(b)}(\boldsymbol{p}_{I^c},p)~.
\end{multline}
Notice that we overcounted by a factor of two by summing over all genera $h$ and subsets $I$, which we compensated by another factor of $\frac{1}{2}$. This reproduces the discontinuity that we derived from the worldsheet in \cite{paper1}.

\paragraph{Poles from recursion.}
The representation (\ref{eq:Rewriting component of stable graph}) that facilitates the analytic continuation of the string amplitudes exhibits more poles in the external momenta than we expect based on the worldsheet analysis discussed in our previous paper \cite{paper1}. In particular, it exhibits poles when
\begin{equation}
    \sum_j \sigma_j p_j=(2g-2+n+2k)\frac{b}{2}+m b^{-1}~, \quad k \in \ZZ_{\ge 0}~, m \in \ZZ+\frac{n}{2}~,
\end{equation}
whereas we know from the worldsheet analysis that the string amplitudes $\mathsf{A}_{g,n}^{(b)}$ should only have poles for
\begin{equation}\label{eq:worldsheet poles}
    \sum_{j} \sigma_j p_j = rb + sb^{-1}, \quad r,s\in\mathbb{Z}+\frac{n}{2},\quad |r|,|s| \geq \frac{2g-2+n}{2}\, .
\end{equation}
That the extra poles must cancel is guaranteed by the duality symmetry but the cancellation mechanism is not at all manifest in this representation. 

To see that we only get the poles that we expect from the worldsheet analysis, it is more straightforward to instead employ the recursive representation of the string amplitudes and proceed inductively. In the recursive representation (\ref{eq:Agn recursion relation}), poles of the string amplitude are generated when singularities of the constituent string amplitudes pinch the contour of integration over the internal momenta $q,q'$. For concreteness, consider the last term in the recursive representation
\begin{equation}
    p_1 \mathsf{A}_{g,n}^{(b)}(p_1,\boldsymbol{p}) \supset - \sum_{j=2}^n\int 2q \d q \, \sum_{\pm} \mathsf{H}_b(q,p_1\pm p_j) \mathsf{A}_{0,3}^{(b)}(p_1,p_j,q) \mathsf{A}_{g,n-1}^{(b)}(q,\boldsymbol{p}\setminus p_j)\, .
\end{equation}
Consider in particular the following singularities of the sphere three-point amplitude $\mathsf{A}_{0,3}^{(b)}$ that appears in the recursion
\begin{equation}
    q = \sigma_1 p_1 + \sigma_j p_j + r b + s b^{-1},\, \quad r,s\in\mathbb{Z}+\frac{1}{2}\, ,\  |r|,|s| \geq \frac{1}{2}\, .
\end{equation}
The recursion kernel $\mathsf{H}_b$ does not contribute any singularities in the internal momentum $q$. On the other hand, the other constituent string amplitude $\mathsf{A}_{g,n-1}^{(b)}$ is assumed to exhibit poles at the following values of the momentum $q$
\begin{equation}
    q = - \boldsymbol{\sigma}'\cdot \boldsymbol{p}' - r' b- s'b^{-1}\, , \quad r',s'\in\mathbb{Z} + \frac{n-1}{2}\, ,\  |r'|,|s'| \geq \frac{2g-3+n}{2}\, .
\end{equation}
Here $\boldsymbol{p}'$ is a stand-in for $\boldsymbol{p}\setminus p_j$ and similarly $\boldsymbol{\sigma}'$ is a vector of signs $\{\sigma_2,\ldots,\sigma_n\}$ with $\sigma_j$ omitted. These singularities pinch the contour of integration over $q$ when $r'$ and $s'$ have the same sign as $r$ and $s$, respectively. This generates poles in the full string amplitude when
\begin{multline}
    \sigma_1 p_1 + \boldsymbol{\sigma}\cdot \boldsymbol{p} = (r+r')b + (s+s')b^{-1}\, ,\\ r+r',s+s'\in\mathbb{Z} + \frac{n}{2}\, , \ |r+r'|, |s+s'| \geq \frac{2g-2+n}{2}\, ,
\end{multline}
exactly as expected from the worldsheet in (\ref{eq:worldsheet poles}). Essentially identical considerations apply to the other terms in the recursive representation of the string amplitudes.

\subsection{Dilaton equation}
We will now check that \eqref{eq:Agn through quantum volumes} satisfies the dilaton equation. We already did this for $\mathsf{A}_{1,1}^{(b)}$ and $\mathsf{A}_{0,4}^{(b)}$ in our previous paper \cite{paper1}. 

For the general case, we translate the dilaton and string equation \eqref{eq:topological recursion dilaton and string equations} of topological recursion to $\mathsf{A}_{g,n}^{(b)}$ using \eqref{eq:Agn omegagn relation}.

\paragraph{Dilaton equation of topological recursion.} We have $\d F_{0,1}(z) = \omega^{(b)}_{0,1}(z)$ (\ref{eq: omega01}) such that
\be 
F_{0,1}(z)=\frac{2}{b} \bigg( \frac{1}{Q} \cos(\pi Q \sqrt{z})+\frac{1}{\hat{Q}} \cos( \pi \hat{Q}\sqrt{z})\bigg)~,
\ee
where we recall that $\hat{Q}=b^{-1}-b$ and $Q=b+b^{-1}$. Thus on the LHS of the dilaton equation \eqref{eq:topological recursion dilaton equation} becomes
\begin{align}
    \frac{1}{b}\sum_{m=1}^\infty \Res_{z_{n+1}=z_m^*} \bigg( \frac{2}{Q} \cos(\pi Q \sqrt{z_{n+1}})+\frac{2}{\hat{Q}} \cos( \pi \hat{Q}\sqrt{z_{n+1}})\bigg) \omega^{(b)}_{g,n+1}(\boldsymbol{z},z_{n+1})~.
\end{align}
We notice that this has precisely the right form for the transformation \eqref{eq:Agn omegagn relation} with $p=\frac{1}{2} Q$ and $\frac{1}{2} \hat{Q}$ for the two terms. The other coordinates $\boldsymbol{z}$ are spectators and can be transformed as in \eqref{eq:Agn omegagn relation} on both sides of the dilaton equation. This leads to
\be 
\mathsf{A}_{g,n+1}^{(b)}(\boldsymbol{p},p=\tfrac{1}{2}Q)+\mathsf{A}_{g,n+1}^{(b)}(\boldsymbol{p},p=\tfrac{1}{2}\hat{Q})=-b \, (2g-2+n) \mathsf{A}_{g,n}^{(b)}(\boldsymbol{p})~. \label{eq:Agn dilaton equation 1}
\ee

\paragraph{First string equation.} Let us first consider the case of $k=0$ in \eqref{eq:topological recursion string equation}. We can take the inverse Laplace transform of \eqref{eq:topological recursion string equation}. Consider first the left hand side. Since only the leg $n+1$ is important, we can momentarily put $n=0$. We have
\begin{align}
    \sum_{m=1}^\infty \Res_{z=z_m^*} 2 \cos(\pi b \sqrt{z})\, \omega^{(b)}_{g,1}(z)=b \, \mathsf{A}_{g,1}^{(b)}(p=\tfrac{b}{2})~,
\end{align}
where we compared to the definition \eqref{eq:Agn omegagn relation}. We know that the $\mathsf{A}_{g,1}^{(b)}(p=\frac{b}{2})=0$ as consequence of the `trivial zeros'. Similarly, the right-hand side of the string equation trivializes when one translates it into $\mathsf{A}_{g,n}^{(b)}$ and confirms the existence of this zero of $\mathsf{A}_{g,n}^{(b)}$.

\paragraph{Second string equation.} Let us now repeat the discussion with the second string equation \eqref{eq:topological recursion string equation} with $k=1$, where we learn something new. For the left-hand side, we may again assume that $n=1$, which yields
\begin{align}\nonumber
    \sum_{m=1}^\infty \Res_{z=z_m^*} \xx(z) \yy(z) \omega^{(b)}_{g,1}(z)&=-2\sum_{m=1}^\infty \Res_{z=z_m^*}\Big(\!\cos\big(\pi Q \sqrt{z}\big)+\cos\big(\pi \hat{Q} \sqrt{z}\big)\!\Big)\omega^{(b)}_{g,1}(z) \\
    &=- \Big(Q\,\mathsf{A}_{g,1}^{(b)}(p=\tfrac{1}{2}Q)+\hat{Q}\, \mathsf{A}_{g,1}^{(b)}(p=\tfrac{1}{2}\hat{Q})\Big)~.
\end{align}
To work out the right-hand side, we can put $n=1$ and only consider the corresponding term. It is more convenient to do this in the $w$-coordinate and use \eqref{eq:Agn omegagn relation w}. This leads to
\begin{align}
    &-\int_\gamma \frac{\d w\, \mathrm{e}^{2\pi i p_1 w}}{4\pi i p}\, \partial_w\left( \frac{\xx(w)}{\d \xx(w)} \, \omega^{(b)}_{g,1}(w) \right)\nonumber\\  \nonumber
    &\quad=-\frac{b}{2\pi} \int_\gamma \mathrm{e}^{2\pi i p_1 w} \cot(b^{-1} \pi w) \omega^{(b)}_{g,1}(w)\\ \nonumber
    &\quad=\frac{b}{2\pi i} \int_\gamma \mathrm{e}^{2\pi i p_1 w} \Big(1+2 \sum_{s=1}^\infty \mathrm{e}^{-2\pi i b^{-1} sw}\Big) \omega^{(b)}_{g,1}(w) \\ \nonumber
    &\quad=\frac{b}{2\pi i} \int_\gamma \mathrm{e}^{2\pi i p_1 w} \omega^{(b)}_{g,1}(w) \\
    &\quad=2b p_1 \, \mathsf{A}_{g,1}^{(b)}(p_1)~.
\end{align}
Here, we first integrated by parts and then expanded the cotangent in an infinite absolutely convergent series. We then assumed that $0<\Re(b p_1)<1$, which allows us to drop all terms but the first one in the infinite sum. In the last line, we recognize the definition \eqref{eq:Agn omegagn relation}. For $\Re(b p_j)<0$ we can infer the result because $\mathsf{A}_{g,n}^{(b)}(\boldsymbol{p})$ is an odd function in all $p_j$'s. We can write the result as
\begin{align}
    Q \mathsf{A}_{g,n+1}^{(b)}(\boldsymbol{p},p=\tfrac{1}{2}Q)+\hat{Q} \mathsf{A}_{g,n+1}^{(b)}(\boldsymbol{p},p_{n+1}=\tfrac{1}{2}\hat{Q})=-2 \sum_{j=1}^n \sqrt{(b p_j)^2} \, \mathsf{A}_{g,n}^{(b)}(\boldsymbol{p})~.\label{eq:Agn string equation}
\end{align}
\paragraph{Equivalence to worldsheet dilaton equations.} We can take appropriate linear combinations of  \eqref{eq:Agn dilaton equation 1} and \eqref{eq:Agn string equation} to obtain 
\begin{subequations}
\begin{align}
    \mathsf{A}_{g,n+1}^{(b)}(\boldsymbol{p},p_{n+1}=\tfrac{1}{2}\hat{Q})&=-\left(\tfrac{1}{2}Q(2g-2+n)-\sqrt{p^2}\right)\mathsf{A}_{g,n}^{(b)}(\boldsymbol{p})~, \\
    \mathsf{A}_{g,n+1}^{(b)}(\boldsymbol{p},p_{n+1}=\tfrac{1}{2}Q)&=\left(\tfrac{1}{2}\hat{Q}(2g-2+n)-\sqrt{p^2}\right)\mathsf{A}_{g,n}^{(b)}(\boldsymbol{p})~.
\end{align}    
\end{subequations}
We used that $b^{-1} \sqrt{(b p)^2}=\sqrt{p^2}$ for $p$ in the neighborhood of the physical spectrum and remain agnostic about the precise location of the branch cut. These are precisely the dilaton equations that were derived from the worldsheet theory in \cite{paper1}.

\section{Conclusion} \label{sec:conclusion}
Let us discuss a few futher applications and generalizations.

\begin{figure}[h]
\centering
\begin{tikzpicture}
    \draw[very thick,<->] (-4,0) -- (4,0);
    \draw[very thick, <->] (0,-1) -- (0,3); 
    \draw[very thick, blue] (.6,-.2) to (.6,.2) node[scale=.8, above, black] {1}; 
    \draw[very thick, red] (1.5,-1) to node[scale=.8, right, black] {$13+i \RR$} (1.5,3);     
    \draw[thick] (3.5,2.6)-- (4,2.6);
    \draw[thick] (3.5,2.6) -- (3.5,3);
\draw[blue, opacity=.2, line width=1.7mm] (.6,0) -- (-4,0);    
\foreach \i in {-3.8,-3.5,...,.5}
{
    \draw[thick] ({\i-.1},.1) to ({\i+.1},-.1); 
    \draw[thick] ({\i-.1},-.1) to ({\i+.1},.1); 
}
\node at (3.8,2.8) {$c_m$};    
\end{tikzpicture}
\caption{The possible values of the matter central charge for the 2d string theories of table~\ref{tab:minimal string theories}. The $c\leq 1$ region contains a discrete set of points corresponding to the $(2,p)$ minimal string, a dense set of rational points corresponding to the $(p,q)$ minimal string, and a continuum spanned by the Virasoro minimal string.} \label{fig:matter central charges}
\end{figure}
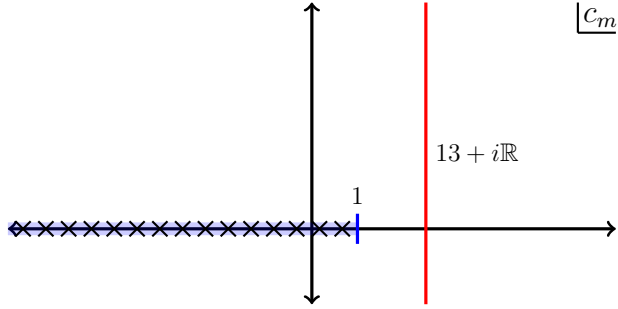

\paragraph{The landscape of minimal string theories.} Bosonic string theories in two (or less than two) target space dimensions are quite special since they don't exhibit a tachyon. They are described on the worldsheet by a CFT with effective central charge $c_{\text{eff}} \le 2$.\footnote{Recall that the effective central charge is defined by $c_{\text{eff}} = c - 24h_{\text{min}}$, where $h_{\text{min}}$ is the conformal weight of the lightest operator in the theory (assumed to be a scalar for the purposes of this discussion). This is the quantity that controls the asymptotic growth of states at high energy.} Only a few examples are known and some are listed in table~\ref{tab:minimal string theories}. Most interestingly, all of these examples admit a dual description. We hope that a full understanding of this two-dimensional landscape will teach us lessons about the more realistic and vast landscape of full superstring theory. 
Another way to plot this landscape at least for the examples involving Liouville theory is in terms of the central charge of the matter theory. They are arranged in the complex plane as indicated in figure~\ref{fig:matter central charges}.

For $c_m=1$, there are many more ``minimal'' string theories, most importantly the $c=1$ string (for reviews, see \cite{Klebanov:1991qa,Moore:1991zv,Ginsparg:1993is,Jevicki:1993qn,Polchinski:1994mb,Martinec:2004td,Nakayama:2004vk}). Its dual description involves matrix quantum mechanics \cite{Moore:1991zv} (rather than a matrix integral), which describes the S-matrix elements of massless bosons scattering off the Liouville wall in a two-dimensional target space. Hence it is a qualitatively distinct holographic duality from the class of models depicted in figure~\ref{fig:matter central charges}, and for this reason we did not include it in the table~\ref{tab:minimal string theories}. It would be very interesting to better understand the relationship between the $c=1$ string and the broader landscape of minimal string theories, as this may represent an instance of the emergence of time, with the dual description transitioning from a matrix integral to matrix quantum mechanics.

The different minimal string theories that we explored are not unrelated. When considering the complex Liouville string as a function of $b^2$, we may analytically continue the perturbative data away from $b^2 \in i \RR$. We may in particular consider the limit $b^2 \to \QQ$. In this limit, one of the Liouville theories has central charge $c\le 1$ and the limiting theory is known as non-analytic Liouville theory \cite{McElgin:2007ak, Ribault:2015sxa}. The case with $c=1$ is also better known as Runkel-Watts theory \cite{Runkel:2001ng}. The structure constants and hence the degenerated string amplitudes $\lim_{b^2 \to \frac{p}{q}}\mathsf{A}_{g,n}^{(b)}(\boldsymbol{p})$ are piecewise analytic functions of the momenta. We also suspect that one can recover from this directly the VMS amplitude by taking $p,q \to \infty$ with $\frac{p}{q} \to b^2 \in \RR$, thanks to the relations summarized in \cite{Ribault:2014hia}. Finally, given that the spectral curve degenerates to the spectral curve of the minimal string, it should also be possible to take a suitable limit and recover the minimal string amplitudes. In other words, it should be possible to recover the amplitudes of these other theories by taking special limits of the complex Liouville string. Thus the complex Liouville string seems to sit at the top of the hierarchy and should in particular still have lessons in stock about the other minimal string theories appearing in table~\ref{tab:minimal string theories}.

\begin{table}[h!]
\centering
\begin{tabular}{m{14em} | m{8.25em} | m{11.0em} } 
Worldsheet & Dual description & Spectral curve  \\
\hline\hline 
 $(2,p)$ minimal string: \newline $(2,p)\,\text{minimal model} \, \oplus$ Liouville  & matrix integral & $\xx(z) = -4z^2~$ \newline $\yy(z) = 2T_p(z)$ \\ 
\hline
$(p,q)$ minimal string: \newline $(p,q)\, \text{minimal model}\, \oplus$ Liouville  & two matrix integral & $\xx(z)= -2T_p(z)~$ \newline $\yy(z) = 2T_q(z)$ \\ 
\hline
Virasoro minimal string: \newline
timelike Liouville $\oplus$ Liouville & matrix integral  & $\xx(z) = -z^2~$ \newline $\yy(z) = \frac{\sin(2\pi b z) \sin(2\pi b^{-1}z)}{z}$ \\ 
 [3ex]
\hline
complex Liouville string: \newline
Liouville $\oplus$ (Liouville)$^*$ & two matrix integral & $\xx(z) = -2\cos(\pi b^{-1}\sqrt{z})~$ \newline $\yy(z)= 2 \cos(\pi b \sqrt{z})$
\end{tabular}
\caption{The landscape of 2d string theories. All theories also exist as unorientable string theories, which maps to the corresponding orthogonal or symplectic matrix integral. } \label{tab:minimal string theories}
\end{table}

\paragraph{Deformations.} The complex Liouville string admits many deformations. They can be described uniformly on the matrix model side by deforming the spectral curve while staying in the same universality class. Such deformations fall into two classes as was discussed in \cite{Seiberg:2003nm} and we expect them to behave similarly in the complex Liouville string: deformations of the locations of the nodal singularities and deformations that open up the nodal singularities to handles. The former deformations are easy to describe on the worldsheet as they correspond to deforming the worldsheet by the on-shell marginal vertex operators $\int \d^2 z\,  \mathfrak{b}_{-1} \tilde{\mathfrak{b}}_{-1} \mathcal{V}_p$. Deformations which open up the nodal singularities are more difficult; they were interpreted in \cite{Seiberg:2003nm} as ZZ-instanton backgrounds. It is not known how to describe them directly in the worldsheet formalism where they likely correspond to a non-local deformation, perhaps by a ground ring operator $\mathcal{O}_{m,n}$. It would be interesting to explore such deformations more systematically.

\paragraph{String field theory perspective.} As mentioned, we can view the graphical representation of the string amplitudes through \eqref{eq:Agn through quantum volumes} as Feynman diagrams of the string field theory description, whose vertices are essentially given by the quantum volumes of the Virasoro minimal string \cite{Collier:2023cyw}.
However, that analogy is only superficial since the string field theory Hilbert space also consists of all worldsheet vertex operators (annihilated by $L_0^-$ and $\mathfrak{b}_0^-$) and additionally comes with a huge gauge redundancy. In \eqref{eq:Agn through quantum volumes} this gauge redundancy seems to be fixed and even the appearing intermediate momenta are all on-shell.
The closed string field theory action $S$ has to satisfy the quantum BV master equation $\{S,S\}+\Delta S=0$, which gives a relation between different vertices \cite{Sen:2024nfd}. We note the structural similarity between this condition and the corresponding quadratic condition on the resolvents in the matrix integral \eqref{eq:analyticity sum over branches quadratic connected}, but we have not been able to make the connection precise.\footnote{We thank Victor Godet for a discussion about this.}
Indeed it has recently been shown in \cite{Firat:2024ajp} that hyperbolic string vertices of closed string field theory obey a version of Mirzakhani's recursion relation. 
A string field theory description that is closer related to the structure observed in eq.~\eqref{eq:Agn through quantum volumes} is via a two-dimensional Kodaira-Spencer theory on the spectral curve \cite{Dijkgraaf:2007sx}. That theory has a cubic vertex which roughly maps to the trivalent nature of degenerated stable graphs.

\paragraph{Uniform transcendentality and symbol alphabet.} The string amplitudes in this theory enjoy certain uniform transcendentality properties. These are most clearly visible in the realization \eqref{eq:Agn second stable graph representation}, which converges for all values of the momenta. Note that we can express the denominator through the function $\Li_0(x)=\frac{x}{1-x}$. We can then simply successively integrate out all the momenta by using the identity
\be 
\int_0^\infty \d y\ \Li_n(a \mathrm{e}^{-y})\,  y^{m-1}=\Gamma(m) \,  \Li_{n+m}(a)~.
\ee
Using that the quantum volumes are polynomials of order $3g-3+n$ and taking also the factors of $\pi$ in \eqref{eq:Agn second stable graph representation}, the amplitude can hence be expressed as a sum of terms
\be 
P_{3g-3+n-\sum_j \ell_j}(\boldsymbol{p})\sum_{\sum \ell_j \le 3g-3+n}\pi^{-2\sum_j \ell_j} \prod_j \Li_{2\ell_j}\big(\mathrm{e}^{2\pi i b (\sum_{i \in I_j} \sigma_i p_i +r_j b+s_j b^{-1})}\big) \label{eq:polylogarithm}
\ee
for different choices of $\ell_j$, $r_j$ and $s_j$ and polynomials $P$ of degree $3g-3+n-\sum_j \ell_j$ in $p_i^2$. These sums are always absolutely convergent and we wrote down the explicit formula for $\mathsf{A}_{0,4}^{(b)}$ and $\mathsf{A}_{1,1}^{(b)}$ in \cite{paper1}.  
This is precisely the polylogarithmic structure of scattering amplitudes that is also found in ordinary QFT scattering amplitudes, except that the arguments in our case are exponentials of momenta. We naturally assign a transcendentality degree to \eqref{eq:polylogarithm} by letting the transcendentality of $\Li_m$ to be the weight $m$ and the transcendentality of $\pi$ to be 1. The transcendentality of every term appearing in $\pi^{2(3g-3+n)} \mathsf{A}_{g,n}^{(b)}(\boldsymbol{p})$ is then $2(3g-3+n)$.
We should note that similar uniform transcendentalities are subject of ongoing research for more realistic string amplitudes \cite{DHoker:2019blr}.

\paragraph{Supersymmetric variant.} We expect that most of the results of this paper as well as \cite{paper1} can be generalized to the type 0B superstring. In that case, one couples two $\mathcal{N}=1$ Liouville theories with complex central charges $\frac{15}{2}+i \RR$. This is natural given that the usual parametrization of the central charge reads
\be 
c=\frac{3}{2}+3(b+b^{-1})^2~,
\ee
which translates again to $b^2 \in i \RR$.
The structure constants for this theory are known \cite{Rashkov:1996np, Poghossian:1996agj}. We expect that there is also a duality similar to the corresponding story for the minimal superstring \cite{Klebanov:2003wg, Seiberg:2003nm, Seiberg:2004ei}, with the $\mathcal{N}=1$ super Virasoro minimal string \cite{Collier:2023cyw,Johnson:2024fkm} volumes appearing as the string vertices in this case. The form of the spectral curve of the minimal superstring suggests that the dual matrix integral is closely related to the bosonic case \cite{Seiberg:2003nm}.

\paragraph{Disordered holography.} The work of Saad, Shenker and Stanford \cite{Saad:2019lba} has initiated a shift in the community, where the old matrix integral is reinterpreted as a disordered quantum mechanical system with the matrix playing the role of the Hamiltonian. One may asked about a corresponding interpretation for the two-matrix integral. The first matrix $M_1$ can be directly interpreted as the Hamiltonian, but the interpretation of the second matrix $M_2$ is less clear. Let us offer one possible interpretation. We can think of $H=M_1 \otimes M_2$ as the Hamiltonian acting on a bipartite system. The two systems are only entangled via the mixing term $\tr(M_1M_2)$ in the matrix integral potential \eqref{eq:2MM}. The boundary matrix integral is then a disorder average over the Hamiltonians of both subsystems. Since the string amplitudes are only computed from the resolvent of the first matrix, we can view the subsystem of the second matrix as a hidden sector that we don't directly have access to.

\paragraph{Relation to $\SL(2,\CC)$ BF-theory.} It is well-known that the first-order formulation JT-gravity is classically equivalent to a mapping class group gauged $\SL(2,\RR)$ BF-theory (see e.g. \cite{Saad:2019lba}). Similarly, the Virasoro minimal string has a quantum group $\mathcal{U}_q(\mathfrak{sl}(2,\RR))$ symmetry and can be understood as a mapping class group gauged version of $\mathcal{U}_q(\mathfrak{sl}(2,\RR))$ BF-theory.
It is tempting to conjecture a similar realization of the complex Liouville string as a BF theory. The natural candidate is BF-theory based on the quantum group $\mathcal{U}_q(\mathfrak{sl}(2,\CC))$. The quantum group $\mathcal{U}_q(\mathfrak{sl}(2,\CC))$ is known as the quantum Lorentz group and consists of two copies of the modular double $\mathcal{U}_q(\mathfrak{sl}(2,\RR))$ \cite{Gaiotto:2024osr}. The ``gravitational'' $\mathcal{U}_q(\mathfrak{sl}(2,\RR))$ is diagonally embedded inside the $\mathcal{U}_q(\mathfrak{sl}(2,\CC))$.
This is in line with the fact that vertex operators are labelled by $\SL(2,\CC)$ representations and the topological sector discussed in section~\ref{subsec:CohFT} is $\SU(2)_q$ Yang-Mills theory in the zero-area limit (which is equivalent to BF-theory).

\section*{Acknowledgements}
We would like to thank Dionysios Anninos, Mattia Biancotto, Alessandro Giacchetto, Victor Godet, Clifford V. Johnson, Raghu Mahajan, Juan Maldacena, Nikita Nekrasov, Hirosi Ooguri, Boris Post, Ashoke Sen, Douglas Stanford, J\"org Teschner, Cumrun Vafa, Nico Valdes-Meller, Herman Verlinde, Edward Witten for for fruitful discussions and comments. We would in particular like to thank Aleksandr Artemev for thoroughly reading a first version of this draft and many helpful commments. 
SC, LE and BM thank l’Institut Pascal
at Universit\'e Paris-Saclay, with the
support of the program ``Investissements d’avenir'' ANR-11-IDEX-0003-01, 
and SC and VAR thank the Kavli Institute for Theoretical Physics (KITP), which is supported in part by grant NSF PHY-2309135, for hospitality during the course of this work. 
VAR is supported in part by the Simons Foundation Grant No. 488653, by the Future Faculty in the Physical Sciences Fellowship at Princeton University, and a DeBenedictis Postdoctoral Fellowship and funds from UCSB. 
SC is supported by the U.S. Department of Energy, Office of Science, Office of High Energy Physics of U.S. Department of Energy under grant Contract Number DE-SC0012567 (High Energy Theory research), DOE Early Career Award DE-SC0021886 and the Packard Foundation Award in Quantum Black Holes and Quantum Computation.
BM gratefully acknowledges funding provided by the Sivian Fund at the Institute for Advanced Study and the National Science Foundation with grant number PHY-2207584.

\appendix

\section{Some background on two-matrix integrals} \label{app:two-matrix models}
We provided a brief introduction to two-matrix integrals in section~\ref{sec:two-matrix models}. Here we fill some of the gaps in the explanation there in the interest of being self-contained.
\subsection{Derivation of the loop equations} \label{subapp:loop equations}
Let us derive the loop equations \eqref{eq:master loop equation}. We do this by using the invariance of the two-matrix integral 
\begin{equation}\label{eq:2MM new}
    \langle R(I) \rangle= \int_{\mathbb{R}^{N^2}} [\d M_1][\d M_2]\, R(I) \, \mathrm{e}^{-N \tr (V_1(M_1) + V_2(M_2) - M_1 M_2)}
\end{equation}
under a change of variables. Here $R(I)$ denotes resolvents for the matrix $M_1$. We separately consider two different change of variables and combine them later.
\paragraph{First loop equation.}
We consider the shift
\begin{equation}\label{eq:shift in M2}
    M_2 \rightarrow M_2 + \varepsilon\,  \frac{1}{x-M_1}
\end{equation}
for infinitesimal $\varepsilon$. Since the shift does not involve $M_2$, there is no Jacobian and terms of order $\varepsilon$ only come from the shifts of the term in the exponent of (\ref{eq:2MM new}). We get 
\begin{equation}
    \left\langle \tr \left(\frac{V_2'(M_2)}{x-M_1} -   \frac{x}{x-M_1} + 1 \right)R(I) \right\rangle =0~. \label{eq:first loop equation}
\end{equation}
\paragraph{Second loop equation.}
For this we consider the change of variables
\begin{equation}
    M_1 \rightarrow M_1+ \varepsilon\,  \frac{1}{x-M_1}\frac{V_2'(y)- V_2'(M_2)}{y-M_2}+ \varepsilon\, \frac{V_2'(y)- V_2'(M_2)}{y-M_2} \frac{1}{x-M_1}~. \label{eq:second change of variables}
\end{equation}
Notice that the symmetrization is necessary to ensure that $M_1$ remains a hermitian matrix after the shift. We now also get contributions from a Jacobian and the resolvents. Let us evaluate them in turn.

For the Jacobian, we notice that for a shift of the type $M_1 \to M_1+ \varepsilon A M_1^m B$ for constant matrices $A$ and $B$ and $m \ge 0$, we get to first order in $\varepsilon$,
\begin{align}\nonumber
    \text{Jac}&=1+\varepsilon \sum_{i,j} \frac{\partial(A M_1^m B)_{ij}}{\partial M_{1,ij}}\\ \nonumber
    &=1+\varepsilon \sum_{i,j,k,\ell} \sum_{n=0}^{m-1} (A M_1^n)_{ik} \, \frac{\partial M_{1,k\ell}}{\partial M_{1,ij}} \, (M_1^{m-n-1} B)_{\ell j} \\ \nonumber
    &=1+\varepsilon \sum_{i,j,k,\ell} \sum_{n=0}^{m-1} (A M_1^n)_{ik} \, \delta_{ik} \delta_{j\ell}\, (M_1^{m-n-1} B)_{\ell j} \\
    &=1+\varepsilon \sum_{n=0}^{m-1} \tr A M_1^n \tr B M_1^{m-n-1}~.
\end{align}
Thus for the change of variables \eqref{eq:second change of variables} we obtain the result by expanding in a Taylor series in $M_1$:
\begin{align}\nonumber
    \text{Jac}&=1+2\varepsilon \sum_{m=0}^\infty x^{-m-1}\sum_{n=0}^{m-1} \tr M_1^n  \frac{V_2'(y)- V_2'(M_2)}{y-M_2} \tr M_1^{m-n-1}\\
    &=1+2 \varepsilon \tr \frac{1}{x-M_1} \frac{V_2'(y)-V_2'(M_2)}{y-M_2} \tr \frac{1}{x-M_1} ~.
\end{align}
Next, we discuss the contribution from the resolvent. We have
\begin{align}
    R(x_i)= \tr \frac{1}{x_i-M_1} 
\rightarrow R(x_i)+2 \varepsilon\, \tr \frac{1}{(x_i-M_1)^2}\frac{1}{x-M_1}\frac{V_2'(y)- V_2'(M_2)}{y-M_2}~,
\end{align}
where we used the cyclicity of the trace. In general we have
\begin{equation}
    R_{M_1}(I)= R_{M_1}(x_1)R_{M_1}(x_2)\cdots R_{M_1}(x_n) = \tr \frac{1}{x_1-M_1} \times \cdots \times \tr \frac{1}{x_n-M_1}~.
\end{equation}
In total we get the loop equation
\begin{align}
    &\bigg\langle \tr \frac{1}{x-M_1} \tr \frac{1}{x-M_1} \frac{V_2'(y)- V_2'(M_2)}{y-M_2} R_{M_1}(I)\bigg\rangle \nonumber\\
    &- N \left\langle \tr \frac{V_1'(M_1)}{x-M_1} \frac{V_2'(y)- V_2'(M_2)}{y-M_2}\,R_{M_1}(I) - \tr \frac{1}{x-M_1}M_2\frac{V_2'(y)- V_2'(M_2)}{y-M_2}R_{M_1}(I)\right\rangle\nonumber\\
    &+ \left\langle \sum_{k=1}^n R_{M_1}(I\, \backslash\, x_k)\,\tr \frac{1}{(x_k-M_1)^2}\frac{1}{x-M_1}\frac{V_2'(y)- V_2'(M_2)}{y-M_2} \right\rangle=0~, \label{eq:second loop equation}
\end{align}
where the first line comes from the Jacobian, the second from the variation of the exponent and the third from the variation of the resolvents.
\paragraph{Master loop equation.} We can use the first loop equation \eqref{eq:first loop equation} to rewrite the first term in the second line of \eqref{eq:second loop equation}. We then recall the definitions \eqref{eq:def U} and \eqref{eq:def P}. It is the straightforward algebra to obtain the master loop equation \eqref{eq:master loop equation}.
\subsection{Analyticity of the resolvents} \label{subapp:analyticity resolvents}
We now explain the analyticity properties of the resolvents that we used in the derivation of topological recursion in section~\ref{subsec:top rec review}. The crucial step for this was taken in \cite{Chekhov:2006vd} and we essentially reproduce their argument.
\paragraph{Main claim.} The main claim of \cite{Chekhov:2006vd} is that
\begin{subequations}
    \begin{align}
    \langle U(\xx(z),y) R(I) \rangle&\overset{!}{=}- b_{d_2} N \bigg\langle\!\!\!\bigg \langle \prod_{i=1}^{d_2} \big(y-V_1'(\xx(z))+\tfrac{1}{N} R(\xx(z^i)) \big) R(I) \bigg\rangle \!\!\! \bigg\rangle_{\!\!\! I}~, \label{eq:main claim Chekhov Eynard Orantin U}\\
    \langle P(\xx(z),y) R(I) \rangle&\overset{!}{=}- b_{d_2} N \bigg\langle\!\!\! \bigg\langle \prod_{i=0}^{d_2} \big(y-V_1'(\xx(z))+\tfrac{1}{N} R(\xx(z^i)) \big) R(I) \bigg\rangle \!\!\! \bigg\rangle_{\!\!\! I}\nonumber\\
    &\quad-\frac{1}{N} \sum_{k=1}^n \partial_{x_k} \frac{\langle U(x_k ,y) R(I \setminus x_k) \rangle}{\xx(z)-x_k}~.
\label{eq:main claim Chekhov Eynard Orantin P}
    \end{align} \label{eq:main claim Chekhov Eynard Orantin}%
\end{subequations}
Notice that the product in the brackets excludes the term $i=0$ in the first line corresponding to the physical sheet, while the second line contains also the physical sheet.

\paragraph{Modified brackets.} Let us explain the double bracket notation in \eqref{eq:main claim Chekhov Eynard Orantin}.\footnote{It was denoted by a quote in \cite{Chekhov:2006vd}.} Let us first motivate it. We want to modify correlators such as to get rid off the singularity in $R_{0,2}$ when $\xx(z)=\xx(z')$, see \eqref{eq:propagator}. We will only define this modified correlator for products of resolvents as appears in \eqref{eq:main claim Chekhov Eynard Orantin}.
To define it, we decompose the correlator into its connected components as in \eqref{eq:disconnected to connected correlators}. Thus we first expand $\langle \!\langle  \cdots \rangle \!\rangle_I$ into connected components as in \eqref{eq:disconnected to connected correlators} with
\be 
\langle \! \langle R(x_1,\dots,x_n) \rangle \! \rangle_{\text{c},I}\equiv \langle R(x_1,\dots,x_n) \rangle_\text{c}+\frac{\delta_{n,2}\delta_{x_1 \not \in I \text{ or } x_2 \not \in I}}{(x_1-x_2)^2}~.
\ee
In other words all connected two-point functions are modified in this way except for $\langle R(x_k,x_\ell) \rangle_\text{c}$ with $\{x_k,x_\ell\} \subseteq I$. Thus we keep the subset $I$ in the notation in \eqref{eq:main claim Chekhov Eynard Orantin} to indicate which propagators are \emph{not} modified. 

Let us note that \eqref{eq:main claim Chekhov Eynard Orantin} is non-singular after the modification since the modified propagator is non-singular for $i \ne j$, see \eqref{eq:propagator}.

\paragraph{Proof.} To prove \eqref{eq:main claim Chekhov Eynard Orantin}, we first notice that the right-hand sides of \eqref{eq:main claim Chekhov Eynard Orantin U} and \eqref{eq:main claim Chekhov Eynard Orantin P} are polynomials of degree $d_2$ and $d_2+1$ in $y$, respectively. The coefficients of $y^{d_2}$ and $y^{d_2+1}$ are trivial to extract and match by construction with \eqref{eq:def U} and \eqref{eq:def P}. 

Let us next check \eqref{eq:main claim Chekhov Eynard Orantin} to leading order in large $N$. The leading term comes from the fully disconnected contribution and the second line of \eqref{eq:main claim Chekhov Eynard Orantin P} is also suppressed. Thus \eqref{eq:main claim Chekhov Eynard Orantin P} reduces at large $N$ to the equality
\be 
P_0(\xx(z),y) \overset{!}{=}-b_{d_2} \prod_{i=0}^{d_2} (y-\yy(z^i))~.
\ee
This is true by construction because it is just the product over the $d_2+1$ roots of $P_0(x,y)$ in $y$. Since also the leading coefficients match, this is a true equality. Similarly, \eqref{eq:main claim Chekhov Eynard Orantin U} gives the expected result at large $N$.

To complete the proof, let us remember from the discussion in section~\ref{subsec:loop equations} that the loop equations \eqref{eq:master loop equation} have a unique solution under the assumption that $\langle P(\xx(z),y)R(I) \rangle_\text{c}$ is a polynomial of degree $\le d_2$ in $y$ (with the exception of the $g=0$ piece for $I=\emptyset$ where it is a degree $d_2+1$ polynomial). This is satisfied for \eqref{eq:main claim Chekhov Eynard Orantin P} since for the connected quantities the leading term $-b_{d_2}N y^{d_2+1}$ only shows up for $g=0$ and $I=\emptyset$. Thus the assumption is satisfied and it only remains to show that \eqref{eq:main claim Chekhov Eynard Orantin} satisfies the loop equations \eqref{eq:master loop equation}. Let us compute $\text{RHS}-\text{LHS}$ of the loop equations \eqref{eq:master loop equation} and insert \eqref{eq:main claim Chekhov Eynard Orantin}. This leads to
\begin{align}
    \text{RHS}-\text{LHS}&=- b_{d_2} N \bigg\langle\!\!\!\bigg \langle \prod_{i=0}^{d_2} \big(y-V_1'(\xx(z))+\tfrac{1}{N} R(\xx(z^i)) \big) R(I) \bigg\rangle \!\!\! \bigg\rangle_{\!\!\! I,\xx(z)} \nonumber\\
    &\quad + b_{d_2} N \bigg\langle\!\!\!\bigg \langle \prod_{i=0}^{d_2} \big(y-V_1'(\xx(z))+\tfrac{1}{N} R(\xx(z^i)) \big) R(I) \bigg\rangle \!\!\! \bigg\rangle_{\!\!\! I} \nonumber\\
    &\quad- \sum_{k=1}^n  \frac{b_{d_2}}{(\xx(z)-x_k)^2} \, \bigg\langle\!\!\!\bigg \langle \prod_{i=1}^{d_2} \big(y-V_1'(\xx(z))+\tfrac{1}{N} R(\xx(z^i)) \big) R(I \setminus x_k) \bigg\rangle \!\!\! \bigg\rangle_{\!\!\! I \setminus x_k}~. \label{eq:Chekhov-Eynard-Orantin plugged in}
\end{align}
The first line comes from the first term on the RHS of \eqref{eq:master loop equation}, while the second line comes from inserting \eqref{eq:main claim Chekhov Eynard Orantin P} for $\langle P(x,y) R(I) \rangle$. The term appearing on the RHS of \eqref{eq:main claim Chekhov Eynard Orantin P} cancels with one of the terms in the second line of the loop equations \eqref{eq:master loop equation}. Clearly, the first and second line of \eqref{eq:Chekhov-Eynard-Orantin plugged in} almost cancel, they just differ in the treatment of the double bracket correlator. The second line also shifts the propagators of the form $\langle R(\xx(z),x_k) \rangle_\text{c}$, while the first one does not. This means that the first two lines combined exactly cancel the third line and one obtains
\be 
\text{LHS}-\text{RHS}=0~.
\ee
This proves \eqref{eq:main claim Chekhov Eynard Orantin}.

\paragraph{Analyticity.} Finally, it is simple to derive analyticity properties of the correlators that are needed for the topological recursion. One expands \eqref{eq:main claim Chekhov Eynard Orantin P} for large $y$. Let us note that the correction term in \eqref{eq:main claim Chekhov Eynard Orantin P} is manifestly single-valued in $\xx(z)$ and only has poles away from the branch points. Extracting the coefficient of $y^{d_2}$ leads to
\be 
\sum_{i=0}^{d_2} \langle \! \langle R(\xx(z^i),I) \rangle \! \rangle_I=\text{analytic in $\xx(z)$}~. \label{eq:analyticity sum over branches linear}
\ee
``Analytic in $\xx(z)$'' means a single-valued function that is analytic around a neighborhood of the branch points. Notice that this statement is quite non-trivial since the resolvents typically have singularities at the branch points of $\xx(z)$, but these cancel out in the sum. One can in fact give an explicit formula for the RHS of \eqref{eq:analyticity sum over branches linear} by keeping track of the other terms of order $y^{d_2}$ in \eqref{eq:main claim Chekhov Eynard Orantin}. We will in the following denote equality up to such analytic terms by $\sim$. 

We can similarly extract the coefficient of order $y^{d_2-1}$ which leads to the statement
\be 
\sum_{i\ne j} \langle \! \langle R(\xx(z^i),\xx(z^j),I) \rangle \! \rangle_I\sim 0~.\label{eq:analyticity sum over branches quadratic}
\ee

\paragraph{Connected parts and genus expansion.} Given \eqref{eq:analyticity sum over branches linear} and \eqref{eq:analyticity sum over branches quadratic}, one can restrict to the connected part. This gives
\begin{subequations}
\begin{align} 
0&\sim \sum_{i=0}^{d_2} \langle\!\langle R(\xx(z^i),\xx(I)) \rangle\!\rangle_\text{c}~, \label{eq:analyticity sum over branches linear connected}\\
0&\sim \sum_{i \ne j} \langle\!\langle R(\xx(z^i),\xx(z^j),\xx(I))  \rangle\!\rangle_\text{c}
+\sum_{J \subseteq I} \langle \!\langle R(\xx(z^i),\xx(J))  \rangle\!\rangle_\text{c}\langle\!\langle R(\xx(z^j),\xx(J^c))  \rangle\!\rangle_\text{c} ~.\label{eq:analyticity sum over branches quadratic connected}
\end{align} \label{eq:analyticity sum over branches connected}%
\end{subequations}
This is proven recursively in the size of the set $I$ from \eqref{eq:analyticity sum over branches linear} and \eqref{eq:analyticity sum over branches quadratic}. For example, if we expand \eqref{eq:analyticity sum over branches linear} into connected components all terms except for the connected part appearing in \eqref{eq:analyticity sum over branches linear connected} are analytic by recursion and thus it also follows for \eqref{eq:analyticity sum over branches linear connected}. A similar argument  demonstrates \eqref{eq:analyticity sum over branches quadratic connected}.

\subsection{Derivation of the topological recursion} \label{subapp:top rec derivation}
We insert the genus expansion and use the definition of $\omega_{g,n}$ \eqref{eq:omegagn definition}, \eqref{eq:omega01 02 definition} into \eqref{eq:analyticity sum over branches connected} to find
\begin{subequations}
\begin{align} 
0&\sim \sum_{i=0}^{d_2} \omega_{g,n}(z^i,I)~, \label{eq:analyticity sum over branches linear omega}\\
0 &\sim \sum_{i \ne j} \bigg(\omega_{g-1,n+1}(z^i,z^j,I)+\sideset{}{'}\sum_{\begin{subarray}{c} 0 \le h \le g \\ J \subseteq I \end{subarray}} \omega_{h,|J|+1}(z^i,J)\omega_{g-h,|J^c|+1}(z^j,J^c)\bigg)  ~.\label{eq:analyticity sum over branches quadratic omega}
\end{align} \label{eq:analyticity sum over branches omega}%
\end{subequations}
To treat the special case $\omega_{0,1}$, we needed to make use of \eqref{eq:analyticity sum over branches linear omega}, which implies that the extra pieces in the definition \eqref{eq:omega01 02 definition} still give something analytic. 

To bring the second equation into a form that is useful to derive the recursion relation, we isolate the terms involving $\omega_{0,1}$ and bring them on the other side
\begin{multline}
    -2\sum_{i \ne j} \omega_{0,1}(z^i) \omega_{g,n}(z^j,I) \\
    \sim \sum_{i \ne j} \bigg(\omega_{g-1,n+1}(z^i,z^j,I)+\sum_{\begin{subarray}{c} 0 \le h \le g \\ J \subseteq I \end{subarray}}  \omega_{h,|J|+1}(z^i,J)\omega_{g-h,|J|+1}(z^j,J^c)\bigg)~, \label{eq:loop equations omega}
\end{multline}
where the prime on the sum means that the terms $h=0$, $J=\emptyset$ as well as $h=g$, $J=I$ are omitted.

We now apply the operation
\be 
\sum_{m\text{ branch points}} \Res_{z=z^*_m} K_m(z_1,z) \label{eq:sum recursion kernel}
\ee 
to both sides of \eqref{eq:loop equations omega}. Analytic terms do not contribute to this operation since they have by assumption no singularities at the branch points.

\paragraph{Right-hand-side.} Let us start by applying \eqref{eq:sum recursion kernel} to the RHS. Only the terms with $\{z^i,z^j\}=\{z,\sigma_m(z)\}$ can contribute since the others all cancel out. For example, if $z^j \ne z,\sigma_m(z)$, the residue at $z^i=z$ and $z^i=\sigma_m(z)$ will cancel because of \eqref{eq:analyticity sum over branches linear omega}. Thus we get
\begin{multline} 
 \text{RHS}=2\sum_{m}\Res_{z=z^*_m} K_m(z_1,z) \bigg(\omega_{g-1,n+1}(z,\sigma_m(z),I)\\
 +\sideset{}{'}\sum_{\begin{subarray}{c} 0 \le h \le g \\ J \subseteq I \end{subarray}}  \omega_{h,|J|+1}(z,J)\omega_{g-h,|J|+1}(\sigma_m(z),J^c)\bigg)~.
\end{multline}
\paragraph{Left-hand-side.} Applying it to the LHS of \eqref{eq:loop equations omega} is more interesting. We find
\begin{align}
    \text{LHS}&=-2 \sum_{m}\Res_{z=z^*_m} K_m(z_1,z) \big(\omega_{0,1}(z)\omega_{g,n}(\sigma_m(z),I)+\omega_{0,1}(\sigma_m(z))\omega_{g,n}(z,I)\big) \label{eq:LHS 1}\\
    &=-2 \sum_{m}\Res_{z=z^*_m} K_m(z_1,z) \big(-\omega_{0,1}(z)\omega_{g,n}(z,I)+\omega_{0,1}(\sigma_m(z))\omega_{g,n}(z,I)\big) \label{eq:LHS 2}\\
    &=\sum_{m}\Res_{z=z^*_m} \int_{z'=\sigma_m(z)}^{z}\omega_{0,2}(z_1,z') \omega_{g,n}(z,I) \label{eq:LHS 3}\\
    &=\sum_{m}\Res_{z=z^*_m} \bigg(\int_{z'=*}^{z}\omega_{0,2}(z_1,z')-\int_{z'=*}^{\sigma_m(z)}\omega_{0,2}(z_1,z')\bigg) \omega_{g,n}(z,I) \label{eq:LHS 4}\\
    &=\sum_{m}\Res_{z=z^*_m} \bigg(\int_{z'=*}^{z}\omega_{0,2}(z_1,z')\omega_{g,n}(z,I)-\int_{z'=*}^{z}\omega_{0,2}(z_1,z') \omega_{g,n}(\sigma_m(z),I)\bigg) \label{eq:LHS 5}\\
    &=2\sum_{m}\Res_{z=z^*_m} \int_{z'=*}^{z}\omega_{0,2}(z_1,z')\omega_{g,n}(z,I)~.\label{eq:LHS 6}
\end{align}
In \eqref{eq:LHS 1} we used that only $\{z^i,z^j\}=\{z,\sigma_m(z)\}$ can contribute to the residue as above. In \eqref{eq:LHS 2}, we used again \eqref{eq:analyticity sum over branches linear omega} along with the observation that the terms $K_m(z_1,z) \omega_{0,1}(z) \omega_{g,n}(z^j,I)$ with $z^j \ne z,\sigma_m(z)$ do not contribute to the residue. We then insert the definition \eqref{eq:definition recursion kernel} in \eqref{eq:LHS 3} and split the integral in \eqref{eq:LHS 4}, where $*$ is an arbitrary reference point on the surface. In \eqref{eq:LHS 5} we change variables $z \to \sigma_m(z)$ in the second term and in \eqref{eq:LHS 6} use \eqref{eq:analyticity sum over branches linear omega} again.

Finally, we can rewrite this as the sum over all the other residues on the surface. On a general spectral curve, one has to be careful since the appearing form is not single-valued in $z$. This turns out to be not an issue because of the property \eqref{eq:A cycle integral trivial}. We thus just restrict to the sphere case for simplicity, where 
\be 
\int_{z'=*}^{z}\omega_{0,2}(z_1,z')=\frac{\d z_1}{*-z_1}-\frac{\d z_1}{z-z_1}
\ee
\emph{is} a single-valued function in $z$. The only other singularity of the integrand in \eqref{eq:LHS 6} is at $z=z_1$ and thus 
\begin{align} 
    \text{LHS}&=-2 \Res_{z=z_1} \int_{z'=*}^{z}\omega_{0,2}(z_1,z')\omega_{g,n}(z,I) \\
    &=2 \omega_{g,n}(z_1,I) ~.
\end{align}
Thus the recursion relation \eqref{eq:topological recursion} follows.

\section{Intersection theory} \label{app:intersection theory}
In this appendix, we derive the expression \eqref{eq:Agn through quantum volumes} of $\mathsf{A}_{g,n}^{(b)}$ as computed from the matrix integral in terms of intersection numbers.
\subsection{General formula}
Let us first fill the gaps in our explanation of \eqref{eq:omegagn general intersection number formula} and define the various parameters that enter it.
We will assume that the spectral curve under consideration has a global rational parametrization. This parametrization doesn't necessarily have to be one-to-one and to emphasize this we will denote the coordinate as $w$. The expressions one gets for the intersection numbers depend very much on the choice of this coordinate $w$ and different choices lead to very different looking expressions. Thus we will make a judicial choice in our example.

$B_{m_1,r,m_2,s}$ is defined in terms of the following expansion coefficients.
Let us expand $w_1$ and $w_2$ in $\omega_{0,2}$ around the branch points $w_{m_1}^*$, $w_{m_2}^*$. To do this, we should first define a local coordinate $\zeta_m(w)$ that satisfies 
\be 
\zeta_m(\sigma_m(w))=-\zeta_m(w)~.
\ee
We can choose $\zeta_m(w)=\sqrt{\xx(w)-\xx(w_m^*)}$. In the case that we discuss below, the local Galois inversion acts linearly and we can simply choose $\zeta_m(w)=w-w_m^*$. 
\begin{multline}
    \omega_{0,2}(w_1,w_2)\sim \bigg[\frac{\delta_{m_1,m_2}}{(\zeta_{m_1}(w_1)-\zeta_{m_2}(w_2))^2}\\
    +2\pi \sum_{r,s=0}^\infty \frac{B_{m_1,r,m_2,s}\, \zeta_{m_1}(w_1)^r \zeta_{m_2}(w_2)^s}{\Gamma(\frac{r+1}{2})\Gamma(\frac{s+1}{2})}\bigg] \d \zeta_{m_1}(w_1)\, \d \zeta_{m_2}(w_2)~ . \label{eq:B definition}
\end{multline}
Notice that this in principle defines $B_{m_1,r,m_2,s}$ for all non-negative integers $r$ and $s$, even though only even integers enter in \eqref{eq:omegagn general intersection number formula}. 
The differentials $\d \eta_{m,\ell}(w)$ are defined as
\begin{align}
    \d \eta_{m,\ell}(w)\overset{w \sim w_{m'}^*}{\sim}\bigg[-\frac{2\delta_{m,m'}\Gamma(\ell+\frac{3}{2})}{\sqrt{\pi} \zeta_{m'}(w)^{2\ell+2}}-2\sqrt{\pi} \sum_{r=0}^\infty \frac{B_{m',r,m,2\ell}\zeta_{m'}(w)^r}{\Gamma(\frac{r+1}{2})}\bigg] \, \d \zeta_{m'}(w)~. \label{eq:eta definition}
\end{align}
The quantities $\hat{t}_{m,k}$ are defined in terms of the Taylor expansion of $\omega_{0,1}$ around the branch points. Let
\be 
\omega_{0,1}(w)=\sum_{k=0,\frac{1}{2},1,\frac{3}{2},\dots}^\infty \frac{\sqrt{\pi}\, t_{m,k}}{2\, \Gamma(k+\frac{3}{2})}\, \zeta_m(w)^{2k+2} \d \zeta_m(w)~. \label{eq:tmk definition}
\ee
The half-integer expansion coefficients cancel out of the recursion kernel \eqref{eq:recursion kernel} and will therefore not appear in the intersection number. formulas. Finally, $\hat{t}_{m,k}$ is defined in terms of $t_{m,k}$ by requiring the equality
\be 
\sum_{k=0}^\infty t_{m,k} u^k=\exp\left(-\sum_{k=0}^\infty \hat{t}_{m,k} u^k \right) \label{eq:hattmk definition}
\ee
as a formal power series in $\CC[u]$. 
\subsection{The spectral curve of the complex Liouville string}
In our case, these three formulas are easily evaluated. We get much more convenient expressions if we use the following parametrization of the spectral curve,
\be 
\xx(w)=-2\cos(\pi b^{-1}w)~, \quad \yy(w)=2\cos(\pi b w)~,
\ee
where we remember that the branch points are located at $w_m^*=b m$ with $m \in \ZZ_{\ge 1}$. $\omega^{(b)}_{0,2}$ takes the following form in these coordinates
\be 
\omega^{(b)}_{0,2}(w_1,w_2)=\left[\frac{1}{(w_1-w_2)^2}-\frac{1}{(w_1+w_2)^2}\right]\, \d w_1 \, \d w_2~. \label{eq:omega02 two term form}
\ee
The Galois inversion has the explicit form $\sigma_m(w)=2w_m^*-w$ and we can hence choose $\zeta_m(w)=w-w_m^*$ above.
For the coefficients $B_{m_1,r,m_2,s}$ we get by inserting \eqref{eq:omega02 two term form} into \eqref{eq:B definition}
\begin{align}
    B_{m_1,r,m_2,s}=\frac{2(-1)^r}{(2b)^{r+s+2}} \left(\frac{\delta_{m_1\ne m_2}}{(m_1-m_2)^{r+s+2}}-\frac{(-1)^s}{(m_1+m_2)^{r+s+2}}\right) \frac{\Gamma(r+s+2)}{\Gamma(\frac{r+2}{2})\Gamma(\frac{s+2}{2})}~.
\end{align}
Hence 
\begin{multline}
    \sum_{r,s=0}^\infty B_{m_\bullet, 2r,m_\circ, 2s} \psi_\bullet^r \psi_\circ^s=\frac{1}{\sqrt{\pi}} \, \sum_{d=0}^\infty \Gamma(d+\tfrac{3}{2}) b^{-2d-2} \\
    \times\left(\frac{\delta_{m_\bullet\ne m_\circ}}{(m_\bullet-m_\circ)^{2d+2}}-\frac{1}{(m_\bullet+m_\circ)^{2d+2}}\right)(\psi_\bullet+\psi_\circ)^d~.
\end{multline}
For the differentials $\d \eta_{m,\ell}$, we get from \eqref{eq:eta definition}
\begin{align}
    \d \eta_{m,\ell}(w)=-\frac{2 \Gamma(\ell+\frac{3}{2})}{\sqrt{\pi}} \left(\frac{1}{(w-w_m^*)^{2\ell+2}}-\frac{1}{(w+w_m^*)^{2\ell+2}}\right) \d w~.
\end{align}
The coefficients $t_{m,k}$ can be evaluated from their definition \eqref{eq:tmk definition} and read for $k \in \ZZ_{\ge 0}$
\be 
t_{m,k}=\frac{8(\frac{\pi}{2})^{2 k+3} (-1)^{k+m} \sin (\pi m b^2)\big((b+b^{-1})^{2 k+2}-(b-b^{-1})^{2 k+2}\big) }{b\, \Gamma(k+2)}~.
\ee
The power series appearing in \eqref{eq:hattmk definition} is hence equal to
\be 
\sum_{k=0}^\infty t_{m,k} u^k=\frac{8\pi (-1)^{m+1} \sin (\pi m  b^2)\, \mathrm{e}^{-\frac{1}{4} (b^2+b^{-2}) \pi^2 u}  \sinh (\frac{1}{2} \pi^2 u)}{b u}~,
\ee
and so
\begin{align}
    \sum_{k=0}^\infty \hat{t}_{m,k} u^k=\log\left(\frac{b(-1)^{m}}{8 \pi \sin(\pi m b^2)}\right)+\frac{b^2+b^{-2}}{4}\, \pi^2 u-\log \left(\frac{\sinh(\frac{1}{2} \pi^2 u)}{u}\right)~.
\end{align}
We can thus read off
\begin{subequations}
\begin{align}
    \hat{t}_{m,0}&=\log\left(\frac{b(-1)^{m}}{4\pi^3\sin(\pi m b^2)}\right)~, \\
    \hat{t}_{m,1}&=\frac{b^2+b^{-2}}{4}\, \pi^2~, \\
    \hat{t}_{m,2k}&=-\frac{B_{2k}\pi^{4k}}{2k(2k)!}~,
\end{align}
\end{subequations}
with $B_{2k}$ the Bernoulli numbers.
Finally, we can use that 
\be 
\kappa_0=2g-2+n
\ee
is just a number. Writing \eqref{eq:omegagn general intersection number formula} out leads to
\begin{align} 
\omega^{(b)}_{g,n}(\boldsymbol{w})&=2^{3g-3+n}(-1)^n\sum_{\Gamma \in \mathcal{G}^\infty_{g,n}} \frac{1}{|\text{Aut}(\Gamma)|} \prod_{v \in \mathcal{V}_\Gamma} \left(\frac{b(-1)^{m_v}}{4\pi^3\sin(\pi m_v b^2)}\right)^{2g_v-2+n_v}  \nonumber\\
&\quad\times\int_{\bM_\Gamma} \prod_{v \in \mathcal{V}_\Gamma}\mathrm{e}^{\frac{b^2+b^{-2}}{4} \pi^2\kappa_1-\sum_k\frac{B_{2k}\pi^{4k}}{(2k)(2k)!}\kappa_{2k}} \nonumber\\
    &\qquad\times\!\prod_{(\bullet,\circ) \in \mathcal{E}_\Gamma} \sum_{d=0}^\infty \frac{\Gamma(d+\tfrac{3}{2})}{\sqrt{\pi}\, b^{2d+2}} \left(\frac{\delta_{m_\bullet\ne m_\circ}}{(m_\bullet-m_\circ)^{2d+2}}-\frac{1}{(m_\bullet+m_\circ)^{2d+2}}\right)(\psi_\bullet+\psi_\circ)^d \nonumber\\
    &\qquad\times \prod_{i=1}^n \sum_{\ell=0}^\infty\frac{2 \Gamma(\ell+\frac{3}{2})}{\sqrt{\pi}} \left(\frac{1}{(w_i-w_{m_i}^*)^{2\ell+2}}-\frac{1}{(w_i+w_{m_i}^*)^{2\ell+2}}\right) \psi_i^\ell \d w~. \label{eq:omegagn in terms of intersection numbers}
\end{align}
\paragraph{Translating to $\mathsf{A}_{g,n}^{(b)}$.} We can now apply eq.~\eqref{eq:Agn omegagn relation} to translate this into a formula for $\mathsf{A}_{g,n}^{(b)}$. Observe that
\be 
\Res_{w=w_m^*} \frac{\cos(2\pi p w)}{p}\, \d \eta_{m,\ell}(w)=\frac{2 \pi  \sin(2\pi m b p) (-1)^\ell (\pi p)^{2\ell}}{\Gamma(\ell+1)}~.
\ee
Thus the corresponding insertion in the integral over $\bM_\Gamma$ becomes
\be 
\sum_{\ell=0}^\infty \frac{2 \pi \sin(2\pi m_i b p_i) (-1)^\ell (\pi p_i)^{2\ell}}{\Gamma(\ell+1)}\, \psi_i^\ell=2 \pi \sin(2\pi m b p)\, \mathrm{e}^{-\pi^2 p_i^2 \psi_i} ~.
\ee
Notice that after these operations, $\pi^2$ appears homogeneously in the cohomology degree. Since we are picking out the top form on moduli space, we get a factor of $\pi^{2 \dim \bM_\Gamma}=\pi^{2(3g-3+n)-2|\mathcal{E}_\Gamma|}$. The number of edges is equal to the difference of the dimension of the total moduli space and the dimension of the moduli space of the graph $\Gamma$, since every edge corresponds to a degenerated cycle of the surface. Using also that the Euler characteristics of the different components of the stable graphs add up to the total Euler characteristic, we can write
\begin{align}
    \mathsf{A}_{g,n}^{(b)}(\boldsymbol{p})&= \sum_{\Gamma \in \mathcal{G}_{g,n}^\infty}\frac{1}{|\text{Aut}(\Gamma)|}\prod_{v \in \mathcal{V}_\Gamma} \left(\frac{b (-1)^{m_v}}{\sqrt{2}\sin(\pi m_v b^2)}\right)^{2g_v-2+n_v} \!\! \int_{\bM_\Gamma} \prod_{v \in \mathcal{V}_\Gamma}\mathrm{e}^{\frac{b^2+b^{-2}}{4} \kappa_1-\sum_k\frac{B_{2k}\kappa_{2k}}{(2k)(2k)!}} \nonumber\\
    &\qquad\times\!\prod_{(\bullet,\circ) \in \mathcal{E}_\Gamma} \sum_{d=0}^\infty \frac{\Gamma(d+\tfrac{3}{2})}{\sqrt{\pi}(\pi b)^{2d+2}}\left(\frac{\delta_{m_\bullet\ne m_\circ}}{(m_\bullet-m_\circ)^{2d+2}}-\frac{1}{(m_\bullet+m_\circ)^{2d+2}}\right)(\psi_\bullet+\psi_\circ)^d \nonumber\\
    &\qquad\times \prod_{i=1}^n \mathrm{e}^{-p_i^2 \psi_i} \sqrt{2}\sin(2\pi m_i b p_i) ~.\label{eq:Agn internal momenta integrated out}
\end{align}
\paragraph{Integration over internal momenta.} As a final step, we notice that for $m_\bullet \ne m_\circ$, we have
\begin{multline}
    \int (-2 p \, \d p) \sqrt{2}\sin(2\pi m_1 b p) \sqrt{2}\sin(2\pi m_2 b p) \mathrm{e}^{-p^2 (\psi_\bullet+\psi_\circ)}\\=\sum_{d=0}^\infty \frac{\Gamma(d+\frac{3}{2})}{\sqrt{\pi}(\pi b)^{2d+2}} \left(\frac{1}{(m_\bullet-m_\circ)^{2d+2}}-\frac{1}{(m_\bullet+m_\circ)^{2d+2}}\right)(\psi_\bullet+\psi_\circ)^d
\end{multline}
as a formal power series in $\psi_\bullet$ and $\psi_\circ$. If $m_\bullet=m_\circ$, we omit by definition of the primed integral \eqref{eq:primed integral} the divergent term. Thus we can write
\begin{align} 
 \mathsf{A}_{g,n}^{(b)}(p_1,\dots,p_n)&=\sum_{\Gamma \in \mathcal{G}_{g,n}^\infty}\frac{1}{|\text{Aut}(\Gamma)|}\prod_{v \in \mathcal{V}_\Gamma} \left(\frac{b(-1)^{m_v}}{\sqrt{2}\sin(\pi m_v b^2)}\right)^{2g_v-2+n_v} \int' \prod_{e \in \mathcal{E}_\Gamma} (-2 p_e \, \d p_e)\nonumber\\
 &\quad \times\prod_{e \in \mathcal{E}_\Gamma} \sqrt{2}\sin(2\pi m_\bullet b p_e)\, \sqrt{2}\sin(2\pi m_\circ b p_e) \prod_{i=1}^n \sqrt{2}\sin(2\pi m_i b p_i)\nonumber\\
 &\quad\times \prod_{v \in \mathcal{V}_\Gamma} \int_{\bM_{g_v,n_v}} \mathrm{e}^{\frac{b^2+b^{-2}}{4} \kappa_1-\sum_k\frac{B_{2k}}{(2k)(2k)!}\kappa_{2k}-\sum_{j \in I_v} p_j^2 \psi_j}~.
\end{align}
Finally, we use that the remaining intersection number is precisely the quantum volume defined in \cite{Collier:2023cyw} to recover \eqref{eq:Agn through quantum volumes}.

\section{Derivation of the recursive representation of the string amplitudes}\label{app:recursion derivation}
Here we fill in the details of the derivation of the recursive reperesentation of the string amplitdues (\ref{eq:Agn recursion relation}) starting from the spectral curve (\ref{eq:spectral curve}) of the complex Liouville string. In this discussion it is again most convenient to parameterize the spectral curve in terms of the $w$ coordinates so that 
\begin{equation}
    \mathsf{x}(w) = -2\cos(\pi b^{-1}w), \quad \mathsf{y}(w) = 2\cos(\pi b w)\,,
\end{equation}
and 
\begin{subequations}
\begin{align}
    \omega_{0,1}^{(b)}(w) &= -\frac{4\pi \cos(\pi b w)\sin(\pi b^{-1} w)}{b}\, \d w ~ ,\\
    \omega_{0,2}^{(b)}(w_1,w_2) &= \left(\frac{1}{(w_1-w_2)^2} - \frac{1}{(w_1+w_2)^2}\right)\d w_1\, \d w_2\, .
\end{align}
\end{subequations}
The branch points of the spectral curve correspond to $w = \pm m b$ for $m\in\mathbb{Z}_{\geq 1}$, with the local Galois inversion given by $\sigma_m(w) = 2mb - w$. The higher resolvent differentials are then determined by the topological recursion (\ref{eq:topological recursion}) with the recursion kernel given by (\ref{eq:recursion kernel w basis}).

To proceed, we will focus on the first term in the recursion relation (\ref{eq:Agn recursion relation}); the other terms follow from nearly identical manipulations. Consider in particular the one-point string amplitude at any genus, $\mathsf{A}_{g,1}^{(b)}(p_1)$ (the analysis is identical for $n \geq 1$, with the other momenta $p_2,\ldots, p_n$ spectating for what follows). From the relation between the string amplitudes and the resolvent differentials (\ref{eq:Agn omegagn relation w}) together with the topological recursion for the latter (\ref{eq:topological recursion}) we have\footnote{Throughout this appendix it is understood that we omit the $\d w_i$ factors from the resolvent differentials $\omega_{g,n}^{(b)}$. This in particular leads to an extra minus sign in front of (\ref{eq:p1Ag1 residue}) compared to (\ref{eq:topological recursion}).}
\begin{align}
    p_1 \mathsf{A}_{g,1}^{(b)}(p_1) &= \sum_{m_1 =1}^\infty \Res_{w_1 = m_1 b}\cos(2\pi p_1 w_1)\omega_{g,1}^{(b)}(w_1) \\
    &\supset -\sum_{m_1=1}^\infty \Res_{w_1 = m_1 b}\cos(2\pi p_1 w_1) \sum_{m=1}^\infty \Res_{w= m b} K_m^{(b)}(w_1,w) \omega_{g-1,2}^{(b)}(w,\sigma_m(w))\label{eq:p1Ag1 residue}\, .
\end{align}
In the second line we have focused on the first term in the topological recursion (\ref{eq:topological recursion}), with the sum over $m$ corresponding to the sum over branch points of the spectral curve. We now swap the order of the sums and their corresponding residues, and use the fact that viewed as a function of $w_1$, the summand only has poles at $w_1 = w$ and $\sigma_m(w)$, to arrive at
\begin{equation}
    p_1\mathsf{A}_{g,1}^{(b)}(p_1) \supset \sum_{m=1}^\infty \Res_{w= m b} \frac{b \sin(2\pi m b p_1)\sin(2\pi p_1(m b- w))}{8\pi \sin(\pi m b^2)\sin(\pi b^{-1}w)\sin(\pi b(m b -w))}\omega_{g-1,2}^{(b)}(w,\sigma_m(w))\, .
\end{equation}
Rewriting $\omega_{g-1,2}^{(b)}$ in terms of $\mathsf{A}_{g-1,2}^{(b)}$\footnote{Recall that the inverse transform that expresses the resolvents in terms of the string amplitudes is given by (\ref{eq:omega in terms of A}).} and $w = m b + 2u$ inside the sum, this then becomes 
\begin{align}
    p_1 \mathsf{A}_{g,1}^{(b)}(p_1) &\supset \sum_{m=1}^\infty\frac{\pi b (-1)^m \sin(2\pi m b p_1)}{\sin(\pi m b^2)} \Res_{u=0}\bigg\{ \frac{\sin(4\pi u p_1)}{\sin(2\pi b u)\sin(2\pi b^{-1} u)}\nonumber\\
    &\quad \times \int 2q \d q\, 2q' \d q' \sin(2\pi q(m b+2u))\sin(2\pi q'(m b -2u))\mathsf{A}_{g-1,2}^{(b)}(q,q')\bigg\}\, . 
\end{align}
This is the first term of the recursive representation as first quoted in the main text in equation (\ref{eq:Agn recursion first pass}). 

\begin{figure}[ht]
    \centering
    \begin{tikzpicture}
        \draw[<->] (-2,0) to (2,0);
        \draw[<->] (0,-2) to (0,2);
        \draw[very thick,red,decoration={markings, mark=at position 0.25 with {\arrow{>}}}, postaction={decorate}] (0,0) circle (1/4);
        \node[thick, cross out, draw=black,scale=3/4] at (0,0) {};
        \node[thick, cross out, draw=black,scale=3/4] at (1/2,1/2) {};
        \node[thick, cross out, draw=black,scale=3/4] at (1,1) {};
        \node[thick, cross out, draw=black,scale=3/4] at (3/2,3/2) {};
        \node[thick, cross out, draw=black,scale=3/4] at (-1/2,-1/2) {};
        \node[thick, cross out, draw=black,scale=3/4] at (-1,-1) {};
        \node[thick, cross out, draw=black,scale=3/4] at (-3/2,-3/2) {};
        \node[thick, cross out, draw=black,scale=3/4] at (2/3,-2/3) {};
        \node[thick, cross out, draw=black,scale=3/4] at (4/3,-4/3) {};
        \node[thick, cross out, draw=black,scale=3/4] at (-2/3,2/3) {};
        \node[thick, cross out, draw=black,scale=3/4] at (-4/3, 4/3) {};

        \node at (2.25,2.25) {$u$};
        \draw (2.1,2.4) to (2.1,2.1) to (2.4,2.1);

        \node[scale=2] at (3,0) {$\Rightarrow$};

        \begin{scope}[shift={(6,0)}]
            \draw[<->] (-2,0) to (2,0);
            \draw[<->] (0,-2) to (0,2);
            \node[thick, cross out, draw=black,scale=3/4] at (0,0) {};
            \node[thick, cross out, draw=black,scale=3/4] at (1/2,1/2) {};
            \node[thick, cross out, draw=black,scale=3/4] at (1,1) {};
            \node[thick, cross out, draw=black,scale=3/4] at (3/2,3/2) {};
            \node[thick, cross out, draw=black,scale=3/4] at (-1/2,-1/2) {};
            \node[thick, cross out, draw=black,scale=3/4] at (-1,-1) {};
            \node[thick, cross out, draw=black,scale=3/4] at (-3/2,-3/2) {};
            \node[thick, cross out, draw=black,scale=3/4] at (2/3,-2/3) {};
            \node[thick, cross out, draw=black,scale=3/4] at (4/3,-4/3) {};
            \node[thick, cross out, draw=black,scale=3/4] at (-2/3,2/3) {};
            \node[thick, cross out, draw=black,scale=3/4] at (-4/3, 4/3) {};

            \draw[very thick,red,decoration={markings, mark=at position 0.25 with {\arrow{<}}}, postaction={decorate}] (-3/2,-3/2) circle (1/4);
            \draw[very thick,red,decoration={markings, mark=at position 0.25 with {\arrow{<}}}, postaction={decorate}] (-1,-1) circle (1/4);
            \draw[very thick,red,decoration={markings, mark=at position 0.25 with {\arrow{<}}}, postaction={decorate}] (-1/2,-1/2) circle (1/4);
            \draw[very thick,red,decoration={markings, mark=at position 0.25 with {\arrow{<}}}, postaction={decorate}] (1/2,1/2) circle (1/4);
            \draw[very thick,red,decoration={markings, mark=at position 0.25 with {\arrow{<}}}, postaction={decorate}] (1,1) circle (1/4);
            \draw[very thick,red,decoration={markings, mark=at position 0.25 with {\arrow{<}}}, postaction={decorate}] (3/2,3/2) circle (1/4);

            \draw[very thick,red,decoration={markings, mark=at position 0.5 with {\arrow{<}}}, postaction={decorate}] (-2,-3/2) to (3/2,2);
            \draw[very thick,red,decoration={markings, mark=at position 0.5 with {\arrow{>}}}, postaction={decorate}] (-3/2,-2) to (2,3/2);

            \node at (2.25,2.25) {$u$};
            \draw (2.1,2.4) to (2.1,2.1) to (2.4,2.1);
            
        \end{scope}
        
    \end{tikzpicture}
    \caption{The deformation of the $u$ contour in defining the residue at $u=0$.}\label{fig:u contour deformation}
\end{figure}
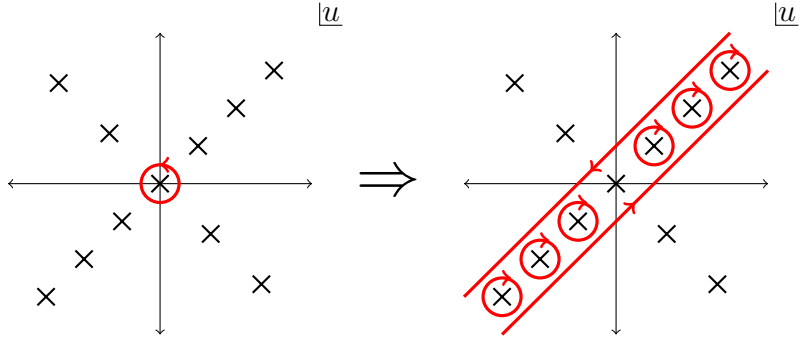

We then use the symmetry properties of the string amplitudes and recognize the sum over $m$ that appears as that which defines $\mathsf{A}_{0,3}^{(b)}$ to recast this as
\begin{align}
    p_1 \mathsf{A}_{g,1}^{(b)}(p_1) &\supset \frac{\pi}{2}\Res_{u=0}\bigg\{\frac{\sin(4\pi u p_1)}{\sin(2\pi b u)\sin(2\pi b^{-1} u)}\nonumber\\
    &\quad\times \int 2q \d q\, 2q' \d q'\, \cos(4\pi u q)\cos(4\pi u q')\mathsf{A}_{0,3}^{(b)}(p_1,q,q')\mathsf{A}_{g-1,2}^{(b)}(q,q')\bigg\}\, .
\end{align}
In order to simplify the representation, we would like to exchange the integral over $u$ that defines the residue with those over $q$ and $q'$. In order to do this, we rewrite
\begin{equation}\label{eq:residue rewrite}
    \Res_{u=0} = -\int_{b(\mathbb{R}+i\varepsilon)}\frac{du}{2\pi i} + \int_{b(\mathbb{R}-i\varepsilon)}\frac{du}{2\pi i} - \sum_{k\in\mathbb{Z}_{\ne 0}}\Res_{u= \frac{kb}{2}}
\end{equation}
as shown in figure \ref{fig:u contour deformation}. On the upper and lower parts of the contour, we can replace $\cos(4\pi u q)\cos(4\pi u q')$ by $\mathrm{e}^{4\pi i u (q+q')}$ and $\mathrm{e}^{-4\pi i u (q+q')}$, respectively. Now all integrals are convergent and we may exchange the $u$ integral with the $q,q'$ integrals. The $u$ integral may then be further simplified with a principal value prescription which picks up part of the residue at $u=0$ and cancels the remaining residues on the right-hand side of (\ref{eq:residue rewrite}). All together this gives
\begin{multline}
    p_1 \mathsf{A}_{g,1}^{(b)}(p_1) \supset  \int 2q \d q \, 2q' \d q'\, \underbrace{\left(\frac{p_1}{2} - \frac{1}{2}\int_\Gamma du\frac{\sin(4\pi u p_1)\sin(4\pi u(q+q'))}{\sin(2\pi b u)\sin(2\pi b^{-1} u)}\right)}_{\mathsf{H}_b(q+q',p_1)}\\
    \times \mathsf{A}_{0,3}^{(b)}(p_1,q,q')\mathsf{A}_{g-1,2}^{(b)}(q,q')\, ,
\end{multline}
where the contour $\Gamma$ is shown in figure \ref{fig:recursion kernel contour}. This is the first term in the final form of the recursive representation as written in (\ref{eq:Agn recursion relation}) in the case $n=1$. The other terms follow from essentially identical considerations that we will not spell out here.

\bibliographystyle{JHEP}
\bibliography{bib}
\end{document}